%\Overfullrule=0pt
 
\newcount\mgnf  %ingrandimento
\mgnf=0
 
\ifnum\mgnf=0
\def\openone{\leavevmode\hbox{\ninerm 1\kern-3.3pt\tenrm1}}%
\def\*{\vglue0.3truecm}\fi
\ifnum\mgnf=1
\def\openone{\leavevmode\hbox{\ninerm 1\kern-3.63pt\tenrm1}}%
\def\*{\vglue0.5truecm}\fi
 
\ifnum\mgnf=0
   \magnification=\magstep0
   \hsize=14truecm\vsize=24.truecm
   \parindent=0.3cm\baselineskip=18pt
\font\titolo=cmbx12
\font\titolone=cmbx10 scaled\magstep 2
\font\cs=cmcsc10
\font\ottorm=cmr8

\font\msytw=msbm10

\font\indbf=cmbx10 scaled\magstep1

\fi
\ifnum\mgnf=1
   \magnification=\magstep1\hoffset=0.truecm
   \hsize=14truecm\vsize=24.truecm
   \baselineskip=18truept plus0.1pt minus0.1pt \parindent=0.9truecm
   \lineskip=0.5truecm\lineskiplimit=0.1pt      \parskip=0.1pt plus1pt
\font\titolo=cmbx12 scaled\magstep 1
\font\titolone=cmbx10 scaled\magstep 3
\font\cs=cmcsc10 scaled\magstep 1
\font\ottorm=cmr8 scaled\magstep 1

\font\msytw=msbm10 scaled\magstep1

\font\indbf=cmbx10 scaled\magstep2
\fi
 
\global\newcount\numsec\global\newcount\numapp
\global\newcount\numfor\global\newcount\numfig\global\newcount\numsub
\numsec=0\numapp=0\numfig=1
\def\veroparagrafo{\number\numsec}\def\veraformula{\number\numfor}
\def\veraappendice{\number\numapp}\def\verasub{\number\numsub}
\def\verafigura{\number\numfig}
 
\def\section(#1,#2){\advance\numsec by 1\numfor=1\numsub=1%
\SIA p,#1,{\veroparagrafo} %
\write15{\string\Fp (#1){\secc(#1)}}%
\write16{ sec. #1 ==> \secc(#1)  }%
\hbox to \hsize{\titolo\hfill \number\numsec. #2\hfill%
\expandafter{\alato(sec. #1)}}\*}
 
\def\appendix(#1,#2){\advance\numapp by 1\numfor=1\numsub=1%
\SIA p,#1,{A\veraappendice} %
\write15{\string\Fp (#1){\secc(#1)}}%
\write16{ app. #1 ==> \secc(#1)  }%
\hbox to \hsize{\titolo\hfill Appendix A\number\numapp. #2\hfill%
\expandafter{\alato(app. #1)}}\*}
 
\def\senondefinito#1{\expandafter\ifx\csname#1\endcsname\relax}
 
\def\SIA #1,#2,#3 {\senondefinito{#1#2}%
\expandafter\xdef\csname #1#2\endcsname{#3}\else
\write16{???? ma #1#2 e' gia' stato definito !!!!} \fi}
 
\def \Fe(#1)#2{\SIA fe,#1,#2 }
\def \Fp(#1)#2{\SIA fp,#1,#2 }
\def \Fg(#1)#2{\SIA fg,#1,#2 }
 
\def\etichetta(#1){(\veroparagrafo.\veraformula)%
\SIA e,#1,(\veroparagrafo.\veraformula) %
\global\advance\numfor by 1%
\write15{\string\Fe (#1){\equ(#1)}}%
\write16{ EQ #1 ==> \equ(#1)  }}
 
\def\etichettaa(#1){(A\veraappendice.\veraformula)%
\SIA e,#1,(A\veraappendice.\veraformula) %
\global\advance\numfor by 1%
\write15{\string\Fe (#1){\equ(#1)}}%
\write16{ EQ #1 ==> \equ(#1) }}
 
\def\getichetta(#1){Fig. \verafigura%
\SIA g,#1,{\verafigura} %
\global\advance\numfig by 1%
\write15{\string\Fg (#1){\graf(#1)}}%
\write16{ Fig. #1 ==> \graf(#1) }}
 
\def\etichettap(#1){\veroparagrafo.\verasub%
\SIA p,#1,{\veroparagrafo.\verasub} %
\global\advance\numsub by 1%
\write15{\string\Fp (#1){\secc(#1)}}%
\write16{ par #1 ==> \secc(#1)  }}
 
\def\etichettapa(#1){A\veraappendice.\verasub%
\SIA p,#1,{A\veraappendice.\verasub} %
\global\advance\numsub by 1%
\write15{\string\Fp (#1){\secc(#1)}}%
\write16{ par #1 ==> \secc(#1)  }}
 
\def\Eq(#1){\eqno{\etichetta(#1)\alato(#1)}}
\def\eq(#1){\etichetta(#1)\alato(#1)}
\def\Eqa(#1){\eqno{\etichettaa(#1)\alato(#1)}}
\def\eqa(#1){\etichettaa(#1)\alato(#1)}
\def\eqg(#1){\getichetta(#1)\alato(fig. #1)}
\def\sub(#1){\0\palato(p. #1){\bf \etichettap(#1)\hskip.3truecm}}
\def\asub(#1){\0\palato(p. #1){\bf \etichettapa(#1)\hskip.3truecm}}
 
\def\equv(#1){\senondefinito{fe#1}$\clubsuit$#1%
\write16{eq. #1 non e' (ancora) definita}%
\else\csname fe#1\endcsname\fi}
\def\grafv(#1){\senondefinito{fg#1}$\clubsuit$#1%
\write16{fig. #1 non e' (ancora) definito}%
\else\csname fg#1\endcsname\fi}
\def\secv(#1){\senondefinito{fp#1}$\clubsuit$#1%
\write16{par. #1 non e' (ancora) definito}%
\else\csname fp#1\endcsname\fi}
 
\def\equ(#1){\senondefinito{e#1}\equv(#1)\else\csname e#1\endcsname\fi}
\def\graf(#1){\senondefinito{g#1}\grafv(#1)\else\csname g#1\endcsname\fi}
\def\secc(#1){\senondefinito{p#1}\secv(#1)\else\csname p#1\endcsname\fi}
\def\sec(#1){{\S\secc(#1)}}
 
\def\BOZZA{
\def\alato(##1){\rlap{\kern-\hsize\kern-1.2truecm{$\scriptstyle##1$}}}
\def\palato(##1){\rlap{\kern-1.2truecm{$\scriptstyle##1$}}}
}
 
\def\alato(#1){}
\def\galato(#1){}
\def\palato(#1){}

{\count255=\time\divide\count255 by 60 \xdef\hourmin{\number\count255}
        \multiply\count255 by-60\advance\count255 by\time
   \xdef\hourmin{\hourmin:\ifnum\count255<10 0\fi\the\count255}}
 
\def\oramin{\hourmin }
 
\def\data{\number\day/\ifcase\month\or gennaio \or febbraio \or marzo \or
aprile \or maggio \or giugno \or luglio \or agosto \or settembre
\or ottobre \or novembre \or dicembre \fi/\number\year;\ \oramin}
\setbox200\hbox{$\scriptscriptstyle \data $}
\footline={\rlap{\hbox{\copy200}}\tenrm\hss \number\pageno\hss}

\let\a=\alpha \let\b=\beta  \let\g=\gamma     \let\d=\delta  \let\e=\varepsilon
  \let\h=\eta   \let\th=\vartheta    \let\l=\lambda
\let\m=\mu    \let\n=\nu            \let\p=\pi      \let\r=\rho
\let\s=\sigma \let\t=\tau        
   \let\o=\omega 
\let\G=\Gamma \let\D=\Delta       
           
\let\O=\Omega 
 
\def\\{\hfill\break} \let\==\equiv

\let\io=\infty 

\let\0=\noindent \def\pagina{{\vfill\eject}}

\def\ie{\hbox{\it i.e.\ }}
\let\dpr=\partial 
\let\bs=\backslash
 
\def\tende#1{\,\vtop{\ialign{##\crcr\rightarrowfill\crcr
 \noalign{\kern-1pt\nointerlineskip}
 \hskip3.pt${\scriptstyle #1}$\hskip3.pt\crcr}}\,}
\def\otto{\,{\kern-1.truept\leftarrow\kern-5.truept\to\kern-1.truept}\,}
\def\fra#1#2{{#1\over#2}}
 
\def\PP{{\cal P}}\def\VV{{\cal V}}
\def\WW{{\cal W}}
\def\TT{{\cal T}}\def\NN{{\cal N}}\def\BB{{\cal B}}
\def\RR{{\cal R}}\def\LL{{\cal L}}
\def\DD{{\cal D}}\def\SS{{\cal S}}
 
\def\T#1{{#1_{\kern-3pt\lower7pt\hbox{$\widetilde{}$}}\kern3pt}}
\def\VVV#1{{\underline #1}_{\kern-3pt
\lower7pt\hbox{$\widetilde{}$}}\kern3pt\,}
\def\W#1{#1_{\kern-3pt\lower7.5pt\hbox{$\widetilde{}$}}\kern2pt\,}

\def\indica{\leaders \hbox to 0.5cm{\hss.\hss}\hfill}
\def\guida{\leaders\hbox to 1em{\hss.\hss}\hfill}
\mathchardef\oo= "0521

\def\pp{{\bf p}}\def\xx{{\bf x}}
\def\yy{{\bf y}}\def\kk{{\bf k}}
\def\dd{{\bf d}}\def\zz{{\bf z}}
 \def\bP{{\bf P}}\def\rr{{\bf r}}
\def\tt{{\bf t}}
\def\ss{{\underline \sigma}}\def\oo{{\underline \omega}}

\def\qed{\raise1pt\hbox{\vrule height5pt width5pt depth0pt}}

 \def\bh{{\bar h}}  

\def\indic{\hbox{\raise-2pt \hbox{\indbf 1}}}

\def\tdh{{\tilde h}}
 
\def\RRR{\hbox{\msytw R}}

 \def\ZZZ{\hbox{\msytw Z}}
 
\def\TTT{\hbox{\msytw T}}

%%% INSERIMENTO FIGURE ( se si usa DVIPS )
%
% Se si vuole utilizzare delle macro postscript personali, contenute
% nel file ini.ps, togliere il commento alla riga seguente
%\special{header=ini.pst}
%
% Il comando seguente inserisce una scatola contenente #3 in modo che
% l'angolo superiore sinistro occupi la posizione (#1,#2)
%
\def\ins#1#2#3{\vbox to0pt{\kern-#2 \hbox{\kern#1 #3}\vss}\nointerlineskip}
%
% Il comando seguente crea una scatola di dimensioni #1x#2 contenente
% il disegno descritto in #4.ps;
% in questo disegno si possono introdurre delle stringhe usando \ins
% e mettendo le istruzioni relative nell'argomento #3.
% Il file #4.ps contiene le istruzioni postscript, che devono essere scritte
% presupponendo che l'origine sia nell'angolo inferiore sinistro della
% scatola, mentre per il resto l'ambiente grafico e' quello standard.
% #5 deve essere della forma \eqg("nome simbolico").
%
% Le istruzioni postscript possono essere inserite nel file che contiene
% l'istruzione \insertplot, racchiudendole fra le istruzioni \initfig{#4}
% e \endfig; inoltre ogni riga deve cominciare con "write13<" e deve finire
% con ">". In questo modo si crea il file #4.ps relativo alla figura.
%
\newdimen\xshift \newdimen\xwidth \newdimen\yshift
 
\def\insertplot#1#2#3#4#5{\par%
\xwidth=#1 \xshift=\hsize \advance\xshift by-\xwidth \divide\xshift by 2%
\yshift=#2 \divide\yshift by 2%
\line{\hskip\xshift \vbox to #2{\vfil%
#3 \includegraphics{#4.ps}}\hfill \raise\yshift\hbox{#5}}}
 
\def\initfig#1{%
\catcode`\%=12\catcode`\{=12\catcode`\}=12
\catcode`\<=1\catcode`\>=2
\openout13=#1.ps}
 
\def\endfig{%
\closeout13
\catcode`\%=14\catcode`\{=1
\catcode`\}=2\catcode`\<=12\catcode`\>=12}
 
%%%%%   FIGURE fig51.ps
 
\initfig{fig51}
\write13<%!>
\write13<%%BoundingBox 0 0 300 150>
\write13<gsave .5 setlinewidth 40 20 260 {dup 0 moveto 140 lineto} for stroke
grestore>
\write13</punto { gsave  % uso: x1 y1 punto>
\write13<2 0 360 newpath arc fill stroke grestore} def>
\write13<40 75 punto>
\write13<60 75 punto>
\write13<80 75 punto>
\write13<100 75 punto 120 68 punto 140 61 punto 160 54 punto 180 47 punto 200
40 punto>
\write13<220 33 punto 240 26 punto 260 19 punto>
\write13<120 82.5 punto>
\write13<140 90 punto>
\write13<160 80 punto>
\write13<160 100 punto>
\write13<180 110 punto>
\write13<180 70 punto>
\write13<200 60 punto>
\write13<200 120 punto>
\write13<220 110 punto>
\write13<220 50 punto>
\write13<240 100 punto>
\write13<240 60 punto>
\write13<120 50 punto>
\write13<260 20 punto>
\write13<240 40 punto>
\write13<240 50 punto>
\write13<260 70 punto>
\write13<200 80 punto>
\write13<260 90 punto>
\write13<260 110 punto>
\write13<220 130 punto>
\write13<40 75 moveto 100 75 lineto 140 90 lineto 200 120 lineto 220 130
lineto>
\write13<200 120 moveto 240 100 lineto 260 110 lineto>
\write13<240 100 moveto 260 90 lineto>
\write13<140 90 moveto 180 70 lineto 200 80 lineto>
\write13<180 70 moveto 220 50 lineto 260 70 lineto>
\write13<220 50 moveto 240 40 lineto>
\write13<220 50 moveto 240 50 lineto>
\write13<100 75 moveto 260 20 lineto>
\write13<100 75 moveto 120 50 lineto stroke>
\write13<grestore>
\endfig

\openin14=\jobname.aux \ifeof14 \relax \else
\input \jobname.aux \closein14 \fi
\openout15=\jobname.aux

%\input fiat
%\BOZZA
{\baselineskip=12pt
\centerline{\titolone Renormalization group, hidden symmetries and}
\vskip.2truecm
\centerline{\titolone approximate Ward identities in the XYZ model, II.}
\vskip1.truecm
\centerline{{\titolo G. Benfatto{${}^\ast$}, V. Mastropietro}%
\footnote{${}^\ast$}{\ottorm Supported by MURST, Italy, and EC HCM contract
number CHRX-CT94-0460.\hfill\break
e-mail: benfatto@mat.uniroma2.it, mastropi@mat.uniroma2.it.}}
\centerline{Dipartimento di Matematica, Universit\`a di Roma ``Tor Vergata''}
\centerline{Via della Ricerca Scientifica, I-00133, Roma}
\vskip1.truecm
\line{\vtop{
\line{\hskip1.5truecm\vbox{\advance \hsize by -3.1 truecm
\0{\cs Abstract.}
{\it An expansion based on renormalization group methods for the spin
correlation function in the $z$ direction of the Heisenberg-Ising $XYZ$ chain
with an external magnetic field directed as the $z$ axis is derived. Moreover,
by using the hidden symmetries of the model, we show that the running coupling
constants are small, if the coupling in the $z$ direction is small enough,
that a critical index appearing in the correlation function is exactly
vanishing (because of an approximate Ward identity) and other properties, so
obtaining a rather detailed description of the $XYZ$ correlation function.}}
\hfill} }}
}
\vskip1.2truecm
\section(4,Introduction)

\sub(4.1) In a preceding paper [BeM1] we have derived an expansion, based
on renormalization group methods, for the ground state energy and the
effective potential of the Heisenberg-Ising $XYZ$ chain, whose hamiltonian
is written in terms of fermionic operators. The expansion is in terms of a
set of {\it running coupling constants} and two {\it renormalization
constants}, related with the spectral gap and the wave function
renormalization; the running coupling constants have to be small enough to
have convergence of the expansion.

In this paper we continue our analysis of the $XYZ$ model by writing an
expansion for the spin correlation function in the direction of the magnetic
field, see \sec(6). With respect to the ground state energy or the effective
potential expansion, two new renormalization constants appear, related with
the (fermionic) density renormalization.

In order to study the asymptotic behaviour of the spin correlation
function, one has to face two main problems. The first one is to show that
the running coupling constants indeed remain small if the coupling $J_3$
between spins in the direction of the magnetic field is small enough. The
second problem is to prove that one of the renormalization constants
corresponding to the density renormalization is almost equal to the square
of the wave function renormalization.
This last property is crucial to obtain the correct asymptotic behaviour of
the correlation function, since it is related to the vanishing of a critical
index.

Such properties are proved by writing the beta function governing the flow
of the renormalization constants or their ratio as the sum of several
terms. One has to prove that one of such terms is exactly vanishing at any
order; once that this is proved, the above properties follow if the
magnetic field is chosen properly, see \sec(5).
One recognizes that such contribution to the Beta function of the
$XYZ$ model is coinciding with the Beta function obtained by applying the
same renormalization group analysis to the Luttinger model. For such model
many symmetry properties are true, and in this sense we can speak of
``hidden symmetries'' for the $XYZ$ model; they are not enjoyed by the $XYZ$
hamiltonian, but the model is close, in a renormalization group sense, to a
model enjoying them.

A crucial role is played in our analysis by the local Gauge invariance, see
\sec(8); note however that, despite the fact that the Luttinger model
Hamiltonian is formally gauge invariant, the ultraviolet and infrared
cutoffs introduced to perform our renormalization group analysis have the
effect that gauge invariance is broken. Nevertheless we can derive an
approximate Ward identity (approximate as the gauge invariance is only
approximately true), which tells us that the ratio between the density
renormalization and the square of the wave function renormalization in the
Luttinger model is approximately one. Note that, if one uses the Ward
identity formally obtained by neglecting the cutoffs, one obtains a ratio
exactly equal to one. This means that the corresponding Luttinger model
beta function is vanishing (but the $XYZ$ beta function is not vanishing)
and we can prove that the related critical index appearing in the
correlation function asymptotic behaviour of the $XYZ$ model is {\it
exactly} vanishing.

We could proceed in a similar way and derive a suitable Ward identity to prove
that the Beta function for the running coupling constants appearing in the
Luttinger model is vanishing; this was done formally in [MD]. However we find
simpler to prove this property by using the explicit expression of the
Luttinger model correlation functions [BGM] based on the exact solution [ML];
this was done in [GS], [BGPS], [BM1].

Finally, in \sec(7) other hidden symmetries are exploited in order to prove
many properties about the correlation function.

The paper is not self-consistent; we use heavily the notations and the
results of [BeM1], to which we refer also for the general introduction on
the $XYZ$ model. We will denote equation (x.y) of [BeM1] by ({\bf I}x.y.).

\vskip1cm
\section(5,The flow of the running coupling constants)
 
\sub(5.1) The convergence of the expansion for the effective potential
is proved by theorems {\bf I}3.12, {\bf I}3.17 under the hypothesis that,
uniformly in $h\ge h^*$, the running coupling constants are small enough
and the bounds ({\bf I}2.98) and ({\bf I}3.88) are satisfied. In this section
we prove that, if $|\l|$ is small enough and $\nu$ is properly chosen,
the above conditions are indeed verified.
 
Let us consider first the bounds in ({\bf I}2.98). They immediately follow
from ({\bf I}3.91) and ({\bf I}3.92), by a simple inductive argument, if the
bounds ({\bf I}3.88) are verified and
$$\e_h\le \bar\e_0\le \bar\e\;,\quad\hbox{for\ }h>h^*\;,\Eq(5.1)$$
with $\bar\e_0$ small enough.
 
Let us now consider the bounds ({\bf I}3.88). By ({\bf I}2.83), ({\bf I}2.84), the
first of ({\bf I}2.89) and the third of ({\bf I}2.98), we get
$$\eqalignno{
{Z_{h-1}\over Z_h} &= 1+ z_h\;,&\eq(5.2)\cr
{\s_{h-1}\over \s_h} &= 1+{s_h/\s_h-z_h\over 1+z_h}\;.&\eq(5.3)\cr}$$
By explicit calculation of the lower order non zero terms contributing
to $z_h$ and $s_h/\s_h$, one can prove that
$$\eqalign{
z_h &= b_1 \l_h^2 + O(\e_h^3)\;,\quad b_1>0\;,\cr
s_h/\s_h &= -b_2 \l_h + O(\e_h^2)\;,\quad b_2>0\;,\cr}\Eq(5.4)$$
which imply ({\bf I}3.88), if $\bar\e_0$ is small enough, with a suitable
constant $c_1$ depending on the constant $c_0$ appearing in Theorem
{\bf I}3.12, since the value of $c_0$ is independent of $c_1$.
 
The equation \equ(5.2) and the definitions ({\bf I}2.109) allow to get the
following representation of the Beta function in terms of the tree
expansions ({\bf I}3.71):
$$\eqalignno{
\l_h &= \l_{h+1} + \left({1\over 1+z_h}\right)^2
\left[-\l_{h+1}(z_h^2+2z_h)+\sum_{n=2}^\io
\sum_{\t\in\TT_{h,n}} l_h(\t)\right]\;,&\eq(5.5)\cr
\d_h &= \d_{h+1} + {1\over 1+z_h} \left[-\d_{h+1}z_h +c^\d_0\l_1\d_{h,0}
+\sum_{n=2}^\io
\sum_{\t\in\TT_{h,n}} \Big( a_h(\t)-z_h(\t)\Big)\right]\;,&\eq(5.6)\cr
\n_h &= \g\n_{h+1} + {1\over 1+z_h} \left[-\g\n_{h+1}z_h+
c^\n_h\g^h\l_{h+1} +\sum_{n=2}^\io
\sum_{\t\in\TT_{h,n}} n_h(\t)\right]\;,&\eq(5.7)\cr}$$
where we have extracted the terms of first order in the running couplings and
we have extended to $h=+1$ the definition of $\l_h$ and $\d_h$, so that,
see ({\bf I}2.81),
$$\l_1=4\l\sin^2(p_F+\p/L)\;,\quad \d_1=-v_0\d^*\;.\Eq(5.8)$$
Note that the first order term proportional to $\l_{h+1}$ in the equation for
$\n_h$ is of size $\g^h$, while the similar term in the equation for $\d_h$ is
equal to zero, if $h<0$; moreover the constants $c^\n_0$ and $c^\l_h$ are
bounded uniformly in $L,\b$.
 
Hence, if we put $\vec a_h=(\d_h,\l_h)$, the Beta function can be written,
if condition \equ(5.1) is satisfied, with $\bar\e_0$ small enough, in the form
$$\eqalignno{
\l_{h-1}&=\l_h+\b^\l_h(\vec a_h,\nu_h;\ldots;\vec a_1,\n_1;u,\d^*)\;,
&\eq(5.9)\cr
\d_{h-1}&=\d_h+\b^\d_h(\vec a_h,\nu_h;\ldots;\vec a_1,\n_1;u,\d^*)\;,
&\eq(5.10)\cr
\nu_{h-1}&=\g\nu_h+\b^\n_h(\vec a_h,\nu_h;\ldots;\vec a_1,\n_1;u,\d^*)\;,
&\eq(5.11)\cr}$$
where $\b^\l_h$, $\b^\d_h$ and $\b_h^\n$ are functions of $\vec
a_h,\n_h,\ldots,\vec a_1,\n_1, u$, which can be easily bounded, by using
Theorem {\bf I}3.12, if the condition \equ(5.1) is verified.
Note that these functions depend on $\vec a_h,\n_h,\ldots,\vec a_1,\n_1, u$,
directly trough the endpoints of the trees, indirectly trough $z_h$ and
the quantities $Z_{h'}/Z_{h'-1}$ and $\s_{h'-1}(\kk')$, $h<h'\le 0$,
appearing in the tree expansions.
 
Let us define
$$\m_h=\sup_{k\ge h} \max\{|\l_k|,|\d_k|\}\;,\quad
\bar\l_h=\sup_{k\ge h} |\l_k|\;.\Eq(5.12)$$
We want to prove the following Lemma.
 
\* \sub(5.2) {\cs Lemma.} {\it Suppose that $u$ satisfies the condition
({\bf I}2.117) and let us consider the equation \equ(5.11) for fixed values of
$\vec a_h$, $Z_{h-1}$ and $\s_{h-1}(\kk')$, $\tdh\le h\le 1$,
satisfying the conditions
$$\eqalignno{
\m_h&\le\bar\e_1\le\bar\e_0\;,&\eq(5.13)\cr
a_0\g^{h-1}&\ge 4|\s_h|\;,&\eq(5.14)\cr
\g^{-c_0\m_h} &\le {\s_{h-1}\over \s_h} \le \g^{+c_0\m_h}\;,&\eq(5.15)\cr
\g^{-c_0\m_h^2} &\le {Z_{h-1}\over Z_h} \le \g^{+c_0\m_h^2}\;,&\eq(5.16)\cr}
$$
for some constant $c_0$.
 
Then, if $\bar\e_0$ is small enough, there exist some constants $\bar\e_1$,
$\h$, $\g'$, $c_1$, $B$, and a family of intervals $I^{(\bh)}$, $\tdh\le
\bh\le 0$, such that $\bar\e_1\le\bar\e_0$, $0<\h<1$, $1<\g'<\g$,
$I^{(\bh)}\subset I^{(\bh+1)}$, $|I^{(\bh)}| \le c_1 \bar\e_1
(\g')^{\bh}$ and, if $\n=\n_1\in I^{(\bh)}$,
$$|\n_h| \le B\bar\e_1 \left[\g^{-\fra12(h-\bh)} +\g^{\h h}
\right]\le\bar\e_0\;,\quad \bh\le h\le 1\;.\Eq(5.17)$$}
 
\*
\sub(5.3) {\it Proof.} Let us consider \equ(5.11),
for fixed values of $\vec a_h$, $Z_h/Z_{h-1}$ (hence of $z_h$) and
$\s_{h-1}(\kk')$, $\tdh\le h\le 1$, satisfying \equ(5.13)-\equ(5.16).
 
Note that, if $|\n_h|\le\bar\e_0$ for $\bh\le h\le 1$ and $\bar\e_0$ is small
enough, the r.h.s. of \equ(5.11) is well defined for $h=\bh$ and we can
write, by using \equ(5.7),
$$\n_{\bh-1}=\g\n_{\bh} + b_\bh + r_\bh\;,\Eq(5.18)$$
where $b_\bh=c^\n_{\bh-1}\g^{\bh-1}\l_\bh$ and $r_\bh$ collects all terms
of second or higher order in $\bar\e_0$.
 
Note also that, in the tree expansion of $n_h(\t)$, the dependence on
$\n_h,\ldots,\n_1$ appears only in the endpoints of the trees and there is no
contribution from the trees with $n\ge 2$ endpoints, which are only of type
$\n$ or $\d$, because of the support properties of the single scale
propagators. It follows, by using ({\bf I}3.91) and \equ(5.14)-\equ(5.16),
that
$$|r_\bh|\le c_2\m_\bh\bar\e_0\;.\Eq(5.19)$$
 
Let us now fix a positive constant $c$, consider the intervals
$$J^{(h)} = [-{b_h\over \g-1}-c\bar\e_1,-{b_h\over \g-1}+c\bar\e_1]\;.
\Eq(5.20)$$
and suppose that there is an interval $I^{(\bh)}$ such that, if $\n_1$
spans $I^{(\bh)}$, then $\n_\bh$ spans the interval $J^{(\bh+1)}$ and
$|\n_h|\le\bar\e_0$ for $\bh\le h\le 1$. Let us call $\tilde J^{(\bh)}$
the interval spanned by $\n_{\bh-1}$ when $\n_1$ spans $I^{(\bh)}$.
Equation \equ(5.18) can be written in the form
$$\n_{\bh-1}+{b_\bh\over \g-1}=\g\Big(\n_{\bh} + {b_\bh\over \g-1}\Big)
+ r_\bh\;.\Eq(5.21)$$
Hence, by using also the definition of $b_h$ and \equ(5.19), we see that
$$\eqalign{
&\min_{\n_1\in I^{(\bh)}} \left[\n_{\bh-1}+{b_\bh\over \g-1}\right] =\cr
&=\g \min_{\n_\bh\in J^{(\bh+1)}} \left[\n_\bh+{b_{\bh+1}\over \g-1}\right]
+ \min_{\n_1\in I^{(\bh)}} \left[r_\bh +{\g\over \g-1}(b_\bh-b_{\bh+1})
\right]\le\cr
&\le -\g c\bar\e_1+ c_2\bar\e_1\bar\e_0 +c_3\g^\bh \bar\e_1\;,\cr}\Eq(5.22)$$
for some constant $c_3$. In a similar way we can show that
$$\max_{\n_1\in I^{(\bh)}} \left[\n_{\bh-1}+{b_\bh\over \g-1}\right]
\ge \g c\bar\e_1- c_2\bar\e_1\bar\e_0- c_3\g^\bh \bar\e_1\;.\Eq(5.23)$$
It follows that, if $c$ is large enough and $\bar\e_0$ is small enough,
$J^{(\bh)}$ is strictly contained in $\tilde J^{(\bh)}$.
On the other hand, it is obvious that there is
a one to one correspondence between $\n_1$ and the sequence $\n_h$,
$\bh-1\le h\le 1$. Hence there is an interval $I^{(\bh-1)}\subset I^{(\bh)}$,
such that, if $\n_1$ spans $I^{(\bh-1)}$, then $\n_{\bh-1}$ spans the
interval $J^{(\bh)}$ and, if $\bar\e_1$ is small enough,
$|\n_h|\le\bar\e_0$ for $\bh-1\le h\le 1$.
 
The previous calculations also imply that the inductive hypothesis is verified
for $\bh=0$, so that we have proved that there exists a decreasing family of
intervals $I^{(\bh)}$, $\tdh\le \bh\le 0$, such that, if $\n=\n_1\in
I^{(\bh)}$, then the sequence $\n_h$ is well defined for $h\ge \bh$ and
satisfies the bound $|\n_h|\le \bar\e_0$.
 
The bound on the size of $I^{(\bh)}$ easily follows \equ(5.18) and
\equ(5.19). Let us denote by $\n_h$ and $\n'_h$, $\bh\le h\le 1$, the
sequences corresponding to $\n_1,\n'_1\in I^{(\bh)}$. We have
$$\n_{h-1}-\n'_{h-1}=\g (\n_h-\n'_h) + r_h-r'_h\;,\Eq(5.24)$$
where $r'_h$ is a shorthand for the value taken from $r_h$ in correspondence of
the sequence $\n'_h$.
Let us now observe that $r_h-r'_h$ is equal to $\g z_{h-1}(1+z_{h-1})^{-1}
(\n'_h-\n_h)$ plus a sum of terms, associated with trees, containing at least
one endpoint of type $\n$, with a difference $\n_k-\n'_k$, $k\ge h$, in place
of the corresponding running coupling, and one endpoint of type $\l$. Then,
if $|\n_k-\n'_k| \le |\n_h-\n'_h|$, $k\ge h$, we have
$$|\n_h-\n'_h| \le {|\n_{h-1}-\n'_{h-1}| \over \g} + C\bar\e_1 |\n_h-\n'_h|
\;.\Eq(5.25)$$
On the other hand, if $h=1$, this bound implies that $|\n_1-\n'_1|\le
|\n_0-\n'_0|$, if $\bar\e_1$ is small enough; hence it allows to show
inductively that, given any $\g'$, such that $1<\g'<\g$, if $\bar\e_1$
is small enough, then
$$|\n_1-\n'_1| \le \g'^{(\bh-1)} |\n_{\bh}-\n'_{\bh}|\;.\Eq(5.26)$$
Since, by definition, if $\n_1$ spans $I^{(\bh)}$, then $\n_\bh$ spans the
interval $J^{(\bh+1)}$, of size $2c\bar\e_1$, the size of $I^{(\bh)}$ is
bounded by $2c\bar\e_1 \g'^{(\bh-1)}$.
 
In order to complete the proof of Lemma \secc(5.2), we have still to prove
the bound \equ(5.17). Note that, if we iterate \equ(5.11), we can write,
if $\bh\le h\le 0$ and $\n_1\in I^{(\bh)}$,
$$ \n_h = \g^{-h+1} \left[ \n_1 + \sum_{k=h+1}^1 \g^{k-2}
\b_k^\n(\n_k,\ldots,\n_1)\right] \;,\Eq(5.27)$$
where now the functions $\b_\nu^{k}$ are thought as functions of $\n_k,\ldots,
\n_1$ only.
 
If we put $h=\bh$ in \equ(5.27), we get the following identity:
$$\n_1 = - \sum_{k=\bh+1}^1\g^{k-2}\b^\n_k(\n_k,\ldots,\n_1)
+\g^{\bh-1}\n_\bh\;.\Eq(5.28)$$
\equ(5.27) and \equ(5.28) are equivalent to
$$\n_h  = -\g^{-h}\sum_{k=\bh+1}^h \g^{k-1} \b^\n_k(\n_k,\ldots,\n_1)
+\g^{-(h-\bh)}\n_\bh\;,\qquad \bh< h\le 1\;.\Eq(5.29)$$
 
The discussion following \equ(5.18) implies that
$$|\b_k^\n(\n_k,\ldots,\n_1)|\le C\m_k\;,\Eq(5.30)$$
if $\bar\e_0$ is small enough. However
this bound it is not sufficient and we have to
analyze in more detail the structure of the functions $\b_h^\n$, by looking
in particular to the trees in the expansion of $n_h(\t)$, which have no
endpoint of type $\n$. Let us suppose that, given a tree with this property,
we decompose the propagators by using ({\bf I}2.99); we get a family of $C^n$
different contributions, which can be bounded as before, by using an argument
similar to that used in \S{\bf I}3.13. However, the terms containing only the
propagators $g^{(h')}_{L,\o}$ cancel out, for simple parity properties. On the
other hand, the terms containing at least one propagator $r_2^{(h_v)}$ or two
propagators $g^{(h_v)}_{\o,-\o}$ (the number of such propagators has to be
even) can be bounded by $(C\e_h)^n(|\s_h|/\g^h)^2$, by using ({\bf I}2.101) and
({\bf I}3.106). Analogously the terms with at least one propagator $r_1^{(h_v)}$
can be bounded by $(C\e_h)^n \g^{\h h}$, with some positive $\h<1$. In fact,
for these terms, by using ({\bf I}2.101), the bound can be improved by a factor
$\g^{h_v}\le \g^{\h h} \g^{\h(h_v-h)}$, for any positive $\h\le 1$, and the
bad factor $\g^{\h(h_v-h)}$ can be controlled by the sum over the scales, if
$\h$ is small enough, thanks to ({\bf I}3.111). Finally, the parity properties of
the propagators imply that the only term linear in the running couplings,
which contributes to $\n_h$, is of order $\g^h$. Hence, we can write
$$\b_h^\n=\m_h\sum_{k=h}^1 \n_k \tilde\beta_{h,k}^\n \g^{-2\h(k-h)}+
\m_h\e_h\left({|\s_h|\over\g^h}\right)^2 \hat\beta_h^\n+ \g^{\h h}
\m_h R^\n_h\;,\Eq(5.31)$$
where $|R^\n_h|,|\hat\beta_h^\n|,|\tilde\beta_{h,k}^\n|\le C$.
 
The factor $\g^{-2\h(k-h)}$ in the r.h.s. of \equ(5.31) follows from the
simple remark that the bound over all the trees contributing to $\n_h$,
which have at least one endpoint of fixed scale $k>h$, can be improved by a
factor $\g^{-\h'(k-h)}$, with $\h'$ positive but small enough. It is
sufficient to use again ({\bf I}3.111), which allows to extract such factor
from the r.h.s. before performing the sum over the scale indices, and to
choose $\h'=2\h$, which is possible if $\h$ is small enough.
 
Let us now observe that the sequence $\n_h$, $\bh< h\le 1$, satisfying
\equ(5.29) can be obtained as the limit as $n\to\io$ of the sequence
$\{\n_h^{(n)}\}$, $\bh< h\le 1$, $n\ge 0$, parameterized by $\n_\bh\in
J^{(\bh+1)}$ and defined recursively in the following way:
$$\eqalign{
\n_h^{(0)}  &= 0\;,\cr
\n_h^{(n)} &= -\g^{-h}\sum_{k=\bh+1}^h \g^{k-1} \b_k^\n(\n_k^{(n-1)},
\ldots,\n_1^{(n-1)})+ \g^{-(h-\bh)}\n_\bh
\;,\qquad n\ge 1\;.\cr}\Eq(5.32)$$
 
In fact, it is easy to show inductively, by using \equ(5.30), that, if
$\bar\e_1$ is small enough, $|\nu_h^{(n)}|\le C\bar\e_1\le \bar\e_0$,
so that \equ(5.32) is meaningful, and
$$\max_{h^*<h\le 1}|\nu_h^{(n)}-\nu_h^{(n-1)}|\le (C\bar\e_1)^n\;.\Eq(5.33)$$
In fact this is true for $n=1$ by \equ(5.30) and the fact that $\nu_h^{(0)}
=0$; for $n>1$ it follows trivially by the fact that
$\b_k^\n(\n_k^{(n-1)},\ldots,\n_1^{(n-1)})-
\b_k^\n(\n_k^{(n-2)},\ldots,\n_1^{(n-2)})$ can be written as a sum of terms
in which there are at least one endpoint of type $\n$, with a difference
$\nu_{h'}^{n-1}-\nu_{h'}^{n-2}$, $h'\ge k$, in place of the corresponding
running coupling, and one endpoint of type $\l$.
Then $\nu_h^{(n)}$ converges as $n\to\io$, for $\bh<h\le 1$, to a limit
$\nu_h$, satisfying \equ(5.29) and the bound $|\nu_h|\le \bar\e_0$, if
$\bar\e_1$ is small enough. Since the solution of the equations \equ(5.29)
is unique, it must coincide with the previous one.
 
Conditions \equ(5.14) and \equ(5.15) imply that
$${|\s_h|\over\g^h}={|\s_\bh|\over\g^\bh} {|\s_h|\over |\s_\bh|}\g^{\bh-h}
\le C\g^{-(h-\bh)(1-c_0\bar\e_1)}\;.\Eq(5.34)$$
Hence, if $\bar\e_1$ is small enough, by \equ(5.31),
$$|\b_k^\n|\le C\bar\e_1 \left[\sum_{m=k}^1 |\n_m| \g^{-2\h(m-k)}+
\bar\e_0 \g^{-\fra12(h-\bh)} + \g^{\h k}\right]\;.\Eq(5.35)$$
Hence, it is easy to show that there exists a constant $\bar c$ such that
$$\eqalign{
|\n_h^{(n)}|&\le \bar c\bar\e_1 \Big[\sum_{m=\bh+1}^h |\n_m^{(n-1)}|
\g^{-(h-m)} + \sum_{m=h+1}^1 |\n_m^{(n-1)}| \g^{-2\h(m-h)}+\cr
&+\bar\e_0 \g^{-\fra12(h-\bh)} + \g^{\h h}
+\g^{-(h-\bh)}\Big]\;.\cr}
\Eq(5.36)$$
 
Let us now suppose that, for some constant $c_{n-1}$,
$$|\n_m^{(n-1)}|\le c_{n-1}\bar\e_1\left[\g^{-\fra12(h-\bh)} + \g^{\h h}
\right]\le \bar\e_0\;,\Eq(5.37)$$
which is true for $n=1$, since $\n_m^{(0)}=0$, if $\bar\e_1$ is small enough.
Then, it is easy to verify that the same bound is verified by $\n_m^{(n)}$,
if $c_{n-1}$ is substituted with
$$c_n=\bar c(1+c_4 c_{n-1}\bar\e_1)\;,\Eq(5.38)$$
where $c_4$ is a suitable constant.
Hence, we can easily prove the bound \equ(5.17) for $\n_h=\lim_{n\to\io}
\n_h^{(n)}$, for $\bar\e_1$ small enough.
  
\*
\sub(5.5)
Let us now consider the equations \equ(5.9) and \equ(5.10), for a fixed,
arbitrary, sequence $\n_h$, $\bh\le h\le 1$, satisfying the bound \equ(5.17).
In order to study the corresponding flow, we compare
our model with an approximate model,
obtained by putting $u=\n=0$ and by substituting all the propagators with the
Luttinger propagator $g^{(k)}_{L,\o}(\xx;\yy)$, see ({\bf I}2.100). It is easy to
see that, in this model, $\s_h(\kk')=\n_h=0$, for any $h\le 1$, so that the
flow of the running couplings is described only by the equations
$$\eqalign{
\l^{(L)}_{h-1}&=\l^{(L)}_h+\b_h^{\l,L}(\vec a^{(L)}_h,\ldots,
\vec a^{(L)}_1;\d^*)\;,\cr
\d^{(L)}_{h-1}&=\d^{(L)}_h+\b_h^{\d,L}(\vec a^{(L)}_h,\ldots,
\vec a^{(L)}_1;\d^*)\;,\cr}\Eq(5.39)$$
where the functions $\b_h^{\l,L}$ and $\b_h^{\d,L}$ can be represented as
in \equ(5.5) and \equ(5.6), by suitably changing the definition of the trees
and of the related quantities $l_h(\t)$, $a_h(\t)$, $z_h(\t)$, which we
shall distinguish by a superscript $L$. Of course
Theorem {\bf I}3.12 applies also to the new model, which differs from the well
known Luttinger model only because the space variables are restricted to the
unit lattice, instead of the real axis.
 
Let us define, for $\a=\l,\d$,
$$r^\a_h(\vec a_h,\nu_h;\ldots;\vec a_1,\n_1;u,\d^*)=
\b^\a_h(\vec a_h,\nu_h;\ldots;\vec a_1,\n_1;u,\d^*)-
\b^{\a,L}_h(\vec a_h,\ldots,\vec a_1;\d^*)\;.\Eq(5.40)$$
Note that, in the r.h.s. of \equ(5.40), the function $\b^{\a,L}_h$ is
calculated at the values of $\vec a_{h'}$, $h\le h'\le 1$, which are the
solutions of the equations \equ(5.9) and \equ(5.10); these values are
of course different from those satisfying the equations \equ(5.39).
We shall prove the following Lemma.
 
\*
\sub(5.6) {\cs Lemma.} {\it Suppose that $u$ satisfies the condition
({\bf I}2.117), the sequence $\n_h$, $\bh\le h\le 1$, satisfies the bound
\equ(5.17) and $\d^*$ satisfies the condition
$$|-\d^* v_0+c_0^\d\l_1| \le |\l_1|\;,\Eq(5.41)$$
$c_0^\d$ being the constant appearing in the r.h.s. of \equ(5.6),
 
Then, if $\h$ is defined as in Lemma \secc(5.2) and $\m_h\le
\bar\e_0$ (hence \equ(5.1) is satisfied) and $\bar\e_0$ is small enough,
$$|r^\l_h| + |r^\d_h| \le C \bar\l_h^2 [\g^{-\fra12 (h-\bh)}+\g^{\h h}]
\;,\quad \bh\le h\le 0\;;\Eq(5.42)$$
$$|r^\l_1|\le C\l_1^2\;,\quad |r^\d_1|\le C|\l_1|\;.\Eq(5.43)$$
}
 
\* \sub(5.7) {\it Sketch of the proof.}
Note that all trees with $n\ge 2$ endpoints, contributing to the expansions in
the r.h.s. of the equations \equ(5.5)-\equ(5.7), may have an endpoint of type
$\n$ or $\d$ only if there are at least two endpoints of type $\l$; this claim
follows from the definition of localization and the support properties of the
single scale propagators.
The bound \equ(5.43) is an easy consequence of this remark,
equations \equ(5.5), \equ(5.6), condition \equ(5.41) and Theorem {\bf I}3.12.
 
We then consider $h\le 0$ and we define
$$\D z_h=z_h-z_h^L= {Z_{h-1}\over Z_h} - {Z^L_{h-1}\over Z^L_h}\;.\Eq(5.44)$$
Remember that all quantities in \equ(5.44) have to be considered as functions
of the same running couplings. Suppose now that
$$|\D z_k| \le c_0\m_k^2 [\g^{-\fra12 (k-\bh)}+\g^{\h k}
]\;,\quad h<k\le 0\;.\Eq(5.45)$$
We want to prove that this bound is verified also for $k=h$, together with
\equ(5.42). Since the proof will also imply that \equ(5.45) is verified
for $k=0$, we shall achieve the proof of Lemma \secc(5.6).
 
By using the decomposition ({\bf I}2.99) of the propagator, it is easy to see
that
$$r^\a_h=\sum_{i=1}^3 r^{\a,i}_h\;,\Eq(5.46)$$
where the quantities $r^{\a,i}_h$ are defined in the following way.
 
\*
\0 1) $r^{\a,1}_h$ is obtained from $\b_h^\a$ by restricting the sum over
the trees in the r.h.s. of \equ(5.5) and \equ(5.6) to those
containing at least one endpoint of type $\n$.
 
\0 2) $r^{\a,2}_h$ is obtained from $\b_h^\a$ by restricting the sum over the
trees to those containing no endpoint of type $\n$, and by substituting, in
each term contributing to the expansions appearing in the r.h.s. of \equ(5.5)
and \equ(5.6), at least one propagator with a propagator of type $r_1^{(h')}$
or $r_2^{(h')}$ (see ({\bf I}2.99)), $h\le h'\le 1$. Note that $z_h$ and all the
ratios $Z_k/Z_{k-1}$, $k>h$, appearing in the expansions are left unchanged.
 
\0 3) $r^{\a,3}_h$ is obtained by subtracting $\b^{\a,L}_h$ from the
expression we get, if we substitute all propagators appearing in the
expansions contributing to $\b^\a_h$ with Luttinger propagators and if we
eliminate all trees containing endpoints of type $\n$.
\*
 
By using \equ(5.17), ({\bf I}2.101) and \equ(5.34), it is easy to prove that
$r^{\a,1}_h$ and $r^{\a,2}_h$ satisfy a bound like \equ(5.42). The main point
is the remark, already used in the proof of Lemma \secc(5.2), that there is an
improvement of order $\g^{-\h'(k-h)}$, $0<\h'<1$, in the bound of the sum over
the trees with a vertex of fixed scale $k>h$. One has also to use a trick
similar to that of \S{\bf I}3.13, in order to keep the bound ({\bf I}3.94) on the
determinants, after the decomposition of the propagators. Finally, one has to
use the remark made at the beginning of this section in order to justify
the presence of $\bar\l_h^2$, instead of $\e_h^2$, in the
r.h.s. of \equ(5.42).
 
In order to prove that a bound like \equ(5.42) is satisfied also by
$r^{\a,3}_h$, one must first prove that the bound in \equ(5.45) is valid for
$k=h$, with the same constant $c_0$. This result can be achieved by decomposing
$\D z_h$ in a way similar to that used for $r^\a_h$; let us call $\D_i z_h$
the three corresponding terms. By proceeding as before, we can show that
$$|\D_1 z_h|+|\D_2 z_h| \le C \bar\l_h^2
[\g^{-\fra12 (h-\bh)}+\g^{\h h}]\;.\Eq(5.47)$$
 
Let us now consider $\D_3 z_h$; we can write $\D_3 z_h=\sum_{n=2}^\io
\sum_{\t\in \TT_{h,n}} \D_3 z_h(\t)$, with $\D_3 z_h(\t)=0$, if $\t$
contains endpoints of type $\n$, and $\D_3 z_h(\t)=\sum_{v\in\t} \bar
z_h(\t,v)$, where $\bar z_h(\t,v)=0$, if $v$ is an endpoint, otherwise $\bar
z_h(\t,v)$ is obtained from $z_h(\t)$ by selecting a family $V$ vertices,
which are not endpoints, containing $v$, and by substituting, for each $v'\in
V$, the factor $Z_{h_{v'}}/Z_{h_{v'}-1}$ with $Z_{h_{v'}}/Z_{h_{v'}-1}-
Z^L_{h_{v'}}/Z^L_{h_{v'}-1}$. By using \equ(5.2), we have
$$|Z_{h_v}/Z_{h_v-1}-Z^L_{h_v}/Z^L_{h_v-1}| \le C |\D
z_{h_v}|^2\;;\Eq(5.48)$$
hence it is easy to show that $\D_3 z_h$ can be bounded as in the proof of
Theorem {\bf I}3.12, by adding a sum over the non trivial vertices (whose
number is proportional to $n$) and, for each term of this sum, a factor
$$C c_0 \bar\l_h^2 [\g^{-\fra12 (h-\bh)}+\g^{\h h}]
\g^{\h(h_{\tilde v}-h)} (h_{\tilde v}-h_{\tilde v'})\;,\Eq(5.49)$$
where $\tilde v$ is the non trivial vertex corresponding to the selected term
and $\tilde v'$ is the non trivial vertex immediately preceding $\tilde v$ or
the root. Hence, we get
$$|\D_3 z_h| \le C c_0 \e_h^2 \bar\l_h^2
[\g^{-\fra12 (h-\bh)}+\g^{\h h}]\;,\Eq(5.50)$$
implying, together with \equ(5.47), the bound \equ(5.45) for $k=h$, if
$\bar\e_0$ is small enough and $c_0$ is large enough.
 
Given this result, it is possible to prove in the same manner that
$r^{\a,3}_h$ satisfies a bound like \equ(5.42). This completes the proof of
Lemma \secc(5.6).
 
\* \sub(5.8) Lemma \secc(5.6) allows to reduce the study of running couplings
flow to the same problem for the flow \equ(5.39). This one, in its turn, can
be reduced to the study of the beta function for the {\it Luttinger model},
see [BGM]. This model is exactly solvable, see [ML], and the Schwinger
functions can be exactly computed, see [BGM].
It is then possible to show, see [BGM], [BGPS], [GS], [BM1], that there
exists $\bar\e>0$, such that, if $|\vec a_h|\le \bar\e$,
$$|\bar\b^{\a,L}_h(\vec a_h,\ldots,\vec a_h)| \le C \m_h^2\g^{\h' h}
\;,\Eq(5.51)$$
where $\bar\b^{\a,L}_h(\vec a_h,...,\vec a_1)$, $\a=\l,\d$, denote the
analogous of the functions $\b^{\a,L}_h(\vec a_h,...,\vec a_1)$ for this model
and $0<\h'<1$. Note that in the l.h.s. of \equ(5.51) all running couplings
$\vec a_k$, $h\le k\le 1$, are put equal to $\vec a_h$ and that $\vec a_h$ can
take any value such that $|\vec a_h|\le \bar\e$, since $\vec a_h$ is a
continuous function of $\vec a_0$ and $\vec a_h=\vec a_0+O(\m_h^2)$, see
[BGPS].
 
We argue now that a bound like \equ(5.51) is valid also for the functions
$\b^{\a,L}_h$. In fact the Luttinger model differs from our approximate model
only because the space coordinates take values on the real axis, instead of
the unit lattice. This implies, in particular, that we have to introduce a
scale decomposition with a scale index $h$ going up to $+\io$. However, as it
has been shown in [GS], the effective potential on scale $h=0$ is well
defined; on the other hand, it differs from the effective potential on scale
$h=0$ of our approximate model only for the non local part of the interaction.
In terms of the representation ({\bf I}2.61) of $\VV^{(0)}(\psi^{(\le 0)})$, this
difference is the same we would get, by changing the kernels of the non local
terms (without qualitatively affecting their bounds) and the delta function,
which in the Luttinger model is defined as $L\b \d_{k,0}\d_{k_0,0}$, instead
of as in ({\bf I}2.62).
 
Note that the difference of the two delta functions has no effect on the local
part of $\VV^{(0)}(\psi^{(\le 0)})$, because of the support properties of
$\hat\psi^{(\le 0)}$, but it slightly affects the non local terms on any scale,
hence it affects the beta function; however, it is easy to show that this is a
negligible phenomenon. Let us consider in fact a particular tree $\t$ and a
vertex $v\in\t$ of scale $h_v$ with $2n$ external fields of space momenta
$k'_r$, $r=1,\ldots,2n$; the conservation of momentum implies that
$\sum_{r=1}^{2n}\s_r k'_r=2\p m$, with $m=0$ in the continuous model, but
$m$ arbitrary integer for the lattice model. On the other hand, $k'_r$ is of
order $\g^{h_v}$ for any $r$, hence $m$ can be different from $0$ only if $n$
is of order $\g^{h_v}$.
Since the number of endpoints following a vertex with $2n$ external
fields is greater or equal to $n-1$ and there is a small factor (of order
$\m_h$) associated with each endpoint, we get an improvement, in the bound of
the terms with $|m|>0$, with respect to the others, of a factor
$\exp(-C\g^{-h_v})$.
Hence, by using the usual arguments, it is easy to show that the
difference between the two beta functions is of order $\m_h^2\g^{\h h}$.
 
The previous considerations prove the following, very important, Lemma.
 
\*
\sub(5.9) {\cs Lemma.}
{\it There are $\bar\e_0$ and $\h'>0$, such that, if $|\m_h|\le \bar\e_0$,
$\a=\l,\d$ and $h\le 0$,
$$|\b^{\a,L}_h(\vec a_h,\ldots,\vec a_h)|\le C\bar\l_h^2 \g^{\h' h}
\;.\Eq(5.52)$$
}
 
\*
We are now ready to prove the following main Theorem on the running couplings
flow.
 
\*
\sub(5.10) {\cs Theorem.} {\it If $u\not=0$ satisfies the condition
({\bf I}2.117) and $\d^*$ satisfies the condition \equ(5.41),
there exist $\bar\e_3$ and a finite integer $h^*\le 0$,
such that, if $|\l_1|\le \bar\e_3$ and $\n$ belongs to a suitable interval
$I^{(h^*)}$, of size smaller than $c|\l_1| \g'^{h^*}$ for some constants $c$
and $\g'$, $1<\g'<\g$, then the running coupling constants are
well defined for $h^*-1\le h\le 0$ and $h^*$ satisfies the definition
({\bf I}2.116).
Moreover, there exist positive constants $c_i$, $i=1,\ldots,5$, such that
$$|\l_h-\l_1|\le c_1|\l_1|^{3/2}\;,\qquad |\d_h|\le c_1|\l_1|\;,
\Eq(5.53)$$
$$\g^{\l_1 c_2 h}< {\s_h\over \s_0}< \g^{\l_1 c_3 h}\;,\Eq(5.54)$$
$$\g^{-c_4\l_1^2 h} < Z_h < \g^{-c_5\l_1^2 h}\;,\Eq(5.55)$$
$$\max\left\{h_{L,\b},
{\log_\g \big({4\g a_0^{-1}\over 1+\d^*}|\sigma_0|\big)\over 1-\l_1 c_2}
\right\} \le h^* \le \max\left\{h_{L,\b},
{\log_\g \big({4\g a_0^{-1}\over 1+\d^*}|\sigma_0|\big)+1-\l_1 c_3\over
1-\l_1 c_3}\right\}\;.\Eq(5.56)$$
 
Finally, it is possible to choose $\d^*$ so that, for a suitable $\h>0$,
$$|\d_h|\le C |\l_1|^{3/2} [\g^{-\h(h-h^*)}+\g^{\h h}]\;.\Eq(5.57)$$
}
 
\*
\sub(5.11) {\it Proof.} We shall proceed by induction.
Equations \equ(5.5), \equ(5.6) and Lemma \secc(5.2) imply that, if $\l_1$
is small enough, there exists an interval $I^{(0)}$, whose size is of
order $\l_1$, such that, if $\n\in I^{(0)}$, then the bound \equ(5.17) is
satisfied, together with
$$|\l_0-\l_1|\le C|\l_1|^2\;,\quad |\d_0-\d_1|=|\d_0|\le C|\l_1|\;.
\Eq(5.58)$$
Let us now suppose that the solution of
\equ(5.9)-\equ(5.11) is well defined for $\bh\le h\le 0$ and satisfies the
conditions \equ(5.14)-\equ(5.17), for any $\n$ belonging to an interval
$I^{(\bh)}$, defined as in Lemma \secc(5.2).
This implies, in particular, that $h^*\le \bh$, see \equ(5.14) and
({\bf I}2.116). Suppose also that there exists a constant $c_0$, such that
$$\bar\l_{\bh}\le 2|\l_1|\;.\Eq(5.59)$$
 
We want to prove that all these conditions are verified also if $\bh$ is
substituted with $\bh-1$, if $\l_1$ is small enough.
The induction will be stopped as soon as the condition
\equ(5.14) is violated for some $\n\in I^{(\bh)}$. We shall put $\n$
equal to one of these values, so defining $h^*$ as equal to $\bh+1$.
 
The fact that the condition on $\n_1$ and the bound \equ(5.17) are verified
also if $\bh-1$ takes the place of $\bh$, follows from Lemma \secc(5.2),
since the condition \equ(5.13) follows from \equ(5.59), if $\l_1$ is
small enough.
There is apparently a problem in using this Lemma, since in its proof
we used the hypothesis that
the values of $\vec a_h$, $Z_{h-1}$ and $\s_{h-1}(\kk')$, $\bh\le h\le 1$,
are independent of $\n_1$. This is not true for the full flow, but the proof
of Lemma \secc(5.6) can be easily extended to cover this case. In fact, the
only part of the proof, where we use the fact that $\vec a_h$ is constant,
is the identity \equ(5.24), which should be corrected by adding to the r.h.s.
the difference $b_h-b'_h$. However, since $\l_1$ is independent of $\n_1$,
it is not hard to prove that $|b_h-b'_h|\le C|\n_h-\n'_h|$ and that the bound
on $r_h-r'_h$ does not change (qualitatively), if we take into account also
the dependence on $\n_1$ of the various quantities, before considered as
constant. Hence, the bound \equ(5.25) is left unchanged.
 
The conditions \equ(5.15) and \equ(5.16) follow immediately from
\equ(5.59) and \equ(5.2)-\equ(5.4). Hence, we still have to show only
that \equ(5.59) is verified also if $\bh$ is
substituted with $\bh-1$, if $\l_1$ is small enough.
 
By using \equ(5.39) and \equ(5.40), we have, if $\a=\l,\d$,
$$\a_{\bh-1}=\a_\bh + \b^{\a,L}_\bh(\vec a_\bh,\ldots,\vec a_\bh)+
\sum_{k=\bh+1}^1 D^\a_{\bh,k} + r^\a_\bh(\vec a_\bh,\nu_\bh;
\ldots;\vec a_1,\n_1;u)\;,\Eq(5.60)$$
where
$$D^\a_{h,k}=\b^{\a,L}_h(\vec a_h,\ldots,\vec a_h,\vec a_k,\vec a_{k+1},
\dots,\vec a_1)-
\b^{\a,L}_h(\vec a_h,\ldots,\vec a_h,\vec a_h,\vec a_{k+1},
\dots,\vec a_1)\;.\Eq(5.61)$$
On the other hand, it is easy to see that $D^\a_{h,k}$ admits a tree expansion
similar to that of $\b^{\a,L}_h(\vec a_h,\ldots,\vec a_1)$, with the property
that all trees giving a non zero contribution must have an endpoint of scale
$h$, associated with a difference $\l_k-\l_h$ or $\d_k-\d_h$. Hence, if $\h$
is the same constant of Lemma \secc(5.2) and Lemma \secc(5.6) and $h\le 0$,
$$|D^\a_{h,k}|\le C |\bar\l_h| \g^{-\h(k-h)} |\vec a_k-\vec a_h|
\;.\Eq(5.62)$$
 
Let us now suppose that $\bh\le h\le 0$ and that
there exists a constant $c_0$, such that
$$|\vec a_{k-1}-\vec a_k|\le c_0 |\l_1|^{3/2} [\g^{-\fra12 (k-\bh)}+
\g^{\th k}]\;,\quad h<k\le 0\;.\Eq(5.63)$$
where $\th=\min\{\h/2,\h'\}$, $\h'$ being defined as in Lemma \secc(5.9).
\equ(5.63) is certainly verified for $k=0$, thanks to \equ(5.5), \equ(5.6);
we want to show that it is verified also if $h$ is substituted with $h-1$,
if $\l_1$ is small enough.
 
By using \equ(5.60), \equ(5.62), \equ(5.42), \equ(5.52) and \equ(5.63),
we get
$$\eqalign{
|\vec a_{h-1}-\vec a_h| &\le C\bar\l_h^2 \g^{\h' h} +
C |\bar\l_h|^2 [\g^{-\fra12 (h-\bh)}+\g^{\h h}]+\cr
& +C c_0 |\bar\l_h|^{5/2}
\sum_{k=h+1}^1 \g^{-\h(k-h)} \sum_{h'=h+1}^k [\g^{-\fra12 (h'-h^*)}+
\g^{\th h'}]\;,\cr}\Eq(5.64)$$
which immediately implies \equ(5.63) with $h\to h-1$ and \equ(5.59)
with $\bh\to \bh-1$.
 
The bound \equ(5.64) implies also \equ(5.53), while the bounds \equ(5.54) and
\equ(5.55) are an immediate consequence of \equ(5.15), \equ(5.16) and an
explicit calculation of the leading terms;
\equ(5.56) easily follows from \equ(5.54) and the definition ({\bf I}2.116)
of $h^*$.
 
All previous results can be obtained uniformly in the value of $\d^*$, under
the condition \equ(5.41). However, by using \equ(5.63) with $\bh=h^*$, it is
not hard to prove, by an implicit function theorem argument (we omit the
details, which are of the same type of those used many times before), that one
can choose $\d^*$ so that
$$|\d_0|\le C|\l_1|^2\;,\quad \d_{h^*/2}=0\;,\Eq(5.65)$$
which easily implies \equ(5.57), for a suitable value of $\h$.
 
\pagina
 
\vskip1.cm
\section(6,The Correlation function)
 
\sub(6.1) The correlation function $\Omega_{L,\b,\xx}^3$, in terms of
fermionic operators, is given by
 
$$\Omega_{L,\b,\xx}^3= <a^+_\xx a^-_\xx a^+_0 a^-_0>_{L,\b}-
<a^+_\xx a^-_\xx>_{L,\b} <a^+_0 a^-_0>_{L,\b}=
{\partial^2 \SS(\phi)\over
\partial \phi(\xx)\phi({\bf 0})}|_{\phi=0}\;,\Eq(6.1)$$
where $\phi(\xx)$ is a bosonic external field, periodic in $x$ and $x_0$,
of period $L$ and $\b$, respectively, and
$$e^{\SS(\phi)}=\int P(d\psi^{(\le 1)})
e^{-\VV^{(1)}(\psi^{(\le 1)})+\int d\xx \phi(\xx)\psi^{(\le 1)+}_\xx
\psi^{(\le 1)-}_\xx}\;.\Eq(6.2)$$
 
Note that, because of the discontinuity at $x_0=0$ of the scale $1$ free
measure propagator $\tilde g^{(1)}_{\o,\o}$ in the limit $M\to\io$ (see
\S{\bf I}2.3), the product $\psi^{(\le 1)+}_\xx \psi^{(\le 1)-}_\xx$
has to be understood as $\psi^{(\le 0)+}_\xx \psi^{(\le 0)-}_\xx+
\lim_{\e\to 0^+} \psi^{(1)+}_{(x,x_0+\e)} \psi^{(\le 1)-}_{(x,x_0)}$.
Since this remark is important only in the explicit calculation of some
physical quantities, but does not produce any problem in the analysis of this
section, we shall in general forget it in the notation.
 
We shall evaluate the integral in the r.h.s. of \equ(6.2) in a way which
is very close to that used for the integration in ({\bf I}2.13). We introduce
the scale decomposition described in \S{\bf I}2.3 and we perform iteratively
the integration of the single scale fields, starting from the field of
scale $1$. The main difference is of course the presence in the interaction
of a new term, that we shall call $\BB^{(1)}(\psi^{(\le 1)},\phi)$; in
terms of the fields $\psi^{(\le 1)\s}_{\xx,\o}$, it can be written as
$$\BB^{(1)}(\psi^{(\le 1)},\phi)=
\sum_{\s_1,\s_2} \int d\xx e^{i\pp_F\xx(\s_1+\s_2)}\phi(\xx)
\psi^{(\le 1)\s_1}_{\xx,\s_1} \psi^{(\le 1)\s_2}_{\xx,-\s_2}\;.\Eq(6.3)$$
 
After integrating the fields $\psi^{(1)},...\psi^{(h+1)}$, $0\ge h\ge h^*$,
we find
$$e^{\SS(\phi)}=e^{-L\b E_h+S^{(h+1)}(\phi)}\int P_{Z_h,\s_h,C_h}(d\psi^{\le
h})e^{-\VV^{(h)}(\sqrt{Z_h}\psi^{(\le h)})+\BB^{(h)}
(\sqrt{Z_h}\psi^{(\le h)},\phi)}\;,\Eq(6.4)$$
where $P_{Z_h,\s_h,C_h}(d\psi^{(\le h)})$ and $\VV^{h}$ are given by ({\bf
I}2.66) and ({\bf I}3.3), respectively, while $S^{(h+1)}$ $(\phi)$, which
denotes the sum over all the terms dependent on $\phi$ but independent of
the $\psi$ field, and $\BB^{(h)}(\psi^{(\le h)}, \phi)$, which denotes the
sum over all the terms containing at least one $\phi$ field and two $\psi$
fields, can be represented in the form
$$S^{(h+1)}(\phi)=\sum_{m=1}^\io\int d\xx_1\cdots d\xx_m
S^{(h+1)}_m(\xx_1,\ldots,\xx_m)
\Big[\prod_{i=1}^m\phi(\xx_i)\Big]\Eq(6.5)$$
$$\eqalign{
&\BB^{(h)}(\psi^{(\le h)},\phi)=\sum_{m=1}^\io\sum_{n=1}^{\io} \sum_{\ss,\oo}
\int d\xx_1\cdots d\xx_m d\yy_1 \cdots d\yy_{2n} \;\cdot\cr
&\qquad\cdot\; B^{(h)}_{m,2n,\ss,\oo}(\xx_1,\ldots,\xx_m;\yy_1,\ldots,\yy_{2n})
\Big[\prod_{i=1}^m\phi(\xx_i)\Big] \Big[\prod_{i=1}^{2n}
\psi^{(\le h)\s_i}_{\yy_i,\o_i}\Big]\;.\cr}\Eq(6.6)$$
 
Since the field $\phi$ is equivalent, from the point of view of dimensional
considerations, to two $\psi$ fields, the only terms in the r.h.s. of
\equ(6.6) which are not irrelevant are those with $m=1$ and $n=1$, which are
marginal. However, if $\sum_{i=1}^2 \s_i\o_i\not=0$, also these terms are
indeed irrelevant, since the dimensional bounds are improved by the presence
of a non diagonal propagator, as for the analogous terms with no $\phi$ field
and two $\psi$ fields, see \S{\bf I}3.14. Hence we extend the definition of the
localization operator $\LL$, so that its action on $\BB^{(h)}(\psi^{(\le
h)},\phi)$ in described in the following way, by its action on the kernels
$B^{(h)}_{m,2n,\ss,\oo}(\xx_1,\ldots,\xx_m;\yy_1,\ldots,\yy_{2n})$:
 
\*
\0 1) if $m=1$, $n=1$ and $\sum_{i=1}^2 \s_i\o_i=0$, then
$$\eqalign{
&\LL B^{(h)}_{1,2,\ss,\oo}(\xx_1;\yy_1,\yy_2)=
\s_1\o_1\d(\yy_1-\xx_1)\d(\yy_2-\xx_1)\;\cdot\cr
&\qquad \cdot\;\int d\zz_1 d\zz_2
c_\b(2x_0-z_{10}-z_{20}) c_L(z_1-z_2) B^{(h)}_{1,2,\ss,\oo}
(\xx_1;\zz_1,\zz_2)\;;\cr}\Eq(6.7)$$
 
\0 2) in all the other cases
$$\LL B^{(h)}_{m,2n,\ss,\oo}(\xx_1,...\xx_m;\yy_1,...,\yy_{2n})=0\;.\Eq(6.8)$$
\*
 
Let us define, in analogy to definition ({\bf I}3.2), the Fourier transform of
$B^{(h)}_{1,2,\ss,\oo}(\xx_1;\yy_1,\yy_2)$ by the
equation
$$\eqalign{
&B^{(h)}_{1,2,\ss,\oo}(\xx_1;\yy_1,\yy_2) = \cr
&={1\over (L\b)^3}\sum_{\pp,\kk'_1,\kk'_2}
e^{i\pp\xx -i\sum_{r=1}^2 \s_r \kk'_r \yy_r}
\hat B^{(h)}_{1,2,\ss,\oo}(\pp,\kk'_1)
\d(\sum_{r=1}^2\s_r(\kk'_r+ \pp_F)-\pp)\;,\cr}\Eq(6.9)$$
where $\pp=(p,p_0)$ is summed over momenta of the form $(2\p n/L,2\p m/\b)$,
with $n,m$ integers. Hence \equ(6.7) can be written in the form
$$\eqalign{
\LL B^{(h)}_{1,2,\ss,\oo}(\xx_1;\yy_1,\yy_2)&=
\s_1\o_1\d(\yy_1-\xx_1)\d(\yy_2-\xx_1) e^{i\pp_F\xx(\s_1+\s_2)}\;\cdot\cr
&\cdot\; \fra14 \sum_{\h,\h'=\pm 1} \hat B^{(h)}_{1,2,\ss,\oo}
(\bar\pp_{\h'}+2\pp_F(\s_1+\s_2),\bar\kk_{\h,\h'})\;,\cr}\Eq(6.10)$$
where $\bar\kk_{\h,\h'}$ is defined as in ({\bf I}2.73) and
$$\bar\pp_{\h'}=\left(0,\h' {2\p\over \b}\right)\;.\Eq(6.11)$$
 
By using the symmetries of the interaction, as in \S{\bf I}2.4, it is easy
to show that
$$\LL \BB^{(h)}(\psi^{(\le h)},\phi)={Z^{(1)}_h\over Z_h} F_1^{(\le h)}
+{Z^{(2)}_h\over Z_h} F_2^{(\le h)}\;,\Eq(6.12)$$
where $Z^{(1)}_h$ and $Z^{(2)}_h$ are real numbers, such that
$Z^{(1)}_1=Z^{(2)}_1=1$ and
$$F_1^{(\le h)}=\sum_{\s=\pm 1}\int d\xx \phi(\xx)e^{2i\s \pp_F\xx}
\psi^{(\le h)\s}_{\xx,\s}\psi^{(\le h)\s}_{\xx,-\s}\;,\Eq(6.13)$$
$$F_2^{(\le h)}=\sum_{\s=\pm 1}\int d\xx \phi(x)
\psi^{(\le h)\s}_{\xx,\s}\psi^{(\le h)-\s}_{\xx,\s}\;.\Eq(6.14)$$
 
By using the notation of \S{\bf I}2.5, we can write the integral in the r.h.s.
of \equ(6.4) as
$$\eqalign{
&e^{-L\b t_h} \int P_{\tilde Z_{h-1},\s_{h-1},C_h}(d\psi^{(\le h)})
e^{-\tilde\VV^{(h)}(\sqrt{Z_h}\psi^{(\le h)})+\BB^{(h)}
(\sqrt{Z_h}\psi^{(\le h)},\phi)}\;=\cr
&= e^{-L\b t_h} \int P_{Z_{h-1},\s_{h-1},C_{h-1}}(d\psi^{(\le h-1)})\;\cdot\cr
&\cdot\; \int P_{Z_{h-1},\s_{h-1},\tilde f_h^{-1}}(d\psi^{(h)})
e^{-\hat\VV^{(h)}(\sqrt{Z_{h-1}}\psi^{(\le h)})+\hat\BB^{(h)}
(\sqrt{Z_{h-1}}\psi^{(\le h)},\phi)}\;,\cr}\Eq(6.15)$$
where $\hat\VV^{(h)}(\sqrt{Z_{h-1}}\psi^{(\le h)})$ is defined as in
({\bf I}2.107) and
$$\hat\BB^{(h)}(\sqrt{Z_{h-1}}\psi^{(\le h)},\phi)=
\BB^{(h)}(\sqrt{Z_{h}}\psi^{(\le h)},\phi)\;.\Eq(6.16)$$
$\BB^{(h-1)}(\sqrt{Z_{h-1}}\psi^{(\le h-1)},\phi)$ and $S^{(h)}(\phi)$
are then defined through the analogous of ({\bf I}2.110), that is
$$\eqalign{
&e^{-\VV^{(h-1)}(\sqrt{Z_{h-1}}\psi^{(\le h-1)})+\BB^{(h-1)}
(\sqrt{Z_{h-1}}\psi^{(\le h-1)},\phi)-L\b\tilde E_h+\tilde S^{(h)}(\phi)}=\cr
&=\int P_{Z_{h-1},\s_{h-1},\tilde f_h^{-1}}(d\psi^{(h)})
e^{-\hat\VV^{(h)}(\sqrt{Z_{h-1}}\psi^{(\le h)})+\hat\BB^{(h)}
(\sqrt{Z_{h-1}}\psi^{(\le h)},\phi)}\;.\cr}\Eq(6.17)$$
 
The definitions \equ(6.16) and \equ(6.12) easily imply that
$${Z^{(i)}_{h-1}\over Z^{(i)}_h} = 1 + z^{(i)}_h\;,\quad i=1,2\;,\Eq(6.18)$$
where $z^{(1)}_h$ and $z^{(2)}_h$ are some quantities of order $\e_h$, which
can be written in terms of a tree expansion similar to that described in
\S{\bf I}3, as we shall explain below.
 
As in \S{\bf I}3, the fields of scale between $h^*$ and $h_{L,\b}$ are integrated
in a single step, so we define, in analogy to ({\bf I}3.125),
$$\eqalign{
&e^{\tilde S^{(h^*)}(\phi)-L\b\tilde E_{h^*}}=\cr
&\int P_{Z_{h^*-1},\s_{h^*-1},C_{h^*}}(d\psi^{(\le h^*)})
e^{-\hat\VV^{(h^*)}(\sqrt{Z_{h^*-1}}\psi^{(\le h^*)})+\hat\BB^{(h^*)}
(\sqrt{Z_{h^*-1}}\psi^{(\le h^*)},\phi)}\;.\cr}\Eq(6.19)$$
It follows, by using ({\bf I}3.126), that
$$S(\phi)= -L\b E_{L,\b}+ S^{(h)}(\phi)=
-L\b E_{L,\b}+ \sum_{h=h^*}^1 \tilde S^{(h)}(\phi)\;;\Eq(6.20)$$
hence, by \equ(6.1)
$$\Omega_{L,\b,\xx}^3 = S^{(h)}_2 (\xx,0)=
\sum_{h=h^*}^1 \tilde S^{(h)}_2 (\xx,0)\;.\Eq(6.21)$$
 
\* \sub(6.2)
The functionals $\BB^{(h)}(\sqrt{Z_{h}}\psi^{(\le h)},\phi)$ and
$S^{(h)}(\phi)$ can be written in terms of a tree expansion similar to the
one described in \S(3.2). We introduce, for each $n\ge 0$ and each $m\ge 1$, a
family $\TT^m_{h,n}$ of trees, which are defined as in \S(3.2), with some
differences, that we shall explain.
 
\*
1) First of all, if $\t\in \TT^m_{h,n}$, the tree has $n+m$ (instead of
$n$) endpoints. Moreover, among the $n+m$ endpoints, there are $n$ endpoints,
which we call {\it normal endpoints}, which are associated with a contribution
to the effective potential on scale $h_v-1$. The $m$ remaining endpoints,
which we call {\it special endpoints}, are associated with a local term of the
form \equ(6.13) or \equ(6.14); we shall say that they are of type $Z^{(1)}$
or $Z^{(2)}$, respectively.
 
2) We associate with each vertex $v$ a new integer $l_v\in[0,m]$, which
denotes the number of special endpoints following $v$, \ie contained in $L_v$.
 
3) We introduce an {\it external field label} $f^\phi$ to distinguish the
different $\phi$ fields appearing in the special endpoints. $I_v^\phi$ will
denote the set of external field labels associated with the endpoints
following the vertex $v$; of course $l_v=|I_v^\phi|$ and $m=|I_{v_0}^\phi|$.
 
\*
 
These definitions allow to represent $\BB^{(h)}(\sqrt{Z_{h}}\psi^{(\le
h)},\phi)+S^{(h+1)}(\phi)$ in a way similar to that described in detail in
\S{\bf I}3.3-3.11. It is sufficient to extend in an obvious way some
notations and some procedures, in order to take into account the presence
of the new terms depending on the external field and the corresponding
localization operation.
 
In particular, if $l_v\not=0$, the $\RR$ operation associated with the vertex
$v$ can be deduced from \equ(6.7) and \equ(6.8) and can be represented as
acting on the kernels or on the fields in a way similar to what we did in
\S{\bf I}3.1. We will not write it in detail; we only remark that such definition
is chosen so that, when $\RR$ is represented as acting on the fields, no
derivative is applied to the $\phi$ field.
 
All the considerations in \S{\bf I}3.2, up to the modifications listed above, can
be trivially repeated. The same is true for the definition of the labels
$r_v(f)$, described in \S{\bf I}3.3. One has only to consider, in addition to the
cases listed there, the case in which $|P_v|=2$ and $l_v=1$; in such a case,
if there is no non trivial vertex $v'$ such that $v_0\le v'<v$, we make an
arbitrary choice, otherwise we put $r_v(f)=1$ for the $\psi$ field which is an
internal field in the nearest non trivial vertex preceding $v$. As in
\S{\bf I}3.2, this is sufficient to avoid the proliferation of $r_v$ indices.
 
Also the considerations in \S{\bf I}3.4-{\bf I}3.7 can be adjusted without any
difficulty. It is sufficient to add to the three items listed after ({\bf I}3.69)
the case $l_{v_0}=1$, $P_{v_0}=(f_1,f_2)$, by noting that in this case the
action of $\RR$ consists in replacing one external $\psi$ field with a
$D^{11}_{\yy,\xx}$ field.
 
\*
\sub(6.3)
Let us consider in more detail the representation we get for the constants
$z^{(l)}_h$, $l=1,2$, defined in \equ(6.18). We have
$$z^{(l)}_h = \sum_{n=1}^\io
\sum_{\t\in\TT^1_{h,n},\bP\in\PP_\t,\rr: P_{v_0}=(f_1,f_2),
\atop \s_1=\o_1=(-1)^{l-1}\s_2=(-1)^l\o_2=+1}
\sum_{T\in {\bf T}} \sum_{\a\in A_T\atop q_\a(P_{v_0})=0}
z^{(l)}_h(\t,\bP,\rr,T,\a)\;,\Eq(6.22)$$
where, if $\xx$ is the space time point associated with the special endpoint,
$$\eqalign{
&z^{(l)}_h(\t,\bP,\rr,T,\a)=
\Big[\prod_{v\,\hbox{\ottorm not e.p.}}
\Big(Z_{h_v}/Z_{h_v-1}\Big)^{|P_v|/2}\Big]\;\cdot\cr
&\cdot \int d(\xx_{v_0}\bs\xx) h_\a(\xx_{v_0})
\Big[\prod_{i=1}^n d_{j_\a(v^*_i)}^{b_\a(v^*_i)}(\xx_i,\yy_i)
K^{h_i}_{v^*_i}(\xx_{v^*_i})\Big]
\Big\{\prod_{v\,\hbox{\ottorm not e.p.}}{1\over s_v!} \int
dP_{T_v}(\tt_v) \;\cdot\cr
&\cdot\; \det G_\a^{h_v,T_v}(\tt_v)
\Big[\prod_{l\in T_v} \hat\partial^{q_\a(f^-_l)}_{j_\a(f^-_l)}
\hat\partial^{q_\a(f^+_l)}_{j_\a(f^+_l)} [d^{b_\a(l)}_{j_\a(l)}(\xx_l,\yy_l)
\bar\dpr_1^{m_l} g^{(h_v)}_{\o^-_l,\o^+_l}(\xx_l-\yy_l)]\Big]\Big\}\;.\cr}
\Eq(6.23)$$
 
The notations are the same as in \S{\bf I}3.10 and we can derive for
$z^{(l)}_h(\t,\bP,\rr,T,\a)$ a bound similar to ({\bf I}3.110), without the
volume factor $L\b$ (the integration over $x_{v_0}$ is done keeping $\xx$
fixed). The only relevant
difference is that the bounds ({\bf I}3.83) and ({\bf I}3.107) have to be modified,
in order to take into account the properties of the extended localization
operation, by substituting $z(P_v)$ and $\tilde z(P_v)$ with $z(P_v,l_v)$ and
$\tilde z(P_v,l_v)$, respectively, with
$$z(P_v,l_v)=\cases{
1 & if $|P_v|=4$, $l_v=0$\cr
1 & if $|P_v|=2$, $l_v=0$ and $\sum_{f\in P_v} \o(f)\not=0\;,$\cr
2 & if $|P_v|=2$, $l_v=0$ and $\sum_{f\in P_v} \o(f)=0\;,$\cr
1 & if $|P_v|=2$, $l_v=1$ and $\sum_{f\in P_v} \s(f)\o(f)=0\;,$\cr
0 & otherwise.\cr}\Eq(6.24)$$
$$\tilde z(P_v,l_v)=\cases{
1 & if $|P_v|=2$, $l_v=0$ and $\sum_{f\in P_v} \o(f)\not=0\;,$\cr
1 & if $|P_v|=2$, $l_v=1$ and $\sum_{f\in P_v} \s(f)\o(f)\not=0\;,$\cr
0 & otherwise.\cr}\Eq(6.25)$$
It follows that
$$\eqalign{
&|z^{(l)}_h(\t,\bP,\rr,T,\a)| \le C^n \e_h^n \g^{-h [D_0(P_{v_0})+l_{v_0}]}
\prod_{v\,\hbox{\ottorm not e.p.}}\Big\{ C^{\sum_{i=1}^{s_v}|P_{v_i}|-|P_v|}
\;\cdot\cr
&\cdot\;{1\over s_v!} \Big(Z_{h_v}/Z_{h_v-1}\Big)^{|P_v|/2}
\g^{-[-2+{|P_v|\over 2}+l_v+z(P_v,l_v)+{\tilde z(P_v,l_v)\over 2}]}
\Big\}\;,\cr}\Eq(6.26)$$
with
$$-2+{|P_v|\over 2}+l_v+z(P_v,l_v)+{\tilde z(P_v,l_v)\over 2}\ge \fra12\;,
\quad\forall v\, \hbox{\ottorm not e.p.}\;.\Eq(6.27)$$
 
Hence, we can proceed as in \S{\bf I}3.14 and, since $D_0(P_{v_0})+l_{v_0}=0$, we
can easily prove the following Theorem.
 
\*
\sub(6.4) {\cs Theorem.} {\it Suppose that $u\not=0$ satisfies the
condition ({\bf I}2.117), $\d^*$ satisfies the condition \equ(5.41),
$\bar\e_3$ is defined as in Theorem \secc(5.10)
and $\n\in I^{(h^*)}$. Then, there exist two constants $\bar\e_4\le
\bar\e_3$ and $c$, independent of $u$, $L$, $\b$,
such that, if $|\l_1|\le \bar\e_4$, then
$$|z^{(l)}_h| \le c|\l_1|\;,\quad 0\le h\le h^*\;.\Eq(6.28)$$
}
 
\* \sub(6.5) Theorem \secc(6.4), the bound \equ(5.55) on $Z_h$, the
definition \equ(6.18) and an explicit first order calculation of
$z^{(1)}_h$ imply that there exist two positive constants $c_1$ and
$c_2$, such that
$$\g^{-c_2\l_1 h} \le {Z^{(1)}_h\over Z_h}\le \g^{-c_1\l_1 h}\;.\Eq(6.29)$$
 
A similar bound is in principle valid also for $Z^{(2)}_h/ Z_h$, but
we shall prove that a much stronger bound is verified, by comparing
our model with the Luttinger model. First of all, we consider
an approximated Luttinger model, which is similar to that introduced in
\sec(5.5). It is obtained from the original model by substituting the free
measure and the potential with the following expressions, where we use the
notation of \S{\bf I}2:
$$\eqalign{
&P^{(L)}(d\psi^{(\le 0)}) =
\prod_{\kk':C_0^{-1}(\kk')>0}\prod_{\o=\pm1}
{d\hat\psi^{(\le 0)+}_{\kk',\o} d\hat\psi^{(\le 0)-}_{\kk',\o}\over
\NN_L(\kk')}\;\cdot\cr
&\cdot\; \exp \left\{-{1\over L\b} \sum_{\o=\pm 1}
\sum_{\kk':C_0^{-1}(\kk')>0} C_0(\kk') \big( -ik_0+\o v_0^* k' \big)
\hat\psi^{(\le 0)+}_{\kk',\o} \hat\psi^{(\le 0)-}_{\kk',\o}\right\}\;,\cr}
\Eq(6.30)$$
$$\eqalign{
V^{(L)}(\psi^{(\le 0)}) &= \l_0^{(L)} \int_{\TTT_{L,\b}} d\xx\;
\psi^{(\le 0)+}_{\xx,+1}\psi^{(\le 0)-}_{\xx,-1}
\psi^{(\le 0)+}_{\xx,-1}\psi^{(\le 0)-}_{\xx,+1}+\cr
&+ \d_0 ^{(L)} \sum_{\o=\pm 1} i\o \int_{\TTT_{L,\b}} d\xx\;
\psi^{(\le h)+}_{\xx,\o} \dpr_x \psi^{(\le h)-}_{\xx,\o}\;,\cr}
\Eq(6.31)$$
where $\NN_L(\kk')=C_0(\kk')(L\b)^{-1}[k_0^2+(v_0^* k')^2]^{1/2}$,
$\l_0^{(L)}$ and $\d_0^{(L)}$ have the role of the running couplings on scale
$0$ of the original model, but are not necessarily equal to them,
$\TTT_{L,\b}$ is the (continuous, as in \S{\bf I}3.15) torus $[0,L]\times [0,\b]$
and $\psi^{(\le 0)}$ is the (continuous) Grassmanian field on $\TTT_{L,\b}$
with antiperiodic boundary conditions. Moreover, the interaction with the
external field $\BB^{(1)}(\psi^{(\le 1)},\phi)$ is substituted with the
corresponding expression on scale $0$, deprived of the irrelevant terms, that
is
$$\BB^{(0)}(\psi^{(\le 0)},\phi)= \sum_{\s=\pm 1}\int d\xx \phi(\xx)
\left( e^{2i\s \pp_F\xx} \psi^{(\le h)\s}_{\xx,\s}\psi^{(\le h)\s}_{\xx,-\s}
+\psi^{(\le h)\s}_{\xx,\s}\psi^{(\le h)-\s}_{\xx,\s}\right)\;.\Eq(6.32)$$
We shall call $Z^{(2,L)}_h$, $z^{(2,L)}_h$, $Z^{(L)}_h$ and $z^{(L)}_h$ the
analogous of $Z^{(2)}_h$, $z^{(2)}_h$, $Z_h$ and $z_h$ for this approximate
Luttinger model.
 
We want to compare the flow of $Z^{(2,L)}_h/Z^{(L)}_h$ with the flow of
$Z^{(2)}_h/Z_h$; hence we write
$${Z^{(2)}_{h-1}\over Z_{h-1}}={Z^{(2)}_h\over Z_h}
\Big[1 + \b^{(2)}(\vec a_h,\nu_h;\ldots;\vec a_1,\n_1;u,\d^*)\Big]
\;,\Eq(6.33)$$
$${Z^{(2,L)}_{h-1}\over Z^{(L)}_{h-1}}={Z^{(2,L)}_h\over Z^{(L)}_h}
\Big[1 + \b^{(2,L)}(\vec a^{(L)}_h,\ldots;\vec a^{(L)}_0,\d^*)\Big]
\;,\Eq(6.34)$$
where $a^{(L)}_h$ are the running couplings in the approximated Luttinger
model (by symmetry $\n^{(L)}_h=0$, since $\n=0$, see \sec(5.5)),
$1 + \b^{(2)}=(1+z^{(2)}_h)/(1+z_h)$ and
$1 + \b^{(2,L)}=(1+z^{(2,L)}_h)/(1+z^{(L)}_h)$.
 
The Luttinger model has a special symmetry, the {\it local
gauge invariance}, which allows to prove many {\it Ward identities}. As we
shall prove in \sec(8), the approximate Luttinger model satisfies some
approximate version of these identities and one of them implies that, if
$|\d_*+(\d_0^{(L)}/v_0)|\le 1/2$,
$$\g^{-C|\l_0^{(L)}|} \le {Z^{(2,L)}_h\over Z^{(L)}_h}\le
\g^{C|\l_0^{(L)}|}\;.\Eq(6.35)$$
 
By proceeding as in the proof of \equ(5.51) (see [BGPS], \S7), one can show
that \equ(6.35) implies that there exists $\bar\e>0$ and $\h'<1$, such that,
if $|\vec a_h|\le \bar\e$,
$$|\b^{(2,L)}_h(\vec a_h,\ldots,\vec a_h,\d^*)| \le C \m_h^2\g^{\h' h}
\;.\Eq(6.36)$$
 
{\bf Remark - }
The analogous bound \equ(5.51) was obtained in [BGPS] by a
comparison with the exact solution of the Luttinger model; this was
possible, thanks to the proof given in [GS] that the effective potential on
scale $0$ is well defined also in the Luttinger model, a non trivial result
because of the ultraviolet problem. This procedure would be much harder in
the case of the bound \equ(6.36), because the density is not well defined
in the Luttinger model, see \S{\bf I}1.3. In any case, the bound \equ(6.35),
whose proof is relatively simple, allows to get very easily the same
result.
 
One can also show, as in the proof of Lemma \secc(5.6), that
$$|\b^{(2)}(\vec a_h,\nu_h;\ldots;\vec a_1,\n_1;u,\d^*)-
\b^{(2,L)}(\vec a_h,,\ldots;\vec a_0,\d^*)|\le
C \bar\l_h^2 [\g^{-\fra12 (h-h^*)}+\g^{\h h}]\;,\Eq(6.37)$$
for any $h\ge h^*$ and for some $\h<1$.
 
Note that, in \equ(6.37), $\b^{(2,L)}$ is evaluated at the values of the
running couplings $\vec a_h$ of the original model; this is meaningful, since
in \equ(6.36) $\vec a_h$ can take any value such that $|\vec a_h|\le \bar\e$;
this follows from the remark, already used in \sec(5.8), that $\vec a_h^{(L)}$
is a continuous function of $\vec a_0^{(L)}$ and $\vec a_h^{(L)}=\vec
a_0^{(L)}+O(\m_h^2)$, see also [BGPS].
 
By using \equ(6.36) and \equ(6.37) and proceeding as in the proof of
Theorem \secc(5.10), one can easily prove the following Theorem.
 
\*
\sub(6.6) {\cs Theorem.} {\it If the hypotheses of Theorem \secc(6.4)
are verified, there exists a positive constant $c_1$, independent of $u$,
$L$, $\b$, such that
$$\g^{-c_1|\l_1|} \le {Z^{(2)}_h\over Z_h}\le \g^{c_1|\l_1|}
\;.\Eq(6.38)$$}
 
\* \sub(6.7) We are now ready to study the expansion of the correlation
function $\O_{L,\b}^3(\xx)$, which follows from \equ(6.21) and the
considerations of \sec(6.2). We have to consider the trees with two special
endpoints, whose space-points we shall denote $\xx$ and $\yy={\bf 0}$;
moreover, we shall denote by $h_\xx$ and $h_\yy$ the scales of the two
special endpoints and by $h_{\xx,\yy}$ the scale of the smallest cluster
containing both special endpoints. Finally $\TT^2_{h,n,l}$ will denote the
family of all trees belonging to $\TT^2_{h,n}$, such that the two special
endpoints are both of type $Z^{(1)}$, if $l=1$, both of type $Z^{(2)}$, if
$l=2$, one of type $Z^{(1)}$ and the other of type $Z^{(2)}$, if $l=3$.
 
If we extract from the expansion the contribution of the trees with one
special endpoint and no normal endpoints, we can write
$$\eqalign{
&\Omega^3_{L,\b}(\xx) =\sum_{h,h'=h^*}^1 \sum_{\s=\pm 1} \Big\{
e^{2 i\s p_F x} \cdot\cr
&\cdot {(Z_{h\vee h'}^{(1)})^2\over Z_{h-1} Z_{h'-1}}
[g_{\s,\s}^{(h)}(-\s\xx) g_{-\s,-\s}^{(h')}(-\s\xx)-
g_{+1,-1}^{(h)}(-\s\xx) g_{-1,+1}^{(h')}(-\s\xx)]+\cr
&+{(Z_{h\vee h'}^{(2)})^2\over Z_{h-1} Z_{h'-1}}
[-g_{\s,\s}^{(h)}(-\s\xx) g_{\s,\s}^{(h')}(\s\xx)+
g_{-1,+1}^{(h)}(-\s\xx) g_{+1,-1}^{(h')}(\s\xx)] \Big\} +\cr
&+\sum_{h=h^*}^1 \left\{ \left({Z_h^{(1)}\over Z_h}\right)^2
G^{(h)}_{1,L,\b}(\xx)+ \left({Z_h^{(2)}\over Z_h}\right)^2
G^{(h)}_{2,L,\b}(\xx) + {Z_h^{(1)} Z_h^{(2)}\over Z_h^2}
G^{(h)}_{3,L,\b}(\xx)\right\}\;,}\Eq(6.39)$$
where $h\vee h'=\max\{h,h'\}$ and $g_{\o_1,\o_2}^{(h^*)}(\xx)$ has to be
understood as $g_{\o_1,\o_2}^{(\le h^*)}(\xx)$; moreover,
$$G^{(h)}_{l,L,\b}(\xx) = \sum_{n=1}^\io \sum_{h_r=h^*-1}^{h-1}
\sum_{\t\in\TT^2_{h_r,n,l}\atop h_{\xx,\yy}=h}
\sum_{\bP\in\PP_\t,\rr\atop P_{v_0}=\emptyset}
\sum_{T\in {\bf T}} \sum_{\a\in A_T}
G^{(h,h_r)}_{l,L,\b}(\xx,\t,\bP,\rr,T,\a)\;,\Eq(6.40)$$
where, if $\hat \xx_{v_0}$ denotes the set of space-time points associated
with the normal endpoints and $i_\xx=i$, if the corresponding special
endpoint is of type $Z^{(i)}$,
$$\eqalign{
&G^{(h,h_r)}_{l,L,\b}(\xx,\t,\bP,\rr,T,\a)=\cr
&=\left({Z_{h_\xx}^{(i_\xx)} Z_h\over Z_{h_\xx-1} Z^{(i_\xx)}_h}\right)
\left({Z_{h_\yy}^{(i_\yy)} Z_h\over Z_{h_\yy-1} Z^{(i_\yy)}_h}\right)
\Big[\prod_{v\,\hbox{\ottorm not e.p.}}
\Big(Z_{h_v}/Z_{h_v-1}\Big)^{|P_v|/2}\Big]\;\cdot\cr
&\cdot \int d\hat \xx_{v_0} h_\a(\hat \xx_{v_0})
\Big[\prod_{i=1}^n d_{j_\a(v^*_i)}^{b_\a(v^*_i)}(\xx_i,\yy_i)
K^{(h_i)}_{v^*_i}(\xx_{v^*_i})\Big]
\Big\{\prod_{v\,\hbox{\ottorm not e.p.}}{1\over s_v!} \int
dP_{T_v}(\tt_v) \;\cdot\cr
&\cdot\; \det G_\a^{h_v,T_v}(\tt_v)
\Big[\prod_{l\in T_v} \hat\partial^{q_\a(f^-_l)}_{j_\a(f^-_l)}
\hat\partial^{q_\a(f^+_l)}_{j_\a(f^+_l)} [d^{b_\a(l)}_{j_\a(l)}(\xx_l,\yy_l)
\bar\dpr_1^{m_l} g^{(h_v)}_{\o^-_l,\o^+_l}(\xx_l-\yy_l)]\Big]\Big\}\;.\cr}
\Eq(6.41)$$
In the r.h.s. of \equ(6.41) all quantities are defined as in \S{\bf I}3, except
the kernels $K^{(h_i)}_{v^*_i}(\xx_{v^*_i})$ associated with the special
endpoints. If $v$ is one of these endpoints, $\xx_v$ is
always a single point and
$$K^{(h_v)}_v(\xx_v)=e^{i\pp_F\xx_v\sum_{f\in I_v}\s(f)}\;.\Eq(6.42)$$
 
We want to prove the following Theorem.
 
\* \sub(6.8) {\cs Theorem.} {\it Suppose that the conditions of Theorem
\secc(6.4) are verified, that $\bar\e_4$ is defined as in that theorem
and that $\d^*$ is chosen so that condition \equ(5.57) is satisfied.
Then, there exist positive constants $\th<1$ and $\bar\e_5\le \bar\e_4$,
independent of $u$, $L$, $\b$,
such that, if $|\l_1|\le \bar\e_5$ and $\g\ge 1+\sqrt{2}$,
given any integer $N\ge 0$,
$$|G^{(h)}_{1,L,\b}(\xx)| + |G^{(h)}_{2,L,\b}(\xx)| + \g^{-\th h}
|G^{(h)}_{3,L,\b}(\xx)|
\le C_N |\l_1|{\g^{2h}\over1+[\g^h|{\bf d}(\xx)|]^{N}}\;,\Eq(6.43)$$
for a suitable constant $C_N$.
 
Moreover, if $h\le 0$, we can write
$$\eqalign{
G^{(h)}_{1,L,\b}(\xx)&=\cos(2 p_F x) \bar G^{(h)}_{1,L,\b}(\xx)+
\sum_{\s=\pm 1} e^{ip_F\s x}s^{(h)}_{1,\s,L,\b}(\xx) +r^{(h)}_{1,L,\b}(\xx)
\;,\cr
G^{(h)}_{2,L,\b}(\xx)&=\bar G^{(h)}_{2,L,\b}(\xx) + s^{(h)}_{2,L,\b}(\xx)
+r^{(h)}_{2,L,\b}(\xx)\;,\cr}\Eq(6.44)$$
so that
$$\bar G^{(h)}_{l,L,\b}(\xx)=\bar G^{(h)}_{l,L,\b}(-\xx)\;,\quad l=1,2\;,
\Eq(6.45)$$
$$|r^{(h)}_{1,L,\b}(\xx)| + |r^{(h)}_{2,L,\b}(\xx)|\le
C_N |\l_1| \g^{2h} {\g^{\th h} \over1+[\g^h|{\bf d}(\xx)|]^{N}}\;,\Eq(6.46)$$
and, if we define $D_{m_0,m_1}= \dpr_0^{m_0} \bar\dpr_1^{m_1}$, given
any integers $m_0,m_1\ge 0$, there exists a constant $C_{N,m_0,m_1}$,
such that
$$\sum_{l=1,2}|D_{m_0,m_1}\bar G^{(h)}_{l,L,\b}(\xx)| \le C_{N,m_0,m_1} |\l_1|
{\g^{2h} \g^{h(m_0+m_1)}\over1+[\g^h|{\bf d}(\xx)|]^{N}}\;,\Eq(6.47)$$
$$\eqalign{
&\sum_{\s=\pm 1} |D_{m_0,m_1}s ^{(h)}_{1,\s,L,\b}(\xx)|
+|D_{m_0,m_1}s ^{(h)}_{2,L,\b}(\xx)|\le\cr &\le C_{N,m_0,m_1} |\l_1|
{\g^{2h} \g^{h(m_0+m_1)}\over1+[\g^h|{\bf d}(\xx)|]^{N}}
[\g^{-\th (h-h^*)} +\g^{\th h}]\;.\cr}\Eq(6.48)$$
 
$\Omega^3_{L,\b}(\xx)$, as well as the functions $\bar G^{(h)}_{l,L,\b}(\xx)$,
$r^{(h)}_{l,L,\b}(\xx)$, $s^{(h)}_{1,\s,L,\b}(\xx)$ and
$s^{(h)}_{2,L,\b}(\xx)$ converge, as $L,\b\to\io$, to continuous bounded
functions on $\ZZZ\times\RRR$, that we shall denote $\Omega^3(\xx)$, $\bar
G^{(h)}_{l}(\xx)$, $r^{(h)}_l(\xx)$, $s^{(h)}_{1,\s}(\xx)$ and
$s^{(h)}_2(\xx)$, respectively. $\bar G^{(h)}_1(\xx)$ and $\bar
G^{(h)}_2(\xx)$ are the restrictions to $\ZZZ\times\RRR$ of two even functions
on $\RRR^2$ satisfying the bound \equ(6.47) with the continuous derivative
$\dpr_1$ in place of the discrete one and $|\xx|$ in place of $|{\bf
d}(\xx)|$.
 
Finally, $\bar G^{(h)}_1(\xx)$, as a function on $\RRR^2$, satisfies
the symmetry relation
$$\bar G^{(h)}_1(x,x_0)=\bar G^{(h)}_1(x_0 v_0^*,{x\over v_0^*})\;.
\Eq(6.49)$$
}

\* \sub(6.9){\it Proof}. As in the proof of Theorem \secc(6.4), we shall try
to mimic as much as possible the proof of the bound ({\bf I}3.110), by only
remarking the relevant differences. Since $D_0(P_{v_0})+l_{v_0}=0$, if the
integral in the r.h.s. of \equ(6.41) were over the set of variables
$x_{v_0}\bs\xx$, we should get for $G^{(h,h_r)}_{l,L,\b}(\xx,\t,\bP,\rr,T,\a)$
the same bound we derived in \sec(6.3) for $z^{(l)}_h(\t,\bP,\rr,T,\a)$.
However, in this case, we have to perform the integration over the set
$x_{v_0}$ by keeping fixed two points ($\xx$ and $\yy$), instead of one; hence
we have to modify the bound ({\bf I}3.102) in a way different from what we did in
the proof of Theorem \secc(6.4).
 
Let us call $\bar v_0$ the higher vertex $v\in \t$, such that both $\xx$
and $\yy$ belong to $\xx_v$; by the definition of $h$, it is a non trivial
vertex and its scale is equal to $h$.
Moreover, given the tree graph $T$ on $x_{v_0}$, let us call $T_{\xx,\yy}$ its
subtree connecting the points of $\xx_{\bar v_0}$ and $\tilde T_{\xx,\yy}=
\cup_{v\ge \bar v_0} \tilde T_v$, $\tilde T_v$ being defined as \S{\bf I}3.15,
after ({\bf I}3.118). We want to bound $\dd(\xx-\yy)$ in terms of the distances
between the points connected by the lines $l\in \tilde T_{\xx,\yy}$.
 
Let us call $\bar v^{(i)}$, $i=1,\ldots,s_{\bar v_0}$ the non trivial vertices
or endpoints following $\bar v_0$. The definition of $\bar v_0$ implies that
$s_{\bar v_0}>1$ and that $\xx$ and $\yy$ belong to two different sets
$\xx_{\bar v^{(i)}}$; note also that $\tilde T_{\bar v_0}$ is an anchored tree
graph between the sets of points $\xx_{\bar v^{(i)}}$. Hence there is an
integer $r$, a family $l_1,\ldots,l_r$ of lines belonging to $\tilde T_{\bar
v_0}$ and a family $v^{(1)},\ldots,v^{(r+1)}$ of vertices to be chosen among
$\bar v^{(1)},\ldots,\bar v^{(s_{\bar v_0})}$, such that $1\le r\le s_{\bar
v_0}-1$ and
$$\eqalign{
|\dd(\xx-\yy)| &\le \sum_{j=1}^r |\dd(\xx'_{l_j}-\yy'_{l_j})|+
\sum_{j=1}^{r+1} |\dd(\xx^{(j)}-\yy^{(j)})|\;\le\cr
&\le \sum_{l\in \tilde T_{\bar v_0}} |\dd(\xx'_l-\yy'_l)|+
\sum_{j=1}^{r+1} |\dd(\xx^{(j)}-\yy^{(j)})|\;,\cr}\Eq(6.50)$$
where $\xx^{(1)}=\xx$, $\yy^{(r+1)}=\yy$, $\xx'_{l_j}$ and $\yy'_{l_j}$ are
defined as in ({\bf I}3.114) and, finally, the couple of points
$(\xx'_{l_j},\yy'_{l_j})$ coincide, up to the order, with the couple
$(\yy^{(j)},\xx^{j+1})$.
 
If no propagator associated with a line $l\in \tilde T_{\xx,\yy}$ is affected
by the regularization, we can iterate in an obvious way the previous
considerations, so getting the bound
$$|\dd(\xx-\yy)|\le \sum_{l\in \tilde T_{\xx,\yy}} |\dd(\xx'_l-\yy'_l)|\;.
\Eq(6.51)$$
However, this is not in general true and
we have to consider in more detail the subsequent steps of the iteration.
 
Let us consider one of the vertices $\xx_{v^{(j)}}$; if $\xx^{(j)}=\yy^{(j)}$,
there is nothing to do. Hence we shall suppose that $\xx^{(j)}\not=\yy^{(j)}$
and we shall say that the propagators associated with the lines $l_j$, if
$1\le j\le r$, and $l_{j-1}$, if $2\le j\le r+1$, are {\it linked} to
$v^{(j)}$. There are two different cases to consider.
 
\*
\0 1) $\xx^{(j)}$ and $\yy^{(j)}$ belong to two different non trivial vertices
or endpoints following $v^{(j)}$ and the propagators linked to $v^{(j)}$ are
not affected by action of $\RR$ on the vertex $v^{(j)}$ or some trivial vertex
$v$, such that $\bar v_0 < v < v^{(j)}$. In this case, we iterate the previous
procedure without any change.
 
\*
\0 2) One of the propagators linked to $v^{(j)}$ is affected by action of
$\RR$ on the vertex $v^{(j)}$ or some trivial vertex $v$, such that $\bar v_0
< v < v^{(j)}$; note that, if there are two linked propagators, only one may
have this property, as a consequence of the regularization procedure described
in \S{\bf I}3. This means that $\xx^{(j)}$ or $\yy^{(j)}$, let us say
$\xx^{(j)}$, is of the form ({\bf I}3.115), with $t_l \not=1$, that is there
are two points $\tilde\xx_l,\xx_l\in \xx_{v^{(j)}}$ and a point $\bar\xx_l\in
\RRR^2$, coinciding with $\xx_l$ modulo $(L,\b)$, such that
$$\xx^{(j)}=\tilde\xx_l+t_l(\bar\xx_l-\tilde\xx_l)\;,\quad
|\bar x_l-x_l|\le 3L/4\;,|\bar x_{l,0}-x_{l,0}|\le 3\b/4\;.\Eq(6.52)$$
By using ({\bf I}2.96), \equ(6.52), the fact that $0\le |t_l|\le 1$ and the
remark that $\dd(\bar\xx_l-\tilde\xx_l)=\dd(\xx_l-\tilde\xx_l)$, we get
$$|\dd(\xx^{(j)}-\yy^{(j)})| \le |\dd(\tilde\xx_l-\yy^{(j)})|+ \sqrt{2}\;
|\dd(\xx_l-\tilde\xx_l)|\;.\Eq(6.53)$$
We can now bound $|\dd(\tilde\xx_l-\yy^{(j)})|$ and
$|\dd(\xx_l-\tilde\xx_l)|$, by proceeding as in the proof of \equ(6.50), since
the points $\tilde\xx_l$, $\xx_l$ and $\yy^{(j)}$ all belong to $v^{(j)}$.
We get
$$|\dd(\xx^{(j)}-\yy^{(j)})| \le (1+\sqrt{2}) \left[
\sum_{l\in \tilde T_{v^{(j)}}} |\dd(\xx'_l-\yy'_l)|+
\sum_{m=1}^{r_j} |\dd(\xx^{'(m)}-\yy^{'(m)})| \right]
\;,\Eq(6.54)$$
where $2\le r_j\le s_{v^{(j)}}$ and the points $\xx^{'(m)}$, $\yy^{'(m)}$
are endpoints of propagators linked to some non trivial vertex or endpoint
following $v^{(j)}$.
 
\*
By iterating the previous procedure we get, instead of \equ(6.51), the bound
$$|\dd(\xx-\yy)|\le \sum_{l\in \tilde T_{\xx,\yy}} (1+\sqrt{2})^{p_l}
|\dd(\xx'_l-\yy'_l)|\;,\Eq(6.55)$$
where, if $l\in T_{v_l}$, $p_l$ is an integer less or equal to the number of
non trivial vertices $v$ such that $\bar v_0\le v<v_l$; note that
$$p_l\le h_{v_l}-h\;.\Eq(6.56)$$
 
Let us now suppose that
$$\g\ge 1+\sqrt{2}\;.\Eq(6.57)$$
Since there are at most $2n+1$ lines in $T$, \equ(6.55), \equ(6.56)
and \equ(6.57) imply that there exists at least one line $l\in
T_{\xx,\yy}$, such that
$$\g^{h_{v_l}}|\dd(\xx'_l-\yy'_l)|
\ge {\g^h |\dd(\xx-\yy)|\over 2n+1}\;.\Eq(6.58)$$
It follows that, given any $N\ge 0$, for the corresponding propagator we can
use, instead of the bound ({\bf I}3.116), the following one:
$$\eqalign{
&\left| \tilde\partial^{q_\a(f^-_l)}_{j_\a(f^-_l)}
\tilde\partial^{q_\a(f^+_l)}_{j_\a(f^+_l)}
[d^{b_\a(l)}_{j_\a(l)}(\xx'_l(t_l),\yy'_l(s_l)) \bar\dpr_1^{m_l}
g^{(h_v)}_{\o^-_l,\o^+_l}(\xx'_l(t_l)-\yy'_l(s_l))]\right| \le\cr
&\le {\g^{h_v[1+q_\a(f^+_l)+q_\a(f^-_l)+m(f^-_l)+m(f^+_l)-b_\a(l)]}\over
1+[\g^{h_v}|\dd(\xx'_l(t_l)-\yy'_l(s_l))|]^3}
\Big({|\s_{h_v}|\over \g^{h_v}}\Big)^{\r_l}
{C_N (2n+1)^N\over 1+[\g^h |\dd(\xx-\yy)|]^N}\;.\cr}\Eq(6.59)$$
For all others propagators we use again the bound ({\bf I}3.116) with $N=3$ and
we proceed as in \S{\bf I}3.15, recalling that we have to substitute in
({\bf I}3.118) $d(\xx_{v_0}\bs\bar\xx)$ with $d\hat\xx_{v_0}$. This implies that,
in the r.h.s. of ({\bf I}3.119), one has to eliminate one $d\rr_l$ factor and,
of course, this can be done in an arbitrary way. We choose to eliminate the
integration over the $\rr_l$ corresponding to a propagator of scale $h$ (there
is at least one of them), so that the bound ({\bf I}3.118) is improved by a
factor $\g^{2h}$.
 
At the end, we get
$$\eqalign{
&|G^{(h,h_r)}_{l,L,\b}(\xx,\t,\bP,\rr,T,\a)| \le (C\e_h)^n C_N (2n+1)^N
{\g^{2h}\over 1+[\g^h {\bf d}(\xx)]^N}\;\cdot\cr
&\cdot\;\left({Z_{h_\xx}^{(i_\xx)} Z_h\over Z_{h_\xx-1} Z^{(i_\xx)}_h}\right)
\left({Z_{h_\yy}^{(i_\yy)} Z_h\over Z_{h_\yy-1}
Z^{(i_\yy)}_h}\right) \prod_{v\,\hbox{\ottorm not e.p.}}
\Big\{ {1\over s_v!} C^{\sum_{i=1}^{s_v}|P_{v_i}|-|P_v|}\;\cdot\cr
&\cdot\;\Big(Z_{h_v}/Z_{h_v-1}\Big)^{|P_v|/2}
\g^{-[-2+{|P_v|\over 2}+l_v+z(P_v,l_v)+{\tilde z(P_v,l_v)\over 2}]}\;.
\Big\}\cr}\Eq(6.60)$$
 
We can now perform as in \S{\bf I}3.14 the various sums in the r.h.s. of
\equ(6.40). There are some differences in the sum over the scale labels, but
they can be easily treated. First of all, one has to take care of the factors
$(Z_{h_\xx}^{(i_\xx)} Z_h)/ (Z_{h_\xx-1} Z^{(i_\xx)}_h)$ and
$(Z_{h_\yy}^{(i_\yy)} Z_h)/ (Z_{h_\yy-1}Z^{(i_\yy)}_h)$. However, by using
\equ(6.29) and \equ(6.38), it is easy to see that these factors have the only
effect to add to the final bound a factor $\g^{C|\l_1|(h_v-h_{v'})}$ for each
non trivial vertex $v$ containing one of the special endpoints and strictly
following the vertex $v_{\xx,\yy}$; this has a negligible effect, thanks to
analogous of the bound ({\bf I}3.111), valid in this case.
The other difference is in the fact that, instead of fixing the scale of the
root, we have now to fix the scale of $v_{\xx,\yy}$. However, this has no
effect, since we bound the sum over the scales with the sum over the the
differences $h_v-h_{v'}$.
 
The previous considerations are sufficient to get the bound \equ(6.43) for
$G_{1,L,\b}^{(h)}(\xx)$ and $G_{2,L,\b}^{(h)}(\xx)$. In order to explain the
factor $\g^{\th h}$ multiplying $G_{3,L,\b}^{(h)}(\xx)$, one has to note that
the trees whose normal endpoints are all of scale lower than $2$ give no
contribution to $G_{3,L,\b}^{(h)}(\xx)$. In fact, these endpoints have the
property that $\sum_{f\in P_v} \s(f)=0$, while this condition is satisfied
from one of the special endpoints but not from the other, in any tree
contributing to $G_{3,L,\b}^{(h)}(\xx)$. It follows, since any propagator
couples two fields with different $\s$ indices, that it is possible to produce
a non zero contribution to $G_{3,L,\b}^{(h)}(\xx)$, only if there is at least
one endpoint of scale $2$; this allows to extract from the bound a factor
$\g^{\th h}$, with $0<\th<1$, as remarked many times before.
 
We now want to show that $G^{(h)}_{1,L,\b}(\xx)$ and $G^{(h)}_{2,L,\b}(\xx)$
can be decomposed as in \equ(6.44), so that the bounds \equ(6.46), \equ(6.47)
and \equ(6.45) are satisfied. To begin with, we define
$r^{(h)}_{i,L,\b}(\xx)$, $i=1,2$, by using the definition \equ(6.40) of
$G^{(h)}_{i,L,\b}(\xx)$, with the constraint that the sum is restricted to the
trees, which contain at least one endpoint of scale $h_v=2$; this implies, in
particular, that $G^{(+1)}_{i,L,\b}(\xx)-r^{(+1)}_{i,L,\b}(\xx)=0$.
Moreover, in the remaining
trees, we decompose the propagators in the following way:
$$g^{(h)}_{\o,\o'}(\xx) = \bar g^{(h)}_{\o,\o'}(\xx) +
\d g^{(h)}_{\o,\o'}(\xx)\;,\Eq(6.61)$$
where $\bar g^{(h)}_{\o,\o'}(\xx)$ is defined by putting, in the r.h.s. of
({\bf I}2.94), $(v^*_0 k')$ in place of $E(k')$, and we absorb in
$r^{(h)}_{i,L,\b}(\xx)$ the terms containing at least one propagator $\d
g^{(h)}_{\o,\o'}(\xx)$, which is of size $\g^{2h}$. The substitution of
$(v^*_0 k')$ in place of $E(k')$ is done also in the definition of the $\RR$
operator, so producing other ``corrections'', to be added to
$r^{(h)}_{i,L,\b}(\xx)$.
An argument similar to that used for $G^{(h)}_{3,L,\b}(\xx)$
easily allows to prove the bound \equ(6.46).
 
$\sum_{\s=\pm 1} \exp(i\s p_F x)s^{(h)}_{1,\s,L,\b}(\xx)$ and
$s^{(h)}_{2,L,\b}(\xx)$ will denote the sum of the trees contributing to
$G^{(h)}_{1,L,\b}(\xx)$ $-r^{(h)}_{1,L,\b}(\xx)$ and $G^{(h)}_{2,L,\b}(\xx)-$
$r^{(h)}_{2,L,\b}(\xx)$, respectively, which have at least one endpoint of
type $\n$ or $\d$.
 
Let us now consider the ``leading'' contribution to $G^{(h)}_{2,L,\b}(\xx)$,
which is defined by the second of the equations \equ(6.44) as $\bar
G^{(h)}_{2,L,\b}(\xx)$ and is obtained by using again \equ(6.40), but with the
constraint that the sum over the trees is restricted to those having only
endpoints with scale $h_v\le 1$ and only normal endpoints of type $\l$.
Moreover we have to use everywhere the propagator $\bar
g^{(h)}_{\o,\o'}(\xx)$, which has well defined parity properties in the $\xx$
variables; it is odd, if $\o=\o'$, and even, if $\o=-\o'$.
 
Note that all the normal endpoints with $h_v\le 1$ are such that $\sum_{f\in
I_v} \s(f)=0$ and that this property is true also for the special endpoints,
which have to be of type $Z^{(2)}$; hence there is no oscillating factor in
the kernels associated with the endpoints, which are suitable constants (the
associated effective potential terms are local). It follows that any graph
contributing to $\bar G^{(h)}_{2,L,\b}(\xx)$ is given, up to a constant, by an
integral over the product of an even number of propagators (we are using here
the fact that there is no endpoint of type $\n$ or $\d$). Moreover, since all
the endpoints satisfy also the condition $\sum_{f\in I_v} \s(f)\o(f)=0$, which
is violated by the set of two lines connected by a non diagonal propagator,
the number of non diagonal propagators has to be even. These remarks
immediately imply that $\bar G^{(h)}_{2,L,\b}(\xx)= \bar
G^{(h)}_{2,L,\b}(-\xx)$.
 
In order to prove the bound \equ(6.47) for $\bar G^{(h)}_{2,L,\b}(\xx)$, we
observe that, since the propagators only couple fields with different $\s$
indices and $\sum_{f\in I_v} \s(f)=0$, given any tree $\t$ contributing to
$\bar G^{(h)}_{2,L,\b}(\xx)$ and any $v\in \t$, we must have
$$\sum_{f\in P_v} \s(f) =0\;.\Eq(6.62)$$
Let us now consider the vertex $\bar v_0$, defined as in \sec(6.9), that is
the higher vertex $v\in \t$, such that both $\xx$ and $\yy={\bf 0}$ belong to
$\xx_v$, and let $v_\xx$ be the vertex immediately following $\bar v_0$, such
that $\xx\in v_\xx$. We can associate with $v_\xx$ a contribution to
$\BB^{h}(\psi^{(\le h)},\phi)$ (recall that $h$ is the scale of $\bar v_0$ and
hence the scale of the external fields of $v_\xx$), with $m=1$ and
$2n=P_{v_\xx}$ (see \equ(6.6)), whose kernel is of the form, thanks to
\equ(6.62)
$$\eqalign{
&B(\xx;\yy_1,\ldots,\yy_{2n})={1\over (L\b)^{2n+1}}
\sum_{\pp,\kk'_1,\ldots,\kk'_{2n}}
e^{i\pp\xx -i\sum_{r=1}^{2n} \s_r \kk'_r \yy_r}\;\cdot\cr
&\quad\cdot\; \hat B(\pp;\kk'_1,\ldots,\kk'_{2n-1})
\d(\sum_{r=1}^{2n}\s_r\kk'_r-\pp)\;.\cr}\Eq(6.63)$$
If we apply the differential operator $\dpr_0^{m_0}$ to $\bar
G^{(h)}_{2,L,\b}(\xx)$, this operator acts on $B(\xx;\yy_1,\ldots,\yy_{2n})$,
so that its Fourier transform is multiplied by $(ip_0)^{m_0}$; since
$p_0=\sum_{r=1}^{2n}\s_r k_{r 0}$ and the external fields of $v_\xx$ are
contracted on a scale smaller or equal to $h$, it is easy to see that there is
an improvement on the bound of $\dpr_0 \bar G^{(h)}_{2,L,\b}(\xx)$, with
respect to the bound of $\bar G^{(h)}_{2,L,\b}(\xx)$, of a factor
$c_{m_0}\g^{h m_0}$, for a suitable constant $c_{m_0}$. We are using here the
fact that $\bar G^{(+1)}_{i,L,\b}(\xx)=0$, so that we can suppose $h\le 0$,
otherwise we would be involved with the singularity of the scale $1$
propagator $g^{(1)}_{\o^-_l,\o^+_l}(\xx_l-\yy_l)$ at $x_l-y_l=0$, which allows
to get uniform bounds on the derivatives only for $|x_l-y_l|$ bounded below, a
condition not verified in general.
 
Let us now consider $\bar\dpr_1^{m_1} \bar G^{(h)}_{2,L,\b}(\xx)$ (see
({\bf I}3.6) for the definition of $\bar\dpr_1$). By using ({\bf I}2.62) and the
conservation of the spatial momentum, we find that $\bar\dpr_1^{m_1}$ acts on
$B(\xx;\yy_1,\ldots,\yy_{2n})$, so that its Fourier transform is multiplied by
$\sin (px)^{m_1}$, with $p=\sum_{r=1}^{2n}\s_r k'_r$ $+2\p m$, where $m$ is an
arbitrary integer and $p$ is chosen so that $|p|\le \p$. If $m=0$, we proceed
as in the case of the time derivative, otherwise we note that the support
properties of the external fields, see \S{\bf I}2.2, implies that
$|\sum_{r=1}^{2n}\s_r k'_r|\le 2n a_0 \g^h$; hence, if $|m|>0$, $2n\ge
(\p/a_0)\g^{-h}$. Since the number of endpoints following $v_\xx$ is
proportional to $2n$ and each endpoint carries a small factor of order $\l_1$,
it is clear that, if $\l_1$ is small enough, we get an improvement in the
bound of the terms with $|m|>0$, with respect to the corresponding
contributions to $\bar G^{(h)}_{2,L,\b}(\xx)$, of a factor $\exp(-C\g^{-h})\le
c_{m_1}\g^{h m_1}$, for some constant $c_{m_1}$. In the same manner, we can
treat the operator $D_{m_0,m_1}$, so proving the bound \equ(6.47) for
$D_{m_0,m_1}\bar G^{(h)}_{2,L,\b}(\xx)$.
 
Let us now consider $G^{(h)}_{1,L,\b}(\xx-\yy)-r^{(h)}_{1,L,\b}(\xx-\yy)$.
In this case the kernels of the two special endpoints $\xx$ and $\yy$ are
equal to $\exp(2i\s_\xx p_F x)$ and $\exp(2i\s_\yy p_F y)$, respectively.
However, since the propagators couple fields with different $\s$ indices and
all the other endpoints satisfy the condition $\sum_{f\in I_v} \s(f)=0$,
$\s_\xx=-\s_\yy$ and we can write
$$G^{(h)}_{1,L,\b}(\xx-\yy)-r^{(h)}_{1,L,\b}(\xx-\yy)=\fra12 \sum_{\s=\pm 1}
e^{2i\s p_F(x-y)} \Big[\bar G^{(h)}_{1,\s}(\xx-\yy)
+2 s^{(h)}_{1,\s,L,\b}(\xx-\yy)\Big]\;,\Eq(6.64)$$
with $\bar G^{(h)}_{1,\s}(\xx)$ having the same properties as $\bar
G^{(h)}_2(\xx)$; in particular it is an even function of $\xx$ and satisfies
the bound \equ(6.47). Moreover, it is easy to see that $\bar
G^{(h)}_{1,+}(\xx-\yy)$ is equal to $\bar G^{(h)}_{1,-}(\yy-\xx)= \bar
G^{(h)}_{1,-}(\xx-\yy)$, hence $\bar G^{(h)}_{1,\s}(\yy-\xx)$ is independent
of $\s$ and we get the decomposition in the first line of \equ(6.44), with
$\bar G^{(h)}_1(\xx-\yy)$ satisfying \equ(6.47) and \equ(6.45).
 
The bound \equ(6.48) is proved in the same way as the bound \equ(6.47). The
factor $[\g^{-\th (h-h^*)} +\g^{\th h}]$ in the r.h.s. comes from the fact
that the trees contributing to $s^{(h)}_{1,\s,L,\b}(\xx)$ and
$s^{(h)}_{2,L,\b}(\xx)$ have at least one vertex of type $\n$ or $\d$, whose
running constants satisfy \equ(5.17) and \equ(5.57).
 
Note that $s^{(h)}_{1,\s,L,\b}(\xx)$ and $s^{(h)}_{2,L,\b}(\xx)$ are not
even functions of $\xx$ and that $s^{(h)}_{1,\s,L,\b}(\xx)$ is not independent
of $\s$.
 
In order to complete the proof of Theorem \secc(6.8), we observe that all the
functions appearing in the r.h.s. of \equ(6.39), as well as those defined in
\equ(6.44), clearly converge, as $L,\b\to\io$, and that their limits can be
represented in the same way as the finite $L$ and $\b$ quantities, by
substituting all the propagators with the corresponding limits. This follows
from the tree structure of our expansions and some straightforward but lengthy
standard arguments; we shall omit the details.
 
Let us consider, in particular, the limits $G^{(h)}_i(\xx)$ of the functions
$G^{(h)}_{i,L,\b}(\xx)$. Their tree expansions contain only trees with
endpoints of scale $h_v\le 1$, which are associated with local terms of type
$\l$ or of the form \equ(6.13) and \equ(6.14), whose $\psi$ fields are of
scale less or equal to $0$. The support properties of the field Fourier
transform imply that the local terms of type $\l$ can be rewritten by
substituting the sum over the corresponding lattice space point with a
continuous integral over $\RRR^1$. We can of course use these new expressions
to build the expansions, since the propagators of scale $h\le 0$, in the limit
$L,\b\to\io$, are well defined smooth functions on $\RRR^2$. For the same
reason, the tree expansions are well defined also if the space points
associated with the special endpoints vary over $\RRR^1$, instead of $\ZZZ^1$;
therefore there is a natural way to extend to $\RRR^2$ the functions
$G^{(h)}_i(\xx)$, which of course satisfy the bound \equ(6.47), with the
continuous derivative $\dpr_1$ in place of the discrete one and $|\xx|$ in
place of $|{\bf d}(\xx)|$, as well as the analogous of identity \equ(6.45).
 
The function $G^{(h)}_{1,L,\b}(\xx)$ satisfies also another symmetry
relation, related with a remarkable property of the propagators $\bar
g^{(h)}_{\o,\o'}$, see \equ(6.61), appearing in its expansion, that is
$$\eqalign{
\bar g^{(h)}_{\o,\o}(x,x_0)&=-i\o \bar g^{(h)}_{-\o,-\o}
\Big(v^*_0 x_0,{x\over v^*_0}\Big)\;,\cr
\bar g^{(h)}_{\o,-\o}(x,x_0)&=-\bar g^{(h)}_{-\o,+\o}
\Big(v^*_0 x_0,{x\over v^*_0}\Big)\;.\cr}\Eq(6.65)$$
On the other hand, each tree contributing to $G^{(h)}_{1,L,\b} (\xx)$ with $n$
normal endpoints (which are all of type $\l$) can be written as a sum of
Feynman graphs (if we use the representation of the regularization operator as
acting on the kernels, see \S{\bf I}3), built by using $4n+4$ $\psi$ fields,
$2n+2$ with $\o=+1$ and $2n+2$ with $\o=-1$, hence containing the same number
of propagators $\bar g^{(h)}_{+1,+1}$ and $\bar g^{(h)}_{-1,-1}$ and, by the
argument used in the proof of \equ(6.45), an even number
of non diagonal propagators. Then, by
using \equ(6.65), we can easily show that the value of any graph, calculated
at $(x,x_0)$, is equal to the value at $(v^*_0 x_0,x/v^*_0)$ of the graph with
the same structure but opposite values for the $\o$-indices of all
propagators, which implies \equ(6.49).
 
\pagina
\vskip1.cm
\section(7, Proof of Theorem {{\bf I}1.5})
 
\sub(7.1)
Theorem {\bf I}3.12 and the analysis performed in \sec(5) and \sec(6) imply
immediately the statements in item a) of Theorem {\bf I}1.5, except the
continuity of $\O^3_{L,\b}(\xx)$ in $x_0=0$, which will be briefly discussed
below. Hence, from now on we shall suppose that all parameters are chosen as
in item a).
 
Let us define
$$\h=\log_\g(1+z^*)\;,\qquad z^*=z_{[h^*/2]}\;,\Eq(7.1)$$
$z_h$ being defined as in \equ(5.2). The analysis performed in \sec(5) allows
to show (we omit the details) that there exists a positive $\th<1$, such that
$$|z_h-z_{h+1}|\le C\l_1^2 [\g^{-\th(h-h^*)}+\g^{\th h}]\;,\quad h^*\le h\le
0\;.\Eq(7.2)$$
We can write
$$\log_\g Z_h = \sum_{h'=h+1}^0 \log_\g[1+z^*+(z_{h'}-z^*)]=
-\h h+\sum_{h'=h+1}^0 r_{h'}\;.\Eq(7.3)$$
On the other hand, if $h>[h^*/2]$, thanks to \equ(7.2), $|r_h|\le C
\sum_{h'=[h^*/2]}^{h-1} |z_{h'}-z_{h'+1}|\le C\l_1^2 \g^{\th h}$ and, if $h\le
[h^*/2]$, $|r_h|\le  C\l_1^2 \g^{-\th (h-h^*)}$; it follows that
$$|r_h|\le  C\l_1^2 [\g^{-\th (h-h^*)}+\g^{\th h}]\;.\Eq(7.4)$$
Hence, if we define
$$c_h={\g^{-\h h}\over Z_{h-1}}\;,\Eq(7.5)$$
we get immediately the bound
$$|c_h-1|\le C \l_1^2\;.\Eq(7.6)$$
 
In a similar way, if we define
$$\tilde\h_1 = \log_\g(1+z^{(1)}_{[h^*/2]})\;,\quad
c^{(1)}_h={\g^{-\tilde\h_1 h}\over Z^{(1)}_h}\;,\Eq(7.7)$$
$z^{(1)}_h$ being defined by \equ(6.18), we get the bound
$$|c^{(1)}_h-1|\le C |\l_1|\;.\Eq(7.8)$$
 
Bounds similar to \equ(7.7) and \equ(7.8) are valid also for the constants
$Z^{(2)}_h$, but in this case Theorem \secc(6.6) implies a stronger result; if
we define
$$c^{(2)}_h = {Z^{(2)}_h\over Z_{h-1}}\;,\Eq(7.9)$$
then
$$|c^{(2)}_h-1|\le C |\l_1|\;.\Eq(7.10)$$
 
Let us now consider the terms in the first three lines of the r.h.s. of
\equ(6.39) and let us call $\O^{3,0}_{L,\b}$ their sum; we can write
$$\O^{3,0}_{L,\b}(\xx) = \bar\O^{3,0}_{L,\b}(\xx) + \d\O^{3,0}_{L,\b}(\xx)
\;,\Eq(7.11)$$
where $\bar\O^{3,0}_{L,\b}$ is obtained from $\O^{3,0}_{L,\b}$ by restricting
the sums over $h$ and $h'$ to the values $\le 0$ and by substituting the
propagators $g^{(h)}_{\o,\o'}$ with the propagators $\bar g^{(h)}_{\o,\o'}$,
defined in \equ(6.61). By using the symmetry relations
$$\eqalign{
\bar g^{(h)}_{\o,\o}(x,x_0) &= -\bar g^{(h)}_{\o,\o}(-x,-x_0)=
\bar g^{(h)}_{+,+}(\o x,x_0)\;,\cr
\bar g^{(h)}_{\o,-\o}(\xx) &= \bar g^{(h)}_{\o,-\o}(-\xx)=
\o\bar g^{(h)}_{+,-}(\xx)\;,\cr}\Eq(7.12)$$
it is easy to show that we can write
$$\bar\O^{3,0}_{L,\b}(\xx) = \cos(2p_F x) \bar\O_{1,L,\b}(\xx)+
\bar\O_{2,L,\b}(\xx)\;,\Eq(7.13)$$
$$\eqalign{
\bar\O_{1,L,\b}(\xx)= 2 \sum_{h^*\le h,h'\le 0}
{(Z_{h\vee h'}^{(1)})^2\over Z_{h-1} Z_{h'-1}}
\Big[ &\bar g_{+,+}^{(h)}(x,x_0) \bar g_{+,+}^{(h')}(-x,x_0)+\cr
&+\bar g_{+,-}^{(h)}(x,x_0) \bar g_{+,-}^{(h')}(x,x_0)\Big]\;,\cr}
\Eq(7.14)$$
$$\eqalign{
\bar\O_{2,L,\b}(\xx)= \sum_{h,h'\le 0}
{(Z_{h\vee h'}^{(2)})^2\over Z_{h-1} Z_{h'-1}}
\Big[&\sum_\o \bar g_{+,+}^{(h)}(\o x,x_0) \bar g_{+,+}^{(h')}(\o x,x_0)-\cr
&-2\bar g_{+,-}^{(h)}(x,x_0) \bar g_{+,-}^{(h')}(x,x_0)\Big]\;.\cr}
\Eq(7.15)$$
 
By using \equ(6.39), \equ(6.44), \equ(7.13) and the fact that
$G^{(+1)}_{i,L,\b}(\xx)-r^{(+1)}_{i,L,\b}(\xx)=0$ for $i=1,2$,
we can decompose $\O^3_{L,\b}$ as in ({\bf I}1.13), by defining
$$\O^{3,a}_{L,\b}(\xx)= \bar\O_{1,L,\b}(\xx) +
\sum_{h=h^*}^0 \left({Z_h^{(1)}\over Z_h}\right)^2
\bar G^{(h)}_{1,L,\b}(\xx)\;,\Eq(7.16)$$
$$\O^{3,b}_{L,\b}(\xx)= \bar\O_{2,L,\b}(\xx) + \sum_{h=h^*}^0
\left({Z_h^{(2)}\over Z_h}\right)^2
\bar G^{(h)}_{2,L,\b}(\xx)\;,\Eq(7.17)$$
$$\eqalign{
\O^{3,c}_{L,\b}(\xx) = \d\O^{3,0}_{L,\b}(\xx) +
\sum_{h=h^*}^1 &\left\{ \left({Z_h^{(1)}\over Z_h}\right)^2
r^{(h)}_{1,L,\b}(\xx)+ \left({Z_h^{(2)}\over Z_h}\right)^2
r^{(h)}_{2,L,\b}(\xx) \right. +\cr &+ \left.{Z_h^{(1)} Z_h^{(2)}\over Z_h^2}
G^{(h)}_{3,L,\b}(\xx)\right\} + s_{L,\b}(\xx)\;,\cr}\Eq(7.18)$$
$$s_{L,\b}(\xx) = \sum_{h=h^*}^0\left\{ \sum_{\s=\pm 1} e^{2i\s p_F x}
\left({Z_h^{(1)}\over Z_h}\right)^2 s^{(h)}_{1,\s,L,\b}(\xx) +
\left({Z_h^{(2)}\over Z_h}\right)^2 s^{(h)}_{2,L,\b}(\xx)\right\}\;.
\Eq(7.19)$$
 
Theorem \secc(6.8) implies that $\O^{3,a}_{L,\b}(\xx)$, $\O^{3,b}_{L,\b}(\xx)$
and $s_{L,\b}(\xx)$ are smooth functions of $x_0$, essentially because their
expansions do not contain any graph with a propagator of scale $+1$ (this
propagator has a discontinuity at $x_0=0$). The function
$\O^{3,c}_{L,\b}(\xx)$ is not differentiable at $x_0=0$, but it is in any case
continuous, since all graphs contributing to it have a Fourier transform
decaying at least as $k_0^{-2}$ as $k_0\to\io$.
 
\*
\sub(7.2)
We want now to prove the bounds in item b) of Theorem {\bf I}1.5. To start
with, we consider the function $\bar\O_{1,L,\b}(\xx)$ defined in \equ(7.14) and
note that it can be written in the form
$$\bar\O_{1,L,\b}(\xx)= \sum_{h=h^*}^0 \left({Z_h^{(1)}\over Z_h}\right)^2
\bar\O^{(h)}_{1,L,\b}(\xx)\;,\Eq(7.20)$$
with $\bar\O^{(h)}_{1,L,\b}(\xx)$ satisfying a bound similar to that proved
for $\bar G^{(h)}_{1,L,\b}(\xx)$, see \equ(6.47), that is
$$|D_{m_0,m_1} \bar\O^{(h)}_{1,L,\b}(\xx)| \le C_{N,m_0,m_1}
{\g^{2h} \g^{h(m_0+m_1)}\over1+[\g^h|{\bf d}(\xx)|]^{N}}\;.\Eq(7.21)$$
This claim easily follows from Lemma {\bf I}2.6, together with \equ(7.5)
and \equ(7.6). Hence we can write, by using \equ(6.47), \equ(7.6),
\equ(7.8) and \equ(7.21),
given any positive integers $n_0$, $n_1$ and putting $n=n_0+n_1$,
$$\eqalign{
|\dpr_{x_0}^{n_0} \bar\dpr_x^{n_1} \O^{3,a}_{L,\b}(\xx)|
&\le C_{N,n} \sum_{h=h^*}^0 {\g^{(2+ 2\h_1 +n)h}
\over [1+(\g^h|\dd(\xx)|)^N]}\le\cr
&\le {C_{N,n}\over |\dd(\xx)|^{2+2\h_1+n}}
H_{N,2+2\h_1+n}(|\dd(\xx)|)\;,\cr}\Eq(7.22)$$
where
$$\h_1=\h-\tilde\h_1\;,\Eq(7.23)$$
$$H_{N,\a}(r)=\sum_{h=h^*}^0 {(\g^h r)^\a \over
1+(\g^h r)^N}\;.\Eq(7.24)$$
 
By using the second of the definitions ({\bf I}2.2), the definition \equ(5.8) and
the bounds \equ(5.16), \equ(6.29), one can see that the constant $\h_1$ can be
represented as in ({\bf I}1.14).
 
On the other hand, it is easy to see that, if $\a\ge 1/2$
and $N-\a\ge 1$, there exists a constant $C_{N,\a}$ such that
$$H_{N,\a}(r) \le {C_{N,\a}\over 1+(\D r)^{N-\a}}\;,\quad
\D=\g^{h^*}\;.\Eq(7.25)$$
The definition ({\bf I}2.40), the first of definitions ({\bf I}2.33), the second
bound in ({\bf I}2.34) and the bound \equ(5.56) easily imply that $\D$ can be
represented as in ({\bf I}1.19), with $\h_2$ satisfying the second of equations
({\bf I}1.14).
 
By using \equ(7.22) and \equ(7.25), one immediately gets the bound
({\bf I}1.16). A similar procedure allows to get also the bound ({\bf I}1.17),
by using \equ(7.10).
 
Let us now consider $\O^{3,c}_{L,\b}(\xx)$. By using \equ(6.43) and
\equ(6.46), as well as the remark that one gains a factor $\g^h$ in the
bound of $g^{(h)}_{\o,\o'}(\xx)- \bar g^{(h)}_{\o,\o'}(\xx)$ with respect
to the bound of $\bar g^{(h)}_{\o,\o'}(\xx)$, we get
$$|\O^{3,c}_{L,\b}(\xx)-s_{L,\b}(\xx)| \le {C_N\over |\dd(\xx)|^2} \left[
{H_{N,2+2\h_1+\th}(|\dd(\xx)|) \over |\dd(\xx)|^{\th+2\h_1}}+
{H_{N,2+\th}(|\dd(\xx)|) \over |\dd(\xx)|^{\th}}
\right]\;,\Eq(7.26)$$
for some positive $\th<1$.
 
The bound of $s_{L,\b}(\xx)$ is slightly different, because of the
$\g^{-\th(h-h^*)}$ in the r.h.s. of \equ(6.48). We get, in addition
to a term of the same form as the r.h.s. of \equ(7.26), another term
of the form
$${C_N\over |\dd(\xx)|^2} (\D|\dd(\xx)|)^{\th} \left[
{H_{N,2+2\h_1-\th}(|\dd(\xx)|) \over |\dd(\xx)|^{2\h_1}}+
H_{N,2-\th}(|\dd(\xx)|)\right]\;.\Eq(7.27)$$
The bounds \equ(7.26) and \equ(7.27) immediately imply ({\bf I}1.18), if $\l$
is so small that, for example, $2|\h_1|\le \th/2$.
 
\*
\sub(7.3) We want now to prove the statements in item c) of Theorem
{\bf I}1.5. The existence of the limit as $L,\b\to\io$ of all functions
follows from Theorem \secc(6.8). The claim that $\O^{3,a}(\xx)$ and
$\O^{3,b}(\xx)$ are even as functions of $\xx$ follows from \equ(6.45) and
\equ(7.14)-\equ(7.18). Moreover $\O^{3,a}(\xx)$ and $\O^{3,b}(\xx)$ are the
restriction to $\ZZZ\times\RRR$ of two functions on $\RRR^2$, that we shall
denote by the same symbols, and $\O^{3,a}(\xx)$ satisfies the
symmetry relation ({\bf I}1.22), since this is true for $\lim_{L,\b\to\io}
\bar\O_{1,L,\b}(\xx)$, as it is easy to check by using \equ(6.65), and for $\bar
G_1^{(h)}(\xx)$, see \equ(6.49).
 
In order to prove ({\bf I}1.20), we suppose that $|\xx|\ge 1$ and
we put $\bar\O_i(\xx)=\lim_{L,\b\to\io} \bar\O_{i,L,\b}(\xx)$; then
we define $\tilde \O_i(\xx)$, $i=1,2$, as the functions
which are obtained by making in the r.h.s. of \equ(7.14) and \equ(7.15),
evaluated in the limit $L,\b\to\io$, the substitutions
$${(Z_{h\vee h'}^{(1)})^2\over Z_{h-1} Z_{h'-1}} \rightarrow
[x^2+(v_0^* x_0)^2]^{-\h_1}\;,
\qquad {(Z_{h\vee h'}^{(2)})^2\over Z_{h-1} Z_{h'-1}} \rightarrow
1\;.\Eq(7.28)$$
Note that the choice of $x^2+(v_0^* x_0)^2$, instead of $x^2+x_0^2$, which is
equivalent for what concerns the following arguments, was done only in order
to have a function $\tilde \O_1(\xx)$ satisfying the same symmetry relation as
$\bar\O_1(\xx)$ in the exchange of $(x,x_0)$ with $(v_0^* x_0, x/v_0^*)$.
 
It is easy to see that
$$\eqalign{
&|\bar\O_1(\xx)- \tilde \O_1(\xx)| \le {C_N\over |\xx|^{2+2\h_1}}
\sum_{h^*\le h,h'\le 0} {\g^h|\xx|\over 1+(\g^h|\xx|)^N }
{\g^{h'}|\xx|\over 1+(\g^{h'}|\xx|)^N }\;\cdot\cr
&\cdot \left| \left({x^2+x_0^2 \over x^2+(v_0^* x_0)^2}\right)^{\h_1}
(\g^h|\xx|)^\h (\g^{h'}|\xx|)^\h (\g^{h\vee h'}|\xx|)^{-2\tilde\h_1}
{c_h c_{h'}\over (c^{(1)}_{h\vee h'})^2 }-1\right|\;.\cr}\Eq(7.29)$$
Note that, if $r>0$ and $\a\in\RRR$
$$|r^\a-1| \le |\a\log r|\left(r^\a+r^{-\a}\right)\;;\Eq(7.30)$$
Hence, by using \equ(7.6), \equ(7.8), \equ(7.25) and ({\bf I}1.14), we get
$$|\bar\O_1(\xx)- \tilde \O_1(\xx)| \le {|J_3|\over |\xx|^{2+2\h_1}}
{C_N \over 1+(\D|\xx|)^N }\;.\Eq(7.31)$$
In the same way, one can show that
$$|\bar\O_2(\xx)- \tilde \O_2(\xx)| \le {|J_3|\over |\xx|^2}
{C_N \over 1+(\D|\xx|)^N }\;.\Eq(7.32)$$
 
Let us now define
$$\O^*_1(\xx) = {2\over [x^2+(v_0^* x_0)^2]^{\h_1}}
{1\over (v^*_0)^2} g_{\cal L}(x/v^*_0,x_0) g_{\cal
L}(-x/v^*_0,x_0)\;,\Eq(7.33)$$
$$\O^*_2(\xx) = {1\over (v^*_0)^2} \sum_{\o=\pm 1}
g_{\cal L}(\o x/v^*_0,x_0) g_{\cal L}(\o x/v^*_0,x_0)\;,\Eq(7.34)$$
where
$$g_{\cal L}(\xx) = {1\over (2\p)^2} \int d\kk e^{i\kk\xx}{\chi_0(\kk)
\over -i k_0+k}\;,\Eq(7.35)$$
$\chi_0(\kk)$ being a smooth function of $\kk$, which is equal to $1$, if
$|\kk|\le t_0$, and equal to $0$, if $|\kk|\ge \g t_0$ (see \S{\bf I}2.3 for the
definition of $t_0$).
 
It is easy to check that $\O^*_i(\xx)$, $i=1,2$, is obtained from $\tilde
\O_i(\xx)$ by making in the $L,\b=\io$ expression of the propagators $\bar
g_{\o,\o'}^{(h)}(\xx)$, which are evaluated from ({\bf I}2.92), if $h^*<h\le 0$,
and ({\bf I}2.120), if $h=h^*$, the following substitutions:
$$\s_{h-1}(\kk') \rightarrow 0\;,\qquad \tilde f_h(\kk') \rightarrow
f_h(\kk')\;.\Eq(7.36)$$
Hence, by using also the remark that, by ({\bf I}2.116) and \equ(5.54),
$|\s_h/\g^h|\le C \g^{-(h-h^*)/2}$, it is easy to show that
$$|\O^*_1(\xx)- \tilde \O_1(\xx)| \le {C_N\over |\xx|^{2+2\h_1}}
H_{N,1}(\D|\xx|) \left[ \l_1^2 H_{N,1}(\D|\xx|)
+ (\D|\xx|)^{1/2} H_{N,1/2}(\D|\xx|)\right]\;.\Eq(7.37)$$
In a similar way, one can show also that
$$|\O^*_2(\xx)- \tilde \O_2(\xx)| \le {C_N\over |\xx|^2}
H_{N,1}(\D|\xx|) \left[ \l_1^2 H_{N,1}(\D|\xx|)
+ (\D|\xx|)^{1/2} H_{N,1/2}(\D|\xx|)\right]\;.\Eq(7.38)$$
 
Moreover, by an explicit calculation, one finds that, if $|\xx|\ge 1$,
$$g_{\cal L}(\xx) = {x_0-ix\over 2\p |\xx|^2} F(\xx)\;,\Eq(7.39)$$
where $F(\xx)$ is a smooth function of $\xx$, satisfying the bound
$$|F(\xx)-1| \le {C_N\over 1+|\xx|^N}\;.\Eq(7.40)$$
 
The bounds \equ(7.31) and \equ(7.32), the similar bounds satisfied by
$|\O^{3,a}(\xx)-\bar\O_1(\xx)|$ and $|\O^{3,b}(\xx)-\bar\O_2(\xx)|$ and the
equations \equ(7.37)-\equ(7.40) allow to prove very easily ({\bf I}1.20) and
({\bf I}1.21).
 
\*
\sub(7.4)
We still have to prove the statements in items d) and e) of Theorem
{\bf I}1.5. By using {\bf I}(1.13), \equ(7.18) and \equ(7.19), we see that
$$\eqalign{
\hat \O^3(\kk) &= \sum_{\s=\pm 1} \left[\fra12 \hat \O^{3,a}(k+2\s p_F,k_0) +
\hat s_{1,\s}(k+2\s p_F,k_0) \right]+\cr
&+\hat \O^{3,b}(\kk) +\hat s_2(\kk) +\hat{\d\O}^{3,c}(\kk)\;,\cr}\Eq(7.41)$$
where we used the definitions
$$s_{1,\s}(\xx) = \sum_{h=h^*}^0 \left({Z_h^{(1)}\over Z_h}\right)^2
s_{1,\s}^{(h)}(\xx)\;,\quad
s_2(\xx)= \sum_{h=h^*}^0 \left({Z_h^{(2)}\over Z_h}\right)^2
s_2{(h)}(\xx)\;,\Eq(7.42)$$
$$\d\O^{3,c}(\xx) = \O^{3,c}(\xx) - s(\xx)\;.\Eq(7.43)$$
 
Since any graph contributing to the expansion of
$\O^{3,a}(\xx-\yy)$ has only two propagators of scale $\le 0$ connected to
$\xx$ or $\yy$, $\hat \O^{3,a}(\kk)$ has support on a set of value of $\kk$
such that $|k|\le 2\g t_0<\p$; hence we can calculate $\hat \O^{3,a}(\kk)$
by thinking $\O^{3,a}(\xx)$ as a function on $\RRR^2$. Let us suppose that
$|\kk|>0$ and $|k|\ge |\kk|/2$; then
$$\hat \O^{3,a}(\kk) = \int d\xx e^{i\kk\xx} \O^{3,a}(\xx)=
{i\over k} \int d\xx \left[e^{i\kk\xx}-1\right] \dpr_x
\O^{3,a}(\xx)\;,\Eq(7.44)$$
since $\O^{3,a}(\xx)$, by ({\bf I}1.16), is a smooth function of fast decrease as
$|\xx|\to\io$. If $|k|< |\kk|/2$, it has to be true that $|k_0|\ge |\kk|/2$
and we write a similar identity, with $k_0$ in place of $k$ and $\dpr_{x_0}$
in place of $\dpr_x$. In both case we can write, by using ({\bf I}1.16),
$$|\hat \O^{3,a}(\kk)| \le {C\over |\kk|} \int_{|\xx|\ge |\kk|^{-1}} {d\xx
\over 1+|\xx|^{3+2\h_1}} + C\int_{|\xx|\le |\kk|^{-1}} d\xx
{|\xx|\over 1+|\xx|^{3+2\h_1}}\;.\Eq(7.45)$$
A even better bound can be proved for $|\hat s_{1,\s}(\kk)|$, $\s=\pm 1$,
by using \equ(6.48). Hence, uniformly for $u\to 0$, $|\hat \O^{3,a}(\kk)|
+|\hat s_{1,\s}(\kk)|\le C |\kk|^{-1}$ for $|\kk|\ge 1$ and
$$\fra12 |\hat \O^{3,a}(\kk)|+|\hat s_{1,\s}(\kk)|
\le C\left[ 1 + {1-|\kk|^{2\h_1}\over 2\h_1}\right]\;,
\quad 0<|\kk|\le 1\;.\Eq(7.46)$$
 
This bound is divergent for $|\kk|\to 0$, if $\h_1<0$, that is if $J_3<0$;
however, if $u\not=0$ and $|\kk|\le \D$, we easily get from ({\bf I}1.16)
(with $n=0$) the better bound
$$\fra12 |\hat \O^{3,a}(\kk)| +|\hat s_{1,\s}(\kk)|\le C \left[
1 + {1-\D^{2\h_1}\over 2\h_1}\right]\;.\Eq(7.47)$$
 
In a similar way, by using ({\bf I}1.17), one can prove that
$$|\hat \O^{3,b}(\kk)|+|\hat s_2(\kk)|\le C\left[ 1 + \log |\kk|^{-1}\right]
\;,\quad 0<|\kk|\le 1\;,\Eq(7.48)$$
$$|\hat \O^{3,b}(\kk)| +|\hat s_2(\kk)|\le C \left[ 1 + \log \D^{-1}
\right]\;.\Eq(7.49)$$
However, a more careful analysis of the Fourier transform of the leading
contribution to $\O^{3,b}(\xx)$, given by $\O^*_2(\xx)$ (see \equ(7.34)),
which takes into account the oddness in the exchange $(x,x_0)\rightarrow
(x_0 v_0^*,x/v_0^*)$, shows that $|\hat\O^*_2(\kk)|\le C$. One can show
that a similar bound is satisfied by the Fourier transform of the terms
contributing to $\tilde\O_2(\xx)$ and proportional to $\s_h/\g^h$.
Therefore, in the bounds \equ(7.48) and \equ(7.49), we can
multiply by $J_3$ both $\log |\kk|^{-1}$ and $\log \D^{-1}$.

Let us now consider $\hat{\d\O}^{3,c}(\kk)$. By using \equ(7.26), we see
immediately that, uniformly in $\kk$ and $u$,
$$|\hat{\d\O}^{3,c}(\kk)-\hat s(\kk)| \le C\;.\Eq(7.50)$$
The bounds \equ(7.46)-\equ(7.50), together with the positivity of the
leading term in ({\bf I}1.20) and the remark after \equ(7.49),
immediately imply all the claims in item d) of Theorem {\bf I}1.5.
 
\*
 
Let us now consider $G(x)\=\O^3(x,0)$, $x\in\ZZZ$. It is easy to see, by using
the previous results and the fact that also $s_{1,\s}(\xx)$ and $s_2(\xx)$
are even functions of $\xx$, that $G(x)$ can be written in the form
$$G(x)= \sum_{\s=\pm 1} e^{2i\s p_F x} G_{1,\s}(x) + G_2(x)+\d G(x)\;,
\Eq(7.51)$$
where $G_{1,\s}(x)$ and $G_2(x)$ are the restrictions to $\ZZZ$ of some even
smooth functions on $\RRR$, satisfying, for any integers $n,N\ge 0$, the bounds
$$|\dpr_x^n G_{1,\s}(x)|\le {C_{n,N}\over [1+|x|^{2+n+2\h_1}]
[1+(\D|x|)^N]}\;,\Eq(7.52)$$
$$|\dpr_x^n G_2(x)|\le {C_{n,N}\over 1+|x|^{2+n}
[1+(\D|x|)^N]}\;,\Eq(7.53)$$
while $\d G(x)$ satisfies the bound
$$|\d G(x)| \le {C\over [1+|x|^{2+\th}][1+(\D|x|)^N]}\;,\Eq(7.54)$$
with some $\th>0$.
 
These properties immediately imply that, uniformly in $k$ and $u$,
$$|\hat G(k)|+ |\dpr_k \d\hat G(k)| \le C\;.\Eq(7.55)$$
Let us now consider $\dpr_k\hat G_{1,\s}(k)$ and note that, if $|k|>0$,
$$\eqalign{
\dpr_k\hat G_{1,\s}(k) &= -{1\over k}\int dx [e^{ikx}-1] \dpr_x[x
G_{1,\s}(x)]=\cr
&= -{1\over k}\int_{|x|\ge |k|^{-1}} dx [e^{ikx}-1] \dpr_x[xG_{1,\s}(x)]-\cr
&-{1\over k}\int_{|x|\le |k|^{-1}} dx [e^{ikx}-1-ikx] \dpr_x[xG_{1,\s}(x)]
\;,\cr}\Eq(7.56)$$
where we used the fact that $\dpr_x[xG_{1,\s}(x)]$ is an even function of $x$,
since $G_{1,\s}(x)$ is even, see \equ(6.45).
Hence, if $|k|\ge 1$, $|\dpr_k\hat G_{1,\s}(k)|\le C |k|^{-1}$, while, if
$0<|k|\le 1$, uniformly in $u$,
$$|\dpr_k\hat G_{1,\s}(k)| \le C[1+|k|^{2\h_1}]\;.\Eq(7.57)$$
In a similar way, we can prove that, uniformly in $k$ and $u$,
$$|\dpr_k\hat G_2(k)| \le C\;.\Eq(7.58)$$
 
The bound \equ(7.57) is divergent for $k\to 0$, if $J_3<0$; however,
if $|u|>0$ and $|k|\le \D$, one can get a better bound, by using
the identity
$$\dpr_k\hat G_{1,\s}(k) = i\int_{|x|\ge \D^{-1}} dx e^{ikx}[xG_{1,\s}(x)]+
i\int_{|x|\le \D^{-1}} dx [e^{ikx}-1][xG_{1,\s}(x)]\;,\Eq(7.59)$$
together with \equ(7.52). One finds
$$|\dpr_k\hat G_{1,\s}(k)| \le C [1+ \D^{2\h_1}]\;.\Eq(7.60)$$
 
The bounds \equ(7.55), \equ(7.58) and \equ(7.60), together with the
identity \equ(7.51), imply ({\bf I}1.24). The statements about the
discontinuities of $\dpr_k \hat G(k)$ at $u=0$ and $k=0,\pm 2p_F$ follow from
an explicit calculation involving the leading contribution, obtained by
putting $A_1(\xx)=A_2(\xx)=0$ in ({\bf I}1.20).

\pagina
\vskip1.cm
 
\section(8, Proof of the approximate Ward identity {\equ(6.35)})

\sub(8.1) In this section we prove the relation \equ(6.35) between the
quantities $Z^{(L)}_h$ and $Z_h^{(2,L)}$, related to the approximate
Luttinger model defined by \equ(6.30) and \equ(6.31).
 
First of all, we move from the interaction to the free measure \equ(5.30)
the term proportional to $\d_0^{(L)}$ and we redefine correspondingly the
interaction. This can be realized by slightly changing the free measure
normalization (which has no effect on the problem we are studying), by
putting $\d_0^{(L)}=0$ in \equ(5.31) and by substituting, in \equ(5.30),
$v_0^*$ with $\bar v_0(\kk')=v_0^*+\d_0^{(L)} C_0^{-1}(\kk')$. However,
since $C_0^{-1}(\kk')=1$ on all scales $h<0$, $Z_h^{(2,L)}$ and $Z_h^{(L)}$
may be modified only by a factor $\g^{C|\l_0^{(L)}|}$, if we substitute
$\bar v_0(\kk')$ with $\bar v_0\=\bar v_0({\bf 0})$. It follows that it is
sufficient to prove the bound \equ(6.35) for the corresponding free
measure
$$\eqalign{
&P^{(L)}(d\psi^{(\le 0)}) =
\prod_{\kk':C_0^{-1}(\kk')>0}\prod_{\o=\pm1}
{d\hat\psi^{(\le 0)+}_{\kk',\o} d\hat\psi^{(\le 0)-}_{\kk',\o}\over
\NN_L(\kk')}\;\cdot\cr
&\cdot\; \exp \left\{-{1\over L\b} \sum_{\o=\pm 1}
\sum_{\kk':C_0^{-1}(\kk')>0} C_0(\kk') \big( -ik_0+\o \bar v_0 k' \big)
\hat\psi^{(\le 0)+}_{\kk',\o} \hat\psi^{(\le 0)-}_{\kk',\o}\right\}\;,\cr}
\Eq(8.1a)$$
by using as interaction the function
$$V^{(L)}(\psi^{(\le 0)}) = \l_0^{(L)} \int_{\TTT_{L,\b}} d\xx\;
\psi^{(\le 0)+}_{\xx,+1}\psi^{(\le 0)-}_{\xx,-1}
\psi^{(\le 0)+}_{\xx,-1}\psi^{(\le 0)-}_{\xx,+1}\;.\Eq(8.1b)$$

Let us consider, instead of the free measure \equ(8.1a), the corresponding
measure with {\it infrared cutoff on scale $h$}, $h\le 0$, given by
$$\eqalign{
&P^{(L,h)}(d\psi^{[h,0]}) =
\prod_{\kk':C_{h,0}^{-1}(\kk')>0}\prod_{\o=\pm1}
{d\hat\psi^{[h,0]+}_{\kk',\o} d\hat\psi^{[h,0]-}_{\kk',\o}\over
\NN_L(\kk')}\;\cdot\cr
&\cdot\; \exp \left\{-{1\over L\b} \sum_{\o=\pm 1}
\sum_{\kk':C_{h,0}^{-1}(\kk')>0} C_{h,0}(\kk') \big( -ik_0+\o \bar v_0
k' \big) \hat\psi^{[h,0]+}_{\kk',\o} \hat\psi^{[h,0]-}_{\kk',\o}\right\}
\;,\cr}\Eq(8.1)$$
where $C_{h,0}^{-1}=\sum_{k=h}^0 f_k$.
 
We will find convenient to write the above integration in terms of the
space-time field variables; if we put
$$\DD\psi^{[h,0]}=\prod_{\kk':C_{h,0}^{-1}(\kk')>0}\prod_{\o=\pm1}
{d\hat\psi^{[h,0]+}_{\kk',\o} d\hat\psi^{[h,0]-}_{\kk',\o}\over
\NN_L(\kk')}\;,\Eq(8.2)$$
we can rewrite \equ(8.1) as
$$P^{(L,h)}(d\psi^{[h,0]})=\DD\psi^{[h,0]}
\exp\Big[-\sum_\o\int_{\TTT_{L,\b}} d\xx\;\psi^{[h,0]+}_{\xx,\o}
D^{[h,0]}_\o\psi^{[h,0]-}_{\xx,\o}\Big]\;,\Eq(8.3)$$
where
$$D^{[h,0]}_\o\psi^{[h,0]\s}_{\xx,\o}={1\over L\b} \sum_{\kk':C_{h,0}^{-1}
(\kk')>0} e^{i\s\kk'\xx} C_{h,0}(\kk') (i\s k_0 - \o\s \bar v_0 k')
\hat\psi^{[h,0]\s}_{\kk',\o}\;.\Eq(8.4)$$
 
$D^{[h,0]}_\o$ has to be thought as a ``regularization'' of the linear
differential operator
$$D_\o = {\dpr \over \dpr x_0} + i\o \bar v_0 {\dpr \over \dpr x}\;.\Eq(8.5)$$
 
Let us now introduce the external field variables $\phi^\s_{\xx,\o}$, $\xx
\in\TTT_{L,\b}$, $\o=\pm 1$, antiperiodic in $x_0$ and $x$ and anticommuting
with themselves and $\psi^{[h,0]\s}_{\xx,\o}$, and let us define
$$U(\phi)=-\log\int P^{(L,h)}(d\psi^{[h,0]})e^{-V^{(L)}(\psi^{[h,0]}
+\phi)}\;.\Eq(8.6)$$
 
If we perform the {\it gauge transformation}
$$\psi^{[h,0]\s}_{\xx,\o}\to e^{i\s\a_\xx}
\psi^{[h,0]\s}_{\xx,\o}\;,\Eq(8.7)$$
and we define $(e^{-i\a}\phi)^\s_{\xx,\o} = e^{-i\s\a_\xx}
\phi^\s_{\xx,\o}$, we get
$$\eqalign{
U(\phi) &= -\log\int P^{(L,h)}(d\psi^{[h,0]})
\exp\Big\{-V^{(L)}(\psi^{[h,0]} + e^{-i\a}\phi) -\cr
&-\sum_\o\int d\xx\; \psi^{[h,0]+}_{\xx,\o}
\Big( e^{i\a_\xx} D^{[h,0]}_\o e^{-i\a_\xx}- D^{[h,0]}_\o \Big)
\psi^{[h,0]-}_{\xx,\o}\Big\}\;.\cr}\Eq(8.8)$$
Since $U(\phi)$ is independent of $\a$, the functional derivative of the
r.h.s. of \equ(8.8) w.r.t. $\a_\xx$ is equal to $0$ for any $\xx\in
\TTT_{L,\b}$. Hence, we find the following identity:
$$\sum_\o\left[-\phi^+_{\xx,\o}{\partial U \over \partial\phi^+_{\xx,\o}}
+{\partial U \over \partial\phi^-_{\xx,\o}}\phi^-_{\xx,\o}+
{1\over Z(\phi)} \int P^{(L,h)}(d\psi^{[h,0]})  T_{\xx,\o}\;
e^{-V^{(L)}(\psi^{[h,0]}+\phi)}\right]=0\;,\Eq(8.9)$$
where
$$Z(\phi) = \int P^{(L,h)}(d\psi^{[h,0]}) e^{-V^{(L)}(\psi^{[h,0]}+\phi)}
\;,\Eq(8.10)$$
$$\eqalign{
&\qquad T_{\xx,\o} = \psi^{[h,0]+}_{\xx,\o}[D^{[h,0]}_\o\psi^{[h,0]-}_{\xx,\o}]
+ [D^{[h,0]}_\o\psi^{[h,0]+}_{\xx,\o}]\psi^{[h,0]-}_{\xx,\o}=\cr
&= {1\over (L\b)^2} \sum_{\pp,\kk} e^{-i\pp\xx}
\hat\psi^{[h,0],+}_{\kk,\o}
[C_{h,0}(\pp+\kk) D_\o(\pp+\kk)-C_{h,0}(\kk)D_\o(\kk)]
\hat\psi^{[h,0],-}_{\pp+\kk,\o}\;,\cr}\Eq(8.11)$$
$$D_\o(\kk)=-ik_0+\o \bar v_0 k\;.\Eq(8.12)$$
Moreover, the sum over $\pp$ and $\kk$ in \equ(8.11) is restricted to the
momenta of the form $\pp=(2\p n/L,2\p m/\b)$ and $\kk=(2\p(n+1/2)/L,
2\p(m+1/2)/\b)$, with $n$ and $m$ integers, such that $|p|$, $|p_0|$, $|k_0|$,
$|k|$ are all smaller or equal to $\p$ and satisfy the constraints
$C^{-1}_{h,0}(\pp+\kk)>0$, $C^{-1}_{h,0}(\kk)>0$.
 
Note that \equ(8.11) can be rewritten as
$$T_{\xx,\o} = D_\o [\psi^{[h,0]+}_{\xx,\o} \psi^{[h,0]-}_{\xx,\o}]+
\d T_{\xx,\o}\;,\Eq(8.13)$$
where
$$\eqalign{
\d T_{\xx,\o} &= {1\over (L\b)^2} \sum_{\pp,\kk} e^{-i\pp\xx}
\hat\psi^{[h,0],+}_{\kk,\o}\;\cdot\cr
&\cdot\;\big\{[C_{h,0}(\pp+\kk)-1] D_\o(\pp+\kk)-[C_{h,0}(\kk)-1]D_\o(\kk)\big\}
\hat\psi^{[h,0],-}_{\pp+\kk,\o}\;.\cr}\Eq(8.14)$$
It follows that, if $C_{h,0}$ is substituted with $1$, that is if we consider
the formal theory without any ultraviolet and infrared cutoff, $T_{\xx,\o} =
D_\o [\psi^{[h,0]+}_{\xx,\o}\psi^{[h,0]-}_{\xx,\o}]$ and we would get the
usual Ward identities. As we shall see, the presence of the cutoffs make the
analysis a bit more involved and adds some corrections to the Ward identities,
which however, for $\l_0$ small enough, can be controlled by the same type of
multiscale analysis, that we used in \sec(6).

\*
\sub(8.2) Let us introduce a new external field $J_\xx$,
$\xx\in\TTT_{L,\b}$, periodic in $x_0$ and $x$ and commuting
with the fields $\phi^\s$ and $\psi^{[h,0]\s}$, and let us
consider the functional
$$\WW(\phi,J)=-\log \int
P^{(L,h)}(d\psi^{[h,0]}) e^{-V^{(L)}(\psi^{[h,0]}+\phi)+\int d\xx J_\xx
\sum_\o \psi^{[h,0]+}_{\xx,\o}\psi^{[h,0]-}_{\xx,\o}}\;.\Eq(8.15)$$
We also define the functions
$$\Sigma_{h,\o}(\xx-\yy)=
{\dpr^2\over\dpr\phi^+_{\xx,\o}\dpr\phi^-_{\yy,\o}}
U(\phi)\Big|_{\phi=0}=
{\dpr^2\over \dpr\phi^+_{\xx,\o} \dpr\phi^-_{\yy,\o}}
\WW(\phi,J)\Big|_{\phi=J=0}\;,\Eq(8.16)$$
$$\G_{h,\o}(\xx;\yy,\zz)={\dpr\over\dpr J_\xx}
{\dpr^2\over\dpr\phi^+_{\yy,\o}\dpr\phi^-_{\zz,\o}}
\WW(\phi,J)|_{\phi=J=0}\;.\Eq(8.17)$$
 
These functions have here the role of the {\it self-energy} and the {\it
vertex part} in the usual treatment of the Ward identities. However, they do
not coincide with them, because the corresponding Feynman graphs expansions
are not restricted to the one particle irreducible graphs. However, their
Fourier transforms at zero external momenta, which are the interesting
quantities in the limit $L,\b\to\io$, are the same; in fact, because of the
support properties of the fermion fields, the propagators vanish at zero
momentum, hence the one particle reducible graphs give no contribution at that
quantities.
 
In the language of this paper, if we did not perform any free measure
regularization, $\Sigma_{h,\o}(\xx-\yy)$ would coincide with the kernel of the
contribution to the effective potential on scale $h-1$ with two external
fields, that is the function $W^{(h-1)}_{2,(+,-),(\o,\o)}$ of equation
({\bf I}3.3). Analogously, $1+\G_{h,\o}(\xx;\yy,\zz)$ would coincide with the
kernel $B^{(h-1)}_{1,2,(+,-),(\o,\o)}$ of equation \equ(6.6).
 
Note that
$$\G_{h,\o}(\xx;\yy,\zz)=\sum_{\tilde\o}\G_{h,\o,\tilde\o}(\xx;\yy,\zz)\;,
\Eq(8.18)$$
where $\G_{h,\o,\tilde\o}(\xx;\yy,\zz)$ is defined as in \equ(8.15), by
substituting $J_\xx\sum_\o\psi^{[h,0]+}_{\xx,\o}
\psi^{[h,0]-}_{\xx,\o}$ with
$J_\xx$ $\psi^{[h,0]+}_{\xx,\tilde\o}$ $\psi^{[h,0]-}_{\xx,\tilde\o}$.
 
If we derive the l.h.s. of \equ(8.9) with respect to $\phi^+_{\yy,\o}$ and
to $\phi^-_{\zz,\o}$ and we put $\phi=0$, we get
$$\eqalignno{
& 0=-\d(\xx-\yy)\Sigma_{h,\o}(\xx-\zz)+\d(\xx-\zz)\Sigma_{h,\o}(\yy-\xx)-
&\eq(8.19)\cr
&<\left[{\dpr^2 V\over\dpr\psi^{[h,0]+}_{\yy,\o}\dpr\psi^{[h,0]-}_{\zz,\o}}-
{\dpr V\over \dpr\psi^{[h,0]+}_{\yy,\o}}
{\dpr V\over \dpr\psi^{[h,0]-}_{\zz,\o}}\right]\;;\;
\sum_{\tilde\o}\Big[D_{\tilde\o}(\psi^{[h,0]+}_{\xx,\tilde\o}
\psi^{[h,0]-}_{\xx,\tilde\o})+\d T_{\xx,\tilde\o}\Big]>^T\;,\cr}$$
where $<\cdot;\cdot>^T$ denotes the truncated expectation w.r.t. the measure
$Z(0)^{-1} P^{(L,h)}(d\psi^{[h,0]})$ $e^{-V^{(L)}(\psi^{[h,0]})}$.
 
By using the definitions \equ(8.16) and \equ(8.17), equation \equ(8.19)
can be rewritten as
$$\eqalign{
0&=-\d(\xx-\yy)\Sigma_{h,\o}(\xx-\zz)+\d(\xx-\zz)\Sigma_{h,\o}(\yy-\xx)-\cr
&-\sum_{\tilde\o} D_{\xx,\tilde\o}\G_{h,\o,\tilde\o}(\xx;\yy,\zz)
-\D_{h,\o}(\xx;\yy,\zz)\;,\cr}\Eq(8.20)$$
where
$$\D_{h,\o}(\xx;\yy,\zz)=
<\left[{\dpr^2 V\over \dpr\psi^+_{\yy,\o}\dpr\psi^-_{\zz,\o}}-
{\dpr V\over \dpr\psi^+_{\yy,\o}}
{\dpr V\over \dpr\psi^-_{\zz,\o}}\right]\;;\;
\sum_{\tilde\o}\d T_{\xx,\tilde\o}>^T\;.\Eq(8.21)$$
 
In terms of the Fourier transforms, defined so that, in agreement with
({\bf I}3.2) and \equ(6.9),
$$\Sigma_{h,\o}(\xx-\yy)={1\over L\b} \sum_{\kk}
e^{-i\kk(\xx-\yy)}\hat\Sigma_{h,\o}(\kk)\;,\Eq(8.22)$$
$$\G_{h,\o,\tilde\o}(\xx;\yy,\zz)={1\over (L\b)^2} \sum_{\pp,\kk}
e^{i\pp(\xx-\zz)}e^{-i\kk(\yy-\zz)}
\hat\G_{h,\o,\tilde\o}(\pp,\kk)\;,\Eq(8.23)$$
$$\D_{h,\o}(\xx;\yy,\zz)={1\over (L\b)^2} \sum_{\pp,\kk}
e^{i\pp(\xx-\zz)}e^{-i\kk(\yy-\zz)}
\hat\D_{h,\o}(\pp,\kk)\;,\Eq(8.24)$$
\equ(8.20) can be written as
$$0=\hat\Sigma_{h,\o}(\kk-\pp)-\hat\Sigma_{h,\o}(\kk)+\sum_{\tilde\o}
(-i p_0+\tilde\o p)
\hat\G_{h,\o,\tilde\o}(\pp,\kk) +\hat\D_{h,\o}(\pp,\kk)\;.\Eq(8.25)$$
 
Let us now define
$$\tilde Z_h^{(2)}=1+{1\over 4}\sum_{\h,\h'=\pm 1}
\hat\G_{h,\o}(\bar\pp_{\h'},\bar\kk_{\h,\h'})\;,\Eq(8.26)$$
$$\tilde Z_h=1+{i\over 4}\sum_{\h,\h'=\pm 1}\h'{\b\over\pi}
\hat\Sigma_{h,\o}(\bar\kk_{\h,\h'})\;,\Eq(8.27)$$
where $\bar\pp_{\h'}$ is defined as in \equ(6.11) and $\bar\kk_{\h,\h'}$ as
in ({\bf I}2.73).
 
If we put in \equ(8.25) $\pp=\bar\pp_{\h'}$ and $\kk=\bar\kk_{\h,\h'}$,
multiply both sides by $(i\h'\b)/(2\p)$ and sum over $\h,\h'$, we get
$$\tilde Z_h = \tilde Z_h^{(2)}+\d\tilde Z_h^{(2)}\;,\Eq(8.28)$$
where
$$\d\tilde Z_h^{(2)}={1\over 4}\sum_{\h,\h'=\pm 1}
{\hat\D_{h,\o}(\bar\pp_{\h'},\bar\kk_{\h,\h'})\over -i \bar p_{\h'0}}
\;.\Eq(8.29)$$
 
\* \sub(8.3) The considerations preceding \equ(8.19) suggest that $\tilde Z_h$
and $\tilde Z^{(2)}_h$ are ``almost equal'' to the quantities $Z^{(L)}_h$ and
$Z^{(2,L)}_h$, related to the full approximate Luttinger model and defined
analogously to $Z_h$ and $Z^{(2)}_h$ for the original model, on the base of a
multiscale analysis. In order to clarify this point, we consider the measure
$P^{(L,h)}(d\psi^{[h,0]}) e^{-V^{(L)}(\psi^{[h,0]}+ \psi^{(<h)})}$, where
$\psi^{(<h)}$ is fixed and has the same role of the external field $\phi$ in
\equ(8.15), and define $\bar E_{h-1}$ and $\bar\VV^{(h-1)}(\psi^{(<h)})$, the
{\it one step effective potential} on scale $h-1$, so that
$\bar\VV^{(h-1)}(0)=0$ and
$$e^{-\bar\VV^{(h-1)}(\psi^{(<h)})-L\b \bar E_{h-1}}=
\int P^{(L,h)}(d\psi^{[h,0]})e^{-V^{(L)}(\psi^{[h,0]}+\psi^{(<h)})}\;.
\Eq(8.30)$$
We want to calculate this quantity, by extending to it
the definitions of effective potentials and running couplings, given in
\S{\bf I}2 for the original model.
 
We start from the scale $0$ with potential $\VV^{'(0)}(\psi^{[h,0]},
\psi^{(<h)})= V^{(L)}(\psi^{[h,0]}+\psi^{(<h)})$ and we introduce, in
analogy to the procedure described in \S{\bf I}2.5, for each $\tdh$ such that
$h\le \tdh\le 0$, two constants $Z'_\tdh$, $E'_\tdh$ and an effective
potential $\VV^{'(\tdh)}(\psi,\psi^{(<h)})$, so that $Z'_0=1$, $E'_0=0$ and
$$e^{-\bar\VV^{(h-1)}(\psi^{(<h)})-L\b \bar E_{h-1}}=
\int P_{Z'_\tdh,C_{h,\tdh}}(d\psi^{[h,\tdh]})e^{-\VV^{'(\tdh)}
(\sqrt{Z'_\tdh}\psi^{[h,\tdh]},\psi^{(<h)})-E'_\tdh}\;,\Eq(8.31)$$
where $P_{Z'_\tdh,C_{h,\tdh}}(d\psi^{[h,\tdh]})$ is obtained from the
analogous definition ({\bf I}2.66), by putting $\s_\tdh(\kk')$ $=0$, $E(\kk')=
\bar v_0\sin k'$, and by substituting $C^{-1}_\tdh$ with $C^{-1}_{h,\tdh}=
\sum_{k=h}^\tdh f_k$. Moreover, we suppose that the localization procedure
is applied also to the field $\psi^{(<h)}$, even if it does appear in the
integration measure and, therefore, can not be involved in the free measure
renormalization.
 
We want to compare these effective potentials with the potentials
$\VV^{(\tdh)}(\psi^{(\le \tdh)})$, related to the approximate Luttinger model
without any infrared cutoff and defined following again the procedure
described in \S{\bf I}2. We shall use for the various objects related to this
model the same notation of \S{\bf I}2, while the corresponding objects of the
model with infrared cutoff will be distinguished with a superscript $'$. The
definitions are such that $\VV^{(0)}(\psi^{(\le 0)}) =
\VV^{'(0)}(\psi^{[h,0]}, \psi^{(<0)})$, $Z_0=1$ and
$$\int P_{Z_0,C_0}(d\psi^{(\le 0)}) \, e^{-\VV^{(0)}
(\sqrt{Z_0}\psi^{(\le 0)})} =
\int P_{Z_\tdh,C_\tdh}(d\psi^{(\le \tdh)}) \, e^{-\VV^{(\tdh)}
(\sqrt{Z_\tdh}\psi^{(\le \tdh)})-L\b E_\tdh}\;.\Eq(8.32)$$
 
Note that the single scale propagators involved in the calculation of
$\VV^{(\tdh)}(\sqrt{Z_\tdh}\psi^{(\le \tdh)})$ and
$\VV^{'(\tdh)}(\sqrt{Z'_\tdh}\psi^{[h,\tdh]},\psi^{(<h)})$, that is those with
scale $\bar h\ge \tdh+1$, may differ only if $Z_{\bar h}\not=Z'_{\bar h}$ or
$z_{\bar h}\not=z'_{\bar h}$. This immediately follows from the observation
that, if $h+1\le\tdh\le 0$, the identity ({\bf I}2.90) is satisfied even if we
substitute in ({\bf I}2.89) $C_\tdh$ with $C_{h,\tdh}$. This implies, in
particular, since $z_0=z'_0=0$, that (see ({\bf I}2.110) and ({\bf I}2.107))
$$\VV^{'(-1)}(\sqrt{Z'_{-1}}\psi^{[h,-1]},\psi^{(<h)})=
\VV^{(-1)}[\sqrt{Z_{-1}}(\psi^{[h,-1]}+\psi^{(<h)})]\;,\Eq(8.33)$$
with $Z'_{-1}=Z_{-1}=1$, and that $z_{-1}=z'_{-1}$, $\d_{-1}=\d'_{-1}$,
$\l_{-1}=\l'_{-1}$, $Z'_{-2}=Z_{-2}$.
 
Let us now compare the effective potentials on scale $-2$. The fact that the
free measure in \equ(8.31) does not depend on the fields with scale less than
$h$ implies that the free measure renormalization does not use all the local
part of $\VV^{'(-1)}$ proportional to $z_{-1}$. Therefore, the
analogous of the potential $\hat\VV^{(-1)}(\sqrt{Z_{-2}}$ $\psi^{(\le -1)})$
for the model with infrared cutoff has to be defined so that (see ({\bf I}2.107))
$$\eqalign{
&\hat\VV^{'(-1)}(\sqrt{Z_{-2}}\psi^{[h,-1]},\psi^{(<h)})=
\hat\VV^{(-1)}[\sqrt{Z_{-2}}(\psi^{[h,-1]}+\psi^{(<h)})]+\cr
&+ z_{-1} Z_{-1} \sum_{\o=\pm 1} \int d\xx\; \left[
-\Big(D_\o \psi^{[h,-1]+}_{\xx,\o}\Big) \psi^{(<h)-}_{\xx,\o} +
\psi^{(<h)+}_{\xx,\o} \Big(D_\o \psi^{[h,-1]-}_{\xx,\o}\Big) \right] +\cr
&+z_{-1} Z_{-1} \sum_{\o=\pm 1} \int d\xx\;
\psi^{(<h)+}_{\xx,\o} D_\o \psi^{(<h)-}_{\xx,\o}
\;.\cr}\Eq(8.34)$$
 
It follows, by using also the remark on the single scale propagators following
\equ(8.32), that $\VV^{'(-2)}(\sqrt{Z_{-2}}\psi^{[h,-2]},\psi^{(<h)})$,
calculated through the analogous of ({\bf I}2.110), can be obtained from
$\VV^{(-2)}[\sqrt{Z_{-2}}(\psi^{[h,-2]}+\psi^{(<h)})]$ by adding some new
terms. First of all, there is the term in the third line of \equ(8.34), which
is independent of the integration variables, and the two terms in the second
line with $\psi^{[h,-2]\s}_{\xx,\o}$ in place of $\psi^{[h,-1]\s}_{\xx,\o}$.
Moreover, in the Feynman graph expansion, we have to add the graphs which are
obtained by inserting, in the external lines of a graph contributing to
$\VV^{(-2)}$, one or more vertices corresponding to the two terms in the
second line of \equ(8.34). These new terms are not irrelevant, if the number
of external lines is $2$ or $4$; hence one could worry about the need of new
running couplings in order to regularize the expansion. However, because of
the support properties of the propagators, these new terms do not give any
contribution to the local part (which is calculated by putting equal to
$\bar\kk_{\h,\h'}$ the external momenta, hence also the momenta of the
internal line propagators of the insertions in the external lines), so that
the only running couplings to consider are those related with $\VV^{(-2)}$ and
their values are the same, that is $z_{-2}=z'_{-2}$, $\d_{-2}=\d'_{-2}$,
$\l_{-2}=\l'_{-2}$, $Z'_{-3}=Z_{-3}$.
 
By iterating the previous considerations, it is easy to show that, if
$h\le\tdh\le -2$, one can calculate
$\VV^{'(\tdh)}(\sqrt{Z_{\tdh}}\psi^{[h,\tdh]},\psi^{(<h)})$ by adding to
$\VV^{(\tdh)}[\sqrt{Z_{\tdh}}(\psi^{[h,\tdh]} +\psi^{(<h)})]$ some
new terms. First of all, there are the local terms of the form of that in the
second and the third line of \equ(8.34), with $\psi^{[h,\tdh]\s}_{\xx,\o}$ in
place of $\psi^{[h,-1]\s}_{\xx,\o}$ and $z_{\bar h} Z_{\bar h}$, $\tdh\le \bar
h\le -1$ in place of $z_{-1} Z_{-1}$. Moreover, in the Feynman graph
expansion, we have to add the graphs, which are obtained by inserting, in the
external lines of a graph contributing to $\VV^{(\tdh)}$, one or more vertices
corresponding to terms similar to those in the second line of \equ(8.34), with
$\psi^{[h,\tdh]\s}_{\xx,\o}$ in place of $\psi^{[h,-1]\s}_{\xx,\o}$ and
$z_{\bar h} Z_{\bar h}$, $\tdh\le \bar h\le -1$ in place of $z_{-1} Z_{-1}$.
Finally
$$\eqalign{
&\LL \VV^{'(\tdh)}(\sqrt{Z_{\tdh}}\psi^{[h,\tdh]},\psi^{(<h)})=
\LL \VV^{(\tdh)}[\sqrt{Z_{\tdh}}(\psi^{[h,\tdh]} +\psi^{(<h)})]+\cr
&+\sum_{\bar h=\tdh+1}^{-1} z_{\bar h} Z_{\bar h} \sum_{\o=\pm 1} \int d\xx\;
\left[-\Big(D_\o \psi^{[h,\tdh]+}_{\xx,\o}\Big) \psi^{(<h)-}_{\xx,\o} +
\psi^{(<h)+}_{\xx,\o} \Big(D_\o \psi^{[h,\tdh]-}_{\xx,\o}\Big) \right] +\cr
&+\sum_{\bar h=\tdh+1}^{-1} z_{\bar h} Z_{\bar h} \sum_{\o=\pm 1} \int d\xx\;
\psi^{(<h)+}_{\xx,\o} D_\o \psi^{(<h)-}_{\xx,\o}
\;,\cr}\Eq(8.35)$$
and all the running couplings, as well as the renormalization constants, are
the same as those defined through $\VV^{(\tdh)}(\sqrt{Z_{\tdh}}\psi^{(\le
\tdh)})$.
 
Equations \equ(8.31) and \equ(8.35) also imply that
$$\eqalign{
&\LL\bar\VV^{(h-1)}(\psi^{(<h)}) = \LL\VV^{'(h-1)}(\psi^{(<h)})+\cr
&+\sum_{\bar h=h+1}^{-1} z_{\bar h} Z_{\bar h} \sum_{\o=\pm 1} \int d\xx\;
\psi^{(<h)+}_{\xx,\o} D_\o \psi^{(<h)-}_{\xx,\o}+
z'_h Z_h \sum_{\o=\pm 1} \int d\xx\;
\psi^{(<h)+}_{\xx,\o} D_\o \psi^{(<h)-}_{\xx,\o}
\;,\cr}\Eq(8.36)$$
where $\VV^{'(h-1)}(\psi^{(<h)})$ is obtained from $\VV^{(h-1)}(\psi^{(<h)})$
``almost'' as before. We still have to add some new graphs with suitable
insertions on the external lines, which do not affect the local part, but we
also have to change the propagators of scale $h$, since the function $\tilde
f'_h(\kk')$, calculated as $\tilde f_h(\kk')$, see ({\bf I}2.90), with
$C^{-1}_{h,h}=f_h$ in place of $C^{-1}_h$, is different from $\tilde
f_h(\kk')$.
 
The definition \equ(8.27) of $\tilde Z_h$ and the definition of $\LL$,
together with \equ(8.36), imply that
$$\tilde Z_h = 1 + \sum_{\bar h=h+1}^{-1} z_{\bar h} Z_{\bar h} +z'_h Z_h=
Z_h (1 + z'_h)\;.\Eq(8.37)$$
Since $Z^{(L)}_h=Z_h$ and $|z'_h|\le C|\l_0|^2$, if $\l_0$ is small enough,
as one can show by using the arguments of \sec(5), we get the bound
$$\left|{\tilde Z_h\over Z^{(L)}_h}-1\right| \le C|\l_0|\;.\Eq(8.38)$$
 
A similar argument can be used for $Z^{(2,L)}_h$, by using the results of
\sec(6), and we get the similar bound
$$\left|{\tilde Z^{(2)}_h\over Z^{(2,L)}_h}-1\right| \le C|\l_0|\;.\Eq(8.39)$$
 
We will prove in \sec(8.3) that
$$|\d\tilde Z_h^{(2)}|\le C Z_h^{(2,L)}|\l_0|\;,\Eq(8.40)$$
so that we finally get
$$|{Z_h^{(L)}\over Z_h^{(2,L)}}-1|\le C|\l_0|\;,\Eq(8.41)$$
implying \equ(6.35).

\*
{\bf Remark} \equ(8.40) shows that the corrections to the {\it exact}
Ward identity $Z^{(L)}_h = Z_h^{(2,L)}$
could diverge as $h\to-\io$. This is not important in our proof, since
we are only interested in the ratio $Z^{(L)}_h/ Z_h^{(2,L)}$,
which is near to $1$, but suggests that it would be difficult to prove the
approximate Ward identity, by directly looking at the cancellations in
presence of the cutoffs.
 
\*
\sub(8.4) In order to prove \equ(8.40), we note that
$$\eqalign{
&[C_{h,0}(\pp+\kk)-1] D_\o(\pp+\kk)-[C_{h,0}(\kk)-1]D_\o(\kk)=\cr
&D_\o(\pp) [C_{h,0}(\pp+\kk)-1] + C_{h,0}(\pp+\kk) D_\o(\kk) C_{h,0}(\kk)
[C^{-1}_{h,0}(\kk)-C^{-1}_{h,0}(\pp+\kk)]\;,\cr}\Eq(8.42)$$
and that
$$\eqalign{
&C_{h,0}(\bar\pp_{\h'}+\kk) {[C^{-1}_{h,0}(\kk)-C^{-1}_{h,0}(\bar\pp_{\h'}
+\kk)]\over -i \bar p_{\h'0}} =\cr
&\left. C_{h,0}(\pp+\kk) {[C^{-1}_{h,0}(\kk)-C^{-1}_{h,0}(\bar\pp_{\h'}
+\kk)]\over -i \bar p_{\h'0}} \right|_{\pp=\bar\pp_{\h'}}\;,\cr}
\Eq(8.43)$$
$$C_{h,0}(\pp+\kk)=1+ [C_{h,0}(\pp+\kk)-1]\;.\Eq(8.44)$$
Hence, by using \equ(8.14) and \equ(8.21), we can write
$${\hat\D_{h,\o}(\bar\pp_{\h'},\bar\kk_{\h,\h'})\over -i \bar p_{\h'0}}=
\hat\D^{(1)}_{h,\o,\h'}(\bar\pp_{\h'},\bar\kk_{\h,\h'})\;,\Eq(8.45)$$
where
$$\D^{(1)}_{h,\o,\h'}(\xx;\yy,\zz) =
<\left[{\dpr^2 V\over \dpr\psi^+_{\yy,\o}\dpr\psi^-_{\zz,\o}}-
{\dpr V\over \dpr\psi^+_{\yy,\o}}
{\dpr V\over \dpr\psi^-_{\zz,\o}}\right]\;;\;
\sum_{\tilde\o}\d^{(1)} T_{\xx,\tilde\o,\h'}>^T \;,\Eq(8.46)$$
with
$$\d^{(1)} T_{\xx,\o,\h'} =\psi^{[h,0]+}_{\xx,\o} \d\psi^{[h,0]-}_{\xx,\o}+
\d\tilde\psi^{[h,0]+}_{\xx,\o,\h'} \d\psi^{[h,0]-}_{\xx,\o}+
\d\tilde\psi^{[h,0]+}_{\xx,\o,\h'}\psi^{[h,0]-}_{\xx,\o}\;,\Eq(8.47)$$
$$\d\psi^{[h,0]-}_{\xx,\o}={1\over L\b} \sum_{\kk:C_{h,0}^{-1}(\kk)>0}
e^{-i\kk\xx} C_{h,0}(\kk)(1-C_{h,0} ^{-1}(\kk))
\hat\psi^{[h,0],-}_{\kk,\o}\;,\Eq(8.48)$$
$$\d\tilde\psi^{[h,0]+}_{\xx,\o,\h'}={1\over L\b}
\sum_{\kk:C_{h,0}^{-1}(\kk)>0} e^{i\kk\xx} D_\o(\kk) C_{h,0}(\kk)
{[C^{-1}_{h,0}(\kk)-C^{-1}_{h,0}(\bar\pp_{\h'}
+\kk)]\over -i \bar p_{\h'0}}\hat\psi^{[h,0],+}_{\kk,\o}\;.\Eq(8.49)$$
 
Note that there is no divergence, in the limit $L,\b\to\io$, associated with
the fields $\d\psi^{[h,0]-}$ and $\d\tilde\psi^{[h,0]+}$, even if the function
$C_{h,0}(\kk)$ diverges on the boundary of the set $\{\kk:
C_{h,0}^{-1}(\kk)>0\}$. In fact, the integration of these fields on scale
$\bar h$, with $h\le \bar h\le 0$, yields a factor $\tilde f'_{\bar h}(\kk)$
proportional to $f_{\bar h}(\kk)$ (see ({\bf I}2.90) and the considerations after
\equ(8.36)), and the functions $f_{\bar h}(\kk)$ are non negative, if we
suitably choose the function ({\bf I}2.30); therefore $C_{h,0}(\kk) \tilde
f'_{\bar h}(\kk)$ is bounded.
 
Note also that, $[C^{-1}_{h,0}(\kk)-C^{-1}_{h,0}(\bar\pp_{\h'} +\kk)]/ -i \bar
p_{\h'0}$ is bounded, uniformly in $\b$, and is equal to $0$, at least if
$|\kk|$ belongs to the interval $[a_0\g^h+2\p/\b, a_0-2\p/\b]$ (see \S{\bf
I}2.3). However, the interval where this function vanishes can contain
the interval $[a_0\g^h, a_0]$, if
the function ({\bf I}2.30) is suitably chosen (by slightly broadening the regions
where it has to be equal to $1$ or $0$) and $\b$ is large enough, which is not
of course an important restriction (the real problem is the uniformity of the
bounds in the limit $\b\to\io$, and in any case the following arguments could
be easily generalized to cover the general case). Hence, it is easy to show
that
$$1-C_{h,0} ^{-1}(\kk)={C^{-1}_{h,0}(\kk)-C^{-1}_{h,0}(\bar\pp_{\h'}
+\kk)\over -i \bar p_{\h'0}}=0\;,\quad \hbox{if}\;
\tilde f'_{\bar h}(\kk)\not=0\;\quad h<\bar h<0\;,\Eq(8.50)$$
so that we can write
$$\d\psi^{[h,0]-}_{\xx,\o} = \d\psi^{(0)-}_{\xx,\o} +\d\psi^{(h)-}_{\xx,\o}\;,
\quad \d\tilde\psi^{[h,0]+}_{\xx,\o,\h'} = \d\tilde\psi^{(0)+}_{\xx,\o,\h'}+
\d\tilde\psi^{(h)+}_{\xx,\o,\h'}\;,\Eq(8.51)$$
where the fields $\d\psi^{(h')-}_{\xx,\o}$ and
$\d\tilde\psi^{(h')+}_{\xx,\o,\h'}$ are defined by substituting, in \equ(8.48)
and \equ(8.49), $\hat\psi^{[h,0],+}_{\kk,\o}$ with
$\hat\psi^{(h'),+}_{\kk,\o}$.
 
Let us now consider the functional
$$e^{\SS_{h,\h'}(\psi^{(<h)},J)}= \int P^{(L,h)}(d\psi^{[h,0]}) e^{-V^{(L)}
(\psi^{[h,0]}+\psi^{(<h)}) + \sum_{\tilde\o} \int d\xx J_\xx
\d^{(1)} T_{\xx,\tilde\o,\h'} }\;.\Eq(8.52)$$
We can write for $\SS_{h,\h'}(\psi^{(<h)},J)$ an expansion similar to that
used in \sec(6) to study the correlation function of the original model.
We introduce, for any $\tdh$ such that $h\le\tdh\le -1$, an effective
potential $\VV^{'(\tdh)}(\psi,\psi^{(<h)})$, defined as in \sec(8.3), and
two functionals $S^{'(\tdh+1)}(J)$, $\BB^{'(\tdh)}(\psi,\psi^{(<h)},J)$, so
that, by using the notation of \sec(8.4),
$$\eqalign{
e^{\SS_{h,\h'}(\psi^{(<h)},J)} &= e^{-L\b E'_\tdh+S^{'(\tdh+1)}(J)}
\int P_{Z'_\tdh,C_{h,\tdh}}(d\psi^{[h,\tdh]}) \;\cdot\cr
&\cdot\; e^{-\VV^{'(\tdh)}
(\sqrt{Z'_\tdh}\psi^{[h,\tdh]},\psi^{(<h)})+\BB^{'(\tdh)}
(\sqrt{Z'_\tdh}\psi^{[h,\tdh]},\psi^{(<h)},J)}\;.\cr}\Eq(8.53)$$
We introduce also the functionals $S^{'(h)}(J)$, $\VV^{'(h-1)}(\psi^{(<h)})$
and $\BB^{'(h-1)}(\psi^{(<h)},J)$, such that
$$\SS_{h,\h'}(\psi^{(<h)},J)= S^{'(h)}(J) -\VV^{'(h-1)}(\psi^{(<h)})+
\BB^{'(h-1)}(\psi^{(<h)},J)\;.\Eq(8.54)$$
 
We can write for $\BB^{'(h-1)}(\psi^{(<h)},J)$ a
representation similar to \equ(6.6), with $J$ in place of $\phi$
and $\psi^{(<h)}$ in place of $\psi^{(\le h)}$.
By \equ(8.46)
$$\D^{(1)}_{h,\o,\h'}(\xx;\yy,\zz)=B^{'(h-1)}_{1,2,(+,-),(\o,\o)}
(\xx;\yy,\zz)\;;\Eq(8.55)$$
hence, in order to prove \equ(8.40), we have to study the flow of the local
part of $\BB^{'(\tdh)} (Z^{'-1/2}_\tdh$ $\psi^{[h,\tdh]},\psi^{(<h)},J)$.
 
To start with, let us consider $\BB^{'(-1)} (Z^{'-1/2}_{-1} \psi^{[h,-1]},
\psi^{(<h)},J)$. By \equ(8.51), the graphs contributing to it may have an
external line of type $\d\psi$ or $\d\tilde\psi$ only if that line is of scale
$h$ and $h<-1$. Moreover, if the graph has an external line of this type and
it is not trivial, that is if it has more than one vertex, the corresponding
local part, defined as in \sec(6), is $0$, even if there are only two external
lines, because of the support properties of the propagators, since there is at
least one internal line with momentum equal to one of the external momenta,
which are of order $\b^{-1}$ for the local part. It follows that these graphs
do not participate in any manner to the flow of $\LL\BB^{'(-1)}
(Z^{'-1/2}_{-1} \psi^{[h,-1]}, \psi^{(<h)},J)$, up to the scale $h$; therefore
we modify the definition of $\LL$, so that they are not included.
 
This modification of the definition of $\LL$ allows to study the flow of
$\LL\BB^{'(\tdh)} (Z'_{\tdh-1} \psi^{[h,\tdh]},$ $\psi^{(<h)},J)$
essentially as in \sec(6), since, as we have
explained in \sec(8.3), the infrared cutoff has no influence on the other
local terms, except on the last scale, so that, if $h\le \tdh\le -1$,
$$\LL\BB^{'(\tdh)} (\sqrt{Z'_\tdh}\psi^{[h,\tdh]},\psi^{(<h)},J)=
\LL\BB^{(\tdh)} (\sqrt{Z_\tdh}(\psi^{[h,\tdh]}+\psi^{(<h)}),J)\;,\Eq(8.56)$$
where $\BB^{(\tdh)} (\sqrt{Z_\tdh}\psi^{(\le \tdh)},J)$ is the expression we
should get in absence of infrared cutoff and we used the fact, proved in
\sec(8.3), that $Z'_\tdh=Z_\tdh$. We can write
$$\LL\BB^{(\tdh)} (\sqrt{Z_\tdh}(\psi^{(\le \tdh)}),J)={Z^{(3)}_\tdh\over
Z_\tdh}\sum_{\o=\pm 1}\int d\xx J_\xx \psi^{(\le \tdh)+}_{\xx,\o}
\psi^{(\le \tdh)-}_{\xx,\o}\;.\Eq(8.57)$$
 
The flow of $Z^{(3)}_\tdh$ can be studied, starting from the scale $h=-1$, as
the flow of the renormalization constants $Z^{(2)}_\tdh$ related to the
analogous of the functional \equ(6.2) for the model defined by \equ(8.1a)
and \equ(8.1b), that is
$$e^{\SS(J)}=\int P^{(L)}(d\psi^{(\le 0)})
e^{-V^{(L)}(\psi^{(\le 0)})+\sum_{\o=\pm 1}\int d\xx J_\xx
\psi^{(\le 0)+}_{\xx,\o} \psi^{(\le 0)-}_{\xx,\o} }\;.\Eq(8.58)$$
 
Note that the values of $Z^{(3)}_{-1}$ and $Z^{(2)}_{-1}$ are very different;
in fact, the previous considerations imply that
$$|Z^{(2)}_{-1}-1|\le C|\l_0|\;.\quad |Z^{(3)}_{-1}| \le C|\l_0|\;.\Eq(8.59)$$
However, since the local part on scale $-1$ is of the same form and the
contribution of the non local terms on scale $-1$ to $Z^{(3)}_\tdh/
Z^{(3)}_{\tdh+1}$ or $Z^{(2)}_\tdh/ Z^{(2)}_{\tdh+1}$ is exponentially
depressed, as $\tdh$ decreases, it is easy to show, by using the arguments of
\sec(5.5)-\sec(5.8), that
$$Z^{(3)}_h= {Z^{(3)}_h\over Z^{(3)}_{-1}} Z^{(3)}_{-1}=
{Z^{(2)}_h\over Z^{(2)}_{-1}} [1+O(\l_0)] Z^{(3)}_{-1}\;.\Eq(8.60)$$
The integration of the fields of scale $h$ can only change this identity by
a factor $[1+O(\l_0)]$, hence \equ(8.59) and \equ(8.60) imply that
$$\left|{ Z^{(3)}_{h-1}\over Z^{(2)}_h }-1\right| \le C|\l_0|\;.\Eq(8.61)$$
 
If $\D^{(1)}_{h,\o,\h'}(\xx;\yy,\zz)$ were independent of $\h'$, $\d\tilde
Z^{(2)}_h$ would be exactly equal to $Z^{(3)}_{h-1}$ and \equ(8.40) would have
been proved. Since this is true only in the limit $\b\to\io$, we have to bound
$\hat\D^{(1)}_{h,\o,\h'}(\bar\pp_{\h'},\bar\kk_{\h,\h'})$ for each $\h,\h'$.
This means that we have to bound even the Fourier transform at momenta of
order $\b^{-1}$ of $\RR B^{'(h-1)}_{1,2,(+,-),(\o,\o)} (\xx;\yy,\zz)$, see
\equ(8.55). However, it is easy to see that we still get the bound \equ(8.40),
on the base of a simple dimensional argument (we skip the details, which
should be by now obvious). In fact, if we consider a term contributing to the
expansion of $\RR B^{'(h-1)}_{1,2,(+,-),(\o,\o)} (\xx;\yy,\zz)$ described in
\sec(6), whose external fields are affected by the regularization so that some
derivative acts on them, the corresponding bound differs from the bound of a
generic term contributing to $\LL B^{'(h-1)}_{1,2,(+,-),(\o,\o)}
(\xx;\yy,\zz)$ in the following way. One has to add a factor $\g^{-h_v}$, for
each ``zero'' produced by the regularization and, at the same time, a factor
$\b^{-1}$ produced by the corresponding derivative on the external momenta.
Since $\b^{-1}\g^{-h_v}\le 1$, we get the same result.

\pagina
\vskip1cm
\centerline{\titolo References}
\*
\halign{\hbox to 1.2truecm {[#]\hss} &
        \vtop{\advance\hsize by -1.25 truecm \0#}\cr
A& {I. Affleck: Field theory methods and quantum critical phenomena.
Proc. of Les Houches summer school on Critical phenomena, Random Systems,
Gauge theories, North Holland (1984). }\cr
 
B& {R.J. Baxter:
Eight-Vertex Model in Lattice Statistics.
{\it Phys. Rev. Lett.} {\bf 26}, 832--833 (1971). }\cr
BG& {G. Benfatto, G. Gallavotti:
Perturbation Theory of the Fermi Surface in Quantum Liquid. A General
Quasiparticle Formalism and One-Dimensional Systems.
{\it J. Stat. Phys.} {\bf 59}, 541--664 (1990). }\cr
BGM& {G. Benfatto, G. Gallavotti, V. Mastropietro:
Renormalization Group and the Fermi Surface in the Luttinger Model.
{\it Phys. Rev. B} {\bf 45}, 5468--5480 (1992). }\cr
BGPS& {G. Benfatto, G. Gallavotti, A. Procacci, B. Scoppola:
Beta Functions and Schwinger Functions for a Many Fermions System in One
Dimension.
{\it  Comm. Math. Phys.} {\bf  160}, 93--171 (1994). }\cr
BM1& {F. Bonetto, V. Mastropietro: Beta Function and Anomaly of the Fermi
Surface for a $d=1$ System of Interacting Fermions in a Periodic Potential.
{\it Comm. Math. Phys.} {\bf  172}, 57--93 (1995). }\cr
BM2& {F. Bonetto, V. Mastropietro:
Filled Band Fermi Systems.
{\it Mat. Phys. Elect. Journal} {\bf 2}, 1--43 (1996). }\cr
BeM1& {G.Benfatto, V. Mastropietro:
Renormalization group, hidden symmetries and approximate Ward identities
in the $XYZ$ model, I. {\it preprint} (2000). }\cr
EFIK& {F.Essler, H.Frahm, A. Izergin, V.Korepin:
Determinant representation for correlation functions of spin-${1\over 2}$
XXX and XXZ Heisenberg Magnets.
{\it Comm. Math. Phys.} {\bf  174}, 191--214 (1995). }\cr
GS& {G. Gentile, B. Scoppola:
Renormalization group and the ultraviolet
problem in the Luttinger model,
{\it Comm. Meth. Phys.} {\bf 154}, 153--179 (1993). }\cr
JKM& {J.D. Johnson, S. Krinsky, B.M.McCoy:
Vertical-Arrow Correlation Length in the Eight-Vertex Model and the
Low-Lying Excitations of the $XYZ$ Hamiltonian.
{\it Phys. Rev. A} {\bf 8}, 2526--2547 (1973). }\cr
LP& {A. Luther, I.Peschel: Calculation of critical exponents in two
dimensions from quantum field theory in one dimension.
{\it Phys. Rev. B} {\bf 12}, 9, 3908--3917 (1975). }\cr
Le& {A. Lesniewski:
Effective action for the Yukawa 2 quantum field Theory.
{\it Comm. Math. Phys.} {\bf 108}, 437-467 (1987). }\cr
LSM& {E. Lieb, T. Schultz, D. Mattis:
Two Soluble Models of an Antiferromagnetic Chain.
{\it Ann. of Phys.} {\bf 16}, 407--466 (1961). }\cr
LSM1& {E. Lieb, T. Schultz, D. Mattis:
Two-dimensional Ising model as a soluble problem of many fermions.
{\it Journ. Math. Phys.} Rev. Modern Phys. 36, 856--871 (1964). }\cr
M1& {V. Mastropietro: Small denominators and anomalous behaviour in
the Holstein-Hubbard model.
{\it Comm. Math. Phys} 201, 1, 81-115 (1999). }\cr
M2& {V. Mastropietro: Renormalization group for the XYZ model,
Renormalization group for the XYZ model.
{\it Letters in Mathematical physics}, 47, 339-352 (1999). }\cr
Mc& {B.M. McCoy:
Spin Correlation Functions of the $X-Y$ Model.
{\it Phys. Rev.} {\bf 173}, 531--541 (1968). }\cr
MD& {W.Metzner, C Di Castro:
Conservation laws and correlation functions in the Luttinger liquids.
{\it Phys. Rev. B} {\bf 47}, 16107--16123 (1993). }\cr
NO& {J.W. Negele, H. Orland:
Quantum many-particle systems.
Addison-Wesley, New York (1988). }\cr
S& {S.B. Suterland:
Two-Dimensional Hydrogen Bonded Crystals.
{\it J. Math. Phys.} {\bf 11}, 3183--3186 (1970). }\cr
Sp& {H.Spohn:
Bosonization, vicinal surfaces and Hydrodynamic fluctuation theory.
\hfill\break {\it cond-mat/9908381} (1999). }\cr
Spe& {T.Spencer: A mathematical approach to universality in
two dimensions. {\it preprint} (1999). }\cr
YY& {C.N. Yang, C.P. Yang:
One dimensional chain of anisotropic spin-spin interactions, I and II.
{\it Phys. Rev.} {\bf 150}, 321--339 (1966). }\cr
}
\bye